\documentclass[aps,prx,twocolumn,superscriptaddress]{revtex4-2}

\usepackage{graphicx,latexsym}
\usepackage{amsmath,amssymb,amsfonts}
\usepackage{bm}
\usepackage{braket}
\usepackage[%
 setpagesize=false,%
 bookmarks=false,%
 bookmarksdepth=tocdepth,%
 bookmarksnumbered=true,%
 colorlinks=true,%
 linkcolor=red,%
 citecolor=blue,%
 urlcolor=magenta%
]{hyperref}
\usepackage{color}

\newcommand{\up}{\uparrow}
\newcommand{\down}{\downarrow}

\begin{document}


\title{Exact eigenstates of multicomponent Hubbard models:\\ SU($N$) magnetic $\eta$ pairing, weak ergodicity breaking, and partial integrability}


\author{Masaya Nakagawa}
\email{nakagawa@cat.phys.s.u-tokyo.ac.jp}
\affiliation{Department of Physics, University of Tokyo, 7-3-1 Hongo, Bunkyo-ku, Tokyo 113-0033, Japan}
\author{Hosho Katsura}
\affiliation{Department of Physics, University of Tokyo, 7-3-1 Hongo, Bunkyo-ku, Tokyo 113-0033, Japan}
\affiliation{Institute for Physics of Intelligence, University of Tokyo, 7-3-1 Hongo, Bunkyo-ku, Tokyo 113-0033, Japan}
\affiliation{Trans-scale Quantum Science Institute, University of Tokyo, 7-3-1, Hongo, Tokyo 113-0033, Japan}
\author{Masahito Ueda}
\affiliation{Department of Physics, University of Tokyo, 7-3-1 Hongo, Bunkyo-ku, Tokyo 113-0033, Japan}
\affiliation{Institute for Physics of Intelligence, University of Tokyo, 7-3-1 Hongo, Bunkyo-ku, Tokyo 113-0033, Japan}
\affiliation{Trans-scale Quantum Science Institute, University of Tokyo, 7-3-1, Hongo, Tokyo 113-0033, Japan}
\affiliation{RIKEN Center for Emergent Matter Science (CEMS), Wako, Saitama 351-0198, Japan}


\date{\today}

\begin{abstract}
We construct exact eigenstates of multicomponent Hubbard models in arbitrary dimensions by generalizing the $\eta$-pairing mechanism. 
Our models include the SU($N$) Hubbard model as a special case. 
Unlike the conventional two-component case, the generalized $\eta$-pairing mechanism permits the construction of eigenstates that feature off-diagonal long-range order and magnetic long-range order. 
These states form fragmented fermionic condensates due to a simultaneous condensation of multicomponent $\eta$ pairs. 
While the $\eta$-pairing states in the SU(2) Hubbard model are based on the $\eta$-pairing symmetry, the exact eigenstates in the $N$-component system with $N\geq 3$ arise not from symmetry of the Hamiltonian but from a spectrum generating algebra defined in a restricted Hilbert space. 
We exploit this fact to show that the generalized $\eta$-pairing eigenstates do not satisfy the eigenstate thermalization hypothesis and serve as quantum many-body scar states. 
This result indicates a weak breakdown of ergodicity in the $N$-component Hubbard models for $N\geq 3$. 
Furthermore, we show that these exact eigenstates constitute integrable subsectors in which the Hubbard Hamiltonian effectively reduces to a non-interacting model. 
This partial integrability causes various multicomponent Hubbard models to weakly break ergodicity. 
We propose a method of harnessing dissipation to distill the integrable part of the dynamics and elucidate a mechanism of non-thermalization caused by dissipation. 
This work establishes the coexistence of off-diagonal long-range order and SU($N$) magnetism in excited eigenstates of the multicomponent Hubbard models, which presents a possibility of novel out-of-equilibrium pairing states of multicomponent fermions. 
These models unveil a unique feature of quantum thermalization of multicomponent fermions, which can experimentally be tested with cold-atom quantum simulators. 
\end{abstract}


\maketitle

\section{Introduction}
The Hubbard model \cite{Hubbard63, Kanamori63, Gutzwiller63} is a quintessential minimal model of interacting fermions. This model obeys two simple rules: spin-1/2 fermions hop between lattice sites and interact with each other via an on-site interaction. 
Despite the simplicity, this model exhibits a rich variety of emergent quantum many-body phenomena such as magnetism and superconductivity \cite{Arovas21, Qin21}. 
Understanding the properties of the Hubbard model provides insights into how order can emerge from the interplay among the Fermi-Dirac statistics, kinetics, and interactions. 

While the original Hubbard model concerns two-component (i.e., spin-$1/2$) fermions, its generalization to multicomponent systems has also been studied extensively \cite{Affleck88, Marston89}. 
The multicomponent Hubbard model (MHM) displays unique long-range order because of the large internal degrees of freedom. 
In particular, $N$-component systems may realize SU($N$) magnetism, in which the ordinary spin SU(2) group is generalized to higher-rank Lie groups \cite{Honerkamp04_1, Hermele09, Cazalilla09, Gorshkov10, Corboz11, Wang14, Capponi16}, and also diversify combination of Cooper pairs of fermions \cite{Modawi97, Ho99, Honerkamp04_2, Paananen06, Zhai07, Cherng07, Capponi08, Inaba09, Yip11, Inaba12, Guan13, Okanami14, Koga17, Rapp07, Rapp08, Ozawa10, Ohara11, Nishida12, Niemann12, Tajima19, Tajima21, Yoshida22}. 
Experimentally, quantum simulations with ultracold atoms in optical lattices provide a highly controllable platform for unveiling the physics of the Hubbard model \cite{Esslinger10, Bohrdt21}. 
Since the MHM has experimentally been realized with cold-atom quantum simulators \cite{Taie12, Hofrichter16,  Ozawa18, Taie21, Tusi21}, these systems will open a new avenue for exploration of quantum many-body physics of multicomponent fermions \cite{Cazalilla14} which is the primary objective of our study.

Although the Hubbard model involves only minimal terms of hopping and an on-site interaction, exact analytic solutions of this model are extremely hard to obtain except for the one-dimensional case \cite{LiebWu68, 1dHubbard_book}. Even in one dimension, the MHM with more than two components does not admit Bethe-ansatz solutions \cite{Choy82}. While some rigorous results in higher dimensions have been obtained for the two-component case \cite{Tasaki98, Tasaki98_2, Lieb04, Tasaki_book}, only a few exact results are known for multicomponent cases \cite{Katsura13, Bobrow18, Liu19, Tamura19, Tamura21, Yoshida21}.

It was pointed out by Yang that special exact eigenstates of the two-component Hubbard model can be constructed through an $\eta$-pairing mechanism \cite{Yang89}. 
These eigenstates, called $\eta$-pairing states, are intimately related to the $\eta$-pairing symmetry of the Hubbard Hamiltonian \cite{YangZhang90, Pernici90}. 
The $\eta$-pairing states are constructed through condensation of on-site fermion pairs with a special center-of-mass momentum, and exhibits off-diagonal long-range order (ODLRO). 
Although the $\eta$-pairing states are excited states and cannot be realized in thermal equilibrium except for limiting cases \cite{Singh91, Shen93}, the $\eta$ pairing has recently seen a resurgence of interest since a number of theoretical works suggested its realization in various non-equilibrium situations \cite{Rosch08, Kantian10, Kaneko19, Kaneko20, Ejima20, Werner19, Li20, Kitamura16, Peronaci20, Cook20, Tindall21, Tindall21_2, Diehl08, Kraus08, Bernier13, Buca19, Tindall19, Tsuji21, Nakagawa21, Murakami21}. 
Whereas pairing combinations in multicomponent systems are richer than those in the two-component case, generalizations of the $\eta$-pairing mechanism to the MHM have been limited \cite{Zhai05,Yoshida22}. In particular, the SU($N$) Hubbard model is excluded in the earlier constructions.

In this paper, we construct exact eigenstates of the MHM by generalizing the $\eta$-pairing mechanism to multicomponent fermions. 
Our models include the SU($N$) Hubbard model as a special case.
The generalized $\eta$-pairing states are formed through simultaneous condensation of $\eta$ pairs in different combinations enabled by large internal degrees of freedom. 
The generalized $\eta$-pairing mechanism has two salient features that are absent in the original Yang's result. 
First, the exact eigenstates of the MHM exhibit not only ODLRO but also SU($N$) magnetism due to the multicomponent nature of fermion pairs. 
The coexistence of ODLRO and magnetism is impossible for the $\eta$ pairing in spin-$1/2$ systems, in which on-site fermion pairs necessarily form spin singlets and cannot show magnetism. 
Second, the generalized $\eta$-pairing operators do not present any symmetry of the Hamiltonian but obey a spectrum generating algebra \cite{Mark20_2, Moudgalya20} defined on a Hilbert subspace. 
This algebraic property is related to non-thermalization and weak ergodicity breaking in the MHM, as explained below. 

Importantly, we find that not all $\eta$-pairing states in $N$-component systems become eigenstates of the MHM. 
We identify a family of $\eta$-pairing states that are eigenstates of the model. 
The key to the construction of eigenstates is to choose appropriate combinations of $\eta$ pairs so that the Pauli exclusion principle acts over different pairs to prohibit multiple occupancy. 
In this regard, the three-component system is special in that all combinations of $\eta$ pairs can constitute eigenstates. 

A simultaneous condensation of multicomponent $\eta$ pairs indicates that the generalized $\eta$-pairing states provide a novel realization of fragmented fermionic condensates \cite{Mueller06}, in which multiple Bose-Einstein condensates of fermion pairs coexist. 
Fragmented condensates are usually based on high symmetry of the system and hence fragile against symmetry-breaking perturbations \cite{Mueller06}. In contrast, fragmented $\eta$-paired condensates can be stable because the symmetry-breaking perturbation is negligible in realistic experimental situations unless introduced intentionally. We will point out that the MHM offers a unique system in which the SU($N$) magnetism emerges in robust fragmented fermionic condensates.

Yang's original $\eta$-pairing states have recently received a renewed interest from a perspective of the foundation of statistical mechanics \cite{Vafek17, Moudgalya20, Mark20, Pakrouski20, Pakrouski21}. 
Recent studies on the eigenstate thermalization hypothesis (ETH) \cite{Deutsch91, Srednicki94, Rigol08} have provided strong pieces of evidence that generic quantum many-body systems undergoing unitary dynamics show thermalization to states that are indistinguishable from the thermodynamic ensemble \cite{DAlessio16, Gogolin16, Mori18, Ueda20}. 
While the ETH is expected to hold in a wide range of non-integrable systems, notable exceptions that fail to thermalize have been found \cite{Abanin19, Serbyn21, Papic21, Moudgalya21}. 
In particular, the discoveries of quantum many-body scars and Hilbert space fragmentation reveal that even non-integrable systems may weakly break ergodicity, implying that some special initial states do not thermalize at all \cite{Shiraishi17, Bernien17, Moudgalya18_1, Turner18, Sala20, Khemani20, Serbyn21, Papic21, Moudgalya21}. 
As Yang's $\eta$-pairing states are excited eigenstates with seemingly athermal properties (such as sub-volume law entanglement entropy), it has been discussed whether or not these states violate the ETH \cite{Vafek17}. 
However, the existence of a conserved quantity due to the $\eta$-pairing symmetry prevents these states from becoming quantum many-body scar states unless the Hubbard model is perturbed by tailored symmetry-breaking terms that leave the $\eta$-pairing states as eigenstates \cite{Vafek17, Moudgalya20, Mark20}.

Building on the fact that the generalized $\eta$-pairing eigenstates do not rely on the $\eta$-pairing symmetry, we show that they form genuine quantum many-body scars of the MHM. 
This result indicates that the ergodicity is weakly broken in the $N$-component Hubbard model for $N\geq 3$. 
We also demonstrate that the generalized $\eta$-pairing states indeed show non-thermalizing dynamics, in which the population of each component oscillates persistently. 

Furthermore, we find that the Hilbert space of the MHM has an intriguing structure that may be regarded as a hybrid of integrable and non-integrable systems. In fact, the MHM has two families of exact eigenstates; the first one is composed of the generalized $\eta$-pairing states, and the second one includes SU($N$) ferromagnetic states. In each subspace spanned by those two types of eigenstates, the Hubbard interaction is constant, and therefore the eigenstates are the same as those of the non-interacting case, which is integrable. 
We refer to this structure of integrable subsectors in a non-integrable model as partial integrablitiy \cite{Sato95, Sato96}. 
The partial integrability implies that some initial states in the integrable subsector do not thermalize. Thus, the MHM can show two types of non-thermalizing dynamics, i.e., persistent oscillations of $\eta$-pairing states and integrable dynamics of effectively non-interacting particles. 

We show that partial integrability emerges in a wide class of multicomponent Hubbard-like models beyond the simple MHM. 
For example, we point out that various multiorbital Hubbard models discussed in literature \cite{Roth66, KugelKhomskii73, KugelKhomskii82, Koga04, Gorshkov10, Wu03, Wu06} weakly break ergordicity due to integrable subsectors. 
Moreover, exploiting this structure, we demonstrate that even interacting integrable models can be embedded into extended Hubbard-like models. 

The partial integrability ensures the existence of non-thermalizing initial states. However, this does not mean that such initial states are experimentally accessible. We employ the ideas of decoherence-free subspace \cite{Lidar98, Lidar03} to propose that the non-thermalizing dynamics due to partial integrability of the Hubbard-like models can be experimentally probed by coupling the system to an environment that induces special dissipative processes. We show that such dissipation brings the system into integrable subsectors and enables observation of non-thermalization for a broad class of initial states. 
We also show that such tailored dissipation is realizable with current experimental setups with ultracold alkaline-earth-like atoms.

\subsection{Summary of the results}
In this work, we introduce a family of generalized $\eta$-pairing states, whose general form is given in Eq.~\eqref{eq_SUNeta_ex} in Sec.~\ref{sec_N-color}. They are exact eigenstates of the MHM described by Eq.~\eqref{eq_Hasym} in Sec.~\ref{sec_model}. The mathematical structure underlying the construction of the exact eigenstates is a spectrum generating algebra discussed in Sec.~\ref{sec_SGA}. We also introduce primary $N$-color $\eta$-pairing states [Eq.~\eqref{eq_SUNeta} in Sec.~\ref{sec_N-color}] as a special case of the generalized $\eta$-pairing states that do not contain unpaired fermions. The simple form of these states allows us to study their various properties analytically. They exhibit ODLRO [Eq.~\eqref{eq_paircorr_SU(N)_norm} in Sec.~\ref{sec_ODLRO}] and long-range magnetic correlations [Eq.~\eqref{eq_spincorr_SU(N)_norm} in Sec.~\ref{sec_magnetism}], and exhibit fragmented condensates (Sec.~\ref{sec_fragmented}). We show that they serve as quantum many-body scar states of the MHM by calculating their sub-volume law entanglement entropy [Eq.~\eqref{eq_EE_SUNeta} in Sec.~\ref{sec_EE}] and non-thermalizing dynamics (persistent oscillations) of the number of doublons (Sec.~\ref{sec_oscil}).

We also consider a family of SU($N$) ferromagnetic states, whose general form is provided in Eq.~\eqref{eq_SUNferro_ex} in Sec.~\ref{sec_sector}. They are mapped to the generalized $\eta$-pairing states by a generalized Shiba transformation (Sec.~\ref{sec_Shiba}). The generalized $\eta$-pairing states and the SU($N$) ferromagnetic states constitute integrable subsectors of the MHM and its variants (Sec.~\ref{sec_sector}). Because of partial integrability of the MHM, initial states taken from the integrable subsectors do not thermalize (Sec.~\ref{sec_nontherm_integrability}). Those states belonging to the integrable subsectors can be prepared by using tailored dissipation, and therefore the non-thermalizing dynamics due to partial integrability can be induced by dissipation (Sec.~\ref{sec_diss}). In particular, the generalized $\eta$-pairing states (including the primary $N$-color $\eta$-pairing states) become dark states for such dissipation, which enables realization of the generalized $\eta$ pairing with dissipation engineering.

The rest of this paper is organized as follows. 
In Sec.~\ref{sec_model}, we introduce the multicomponent Hubbard model. 
In Sec.~\ref{sec_construction}, we show how to construct generalized $\eta$-pairing eigenstates and discuss algebraic structures that permit such construction. 
In Sec.~\ref{sec_magSF}, we show that the generalized $\eta$-pairing eigenstates exhibit ODLRO coexisting with SU($N$) magnetism. We also show that these states are fragmented condensates. 
The subsequent two sections are devoted to ergodicity breaking of the model.  
We show in Sec.~\ref{sec_scar} that the generalized $\eta$-pairing eigenstates are quantum many-body scars, which show persistent oscillations in their dynamics. 
In Sec.~\ref{sec_partial_integrability}, we elucidate the structure of partial integrability of the MHM and generalized Hubbard-like models, and exploit it to demonstrate non-thermalization in isolated and dissipative setups. 
Finally, we conclude this paper in Sec.~\ref{sec_conclusion}. 
Some detailed discussions and derivations are relegated to the Appendices. 
Appendix~\ref{sec_SUN_Hubbard} discusses exact eigenstates of the SU($N$) Hubbard model as a special case of our model. 
In Appendix~\ref{sec_corr_calc}, we present a detailed calculation of correlation functions for the exact eigenstates. 
A bound on ODLRO in multicomponent systems is discussed in Appendix~\ref{sec_ODLRO_bound}. 
In Appendix~\ref{sec_EE_calc}, we calculate entanglement entropy of the generalized $\eta$-pairing eigenstates. 
In Appendix~\ref{sec_scar_app}, we discuss a sufficient condition under which the generalized $\eta$-pairing eigenstates become quantum many-body scar states. 
In Appendix~\ref{sec_PAM}, we generalize the construction of exact eigenstates to other Hubbard-like models, which include a periodic Anderson model. 
A detailed calculation of momentum distributions related to non-thermalization due to partial integrability is given in Appendix~\ref{sec_mom_dist_integ}.

\section{Multicomponent Hubbard model\label{sec_model}}

We consider an $N$-component Hubbard model described by the following Hamiltonian:
\begin{align}
H&=T+V,\label{eq_Hasym}\\
T&=-\sum_{\langle i,j\rangle}\sum_{\alpha=1}^N t_{i,j}(c_{i,\alpha}^\dag c_{j,\alpha}+c_{j,\alpha}^\dag c_{i,\alpha}),\\
V&= \frac{1}{2}\sum_j\sum_{\alpha\neq\beta}U_{\alpha,\beta}n_{j,\alpha}n_{j,\beta},
\end{align}
where $c_{j,\alpha}^\dag$ and $c_{j,\alpha}$ denote the fermionic creation and annihilation operators of the $\alpha$th component at site $j$, and $n_{j,\alpha}\equiv c_{j,\alpha}^\dag c_{j,\alpha}$ is its number operator. 
The fermion operators obey the anticommutation relations $\{ c_{i,\alpha},c_{j,\beta}^\dag\}=\delta_{i,j}\delta_{\alpha,\beta}$ and $\{ c_{i,\alpha},c_{j,\beta}\}=0$. 
The fermions hop between nearest-neighbor sites $i,j$ with a tunneling amplitude $t_{i,j}\in\mathbb{R}$. 
For simplicity, we consider a $d$-dimensional hypercubic lattice with even number $N_{\mathrm{s}}=L^d$ of sites, but the following results can straightforwardly be generalized to any bipartite lattices.  
The interaction term $V$ represents an on-site density-density interaction between fermions in components $\alpha$ and $\beta$ with strength $U_{\alpha,\beta}$, and we set $U_{\beta,\alpha}=U_{\alpha,\beta}$ without loss of generality. 
The hopping (kinetic) term $T$ is assumed to be SU($N$) symmetric (i.e., $t_{i,j}$ does not depend on $\alpha$), while the interaction term $V$ does not necessarily have this symmetry. 
Throughout this paper, we set $\hbar=1$.

The MHM \eqref{eq_Hasym} has an internal symmetry $\mathrm{U}(1)^N$, which ensures the conservation of the particle number of each component. When the interaction strength is fine-tuned, the internal symmetry may be enhanced. 
For example, if the interaction strength is independent of components (i.e., $U_{\alpha,\beta}=U$), then the model \eqref{eq_Hasym} reduces to the SU($N$) Hubbard model \cite{Honerkamp04_1}. 
When the interaction strength can be written as
\begin{align}
U_{\alpha,\beta}=
\begin{cases}
U & (\alpha,\beta \leq M); \\
U^{\prime} & (M<\alpha,\beta); \\
U^{\prime\prime} & (\text{otherwise}),
\end{cases}
\end{align}
the model \eqref{eq_Hasym} has $\mathrm{SU}(M)\times\mathrm{SU}(N-M)$ symmetry for general values of $U,U^{\prime}$, and $U^{\prime\prime}$. 
Symmetries described by a product of several special unitary groups can similarly be realized. 
We discuss such fine-tuned cases with high internal symmetries in Appendix \ref{sec_SUN_Hubbard}.  
In addition, the model may have translational invariance or other spatial symmetries depending on the hopping amplitudes $t_{i,j}$, but we do not assume them.

The MHM \eqref{eq_Hasym} includes many important cases relevant to experiments. For example, the model \eqref{eq_Hasym} for $N=3$ can be realized with ultracold three-component $^{6}$Li atoms in an optical lattice, where the interaction parameters $U_{\alpha,\beta}$ can be tuned through a Feshbach resonance \cite{Bartenstein05, Ottenstein08, Huckans09, Williams09}. 
The SU($N$) Hubbard model has experimentally been realized with ultracold alkaline-earth-like atoms such as $^{173}$Yb for $N\leq 6$ and $^{87}$Sr for $N\leq 10$ \cite{Taie12, Hofrichter16, Ozawa18, Taie21, Tusi21}. 
The model with SU($M$)$\times$SU($N-M$) symmetry can be realized with a $^{171}$Yb-$^{173}$Yb mixture \cite{Taie10}, where $M=2$, $N-M=6$, and almost equal masses of the two isotopes effectively render the hopping amplitudes to be SU($N$) symmetric within experimental accuracy \cite{Sugawa11}. 
In typical experimental situations with ultracold atoms, a trap potential term is added to the model \eqref{eq_Hasym}. Although the absence of the trap potential term may be regarded as simplification for the construction of exact eigenstates, we note that recent experiments have realized the Hubbard model with an almost box-like potential, which can be well described by a spatially uniform model \cite{Sompet21}.

Since multicomponent systems are a generalization of two-component systems of spin-1/2 fermions, the number $N$ of different components of fermions may be regarded as ``spin'' degrees of freedom, which are useful in the context of the SU($N$) magnetism \cite{Honerkamp04_1, Hermele09, Cazalilla09, Gorshkov10, Corboz11, Wang14, Capponi16}. However, the different components may be labeled by different colors. The color degrees of freedom are often employed for the description of pairing states in multicomponent systems \cite{Honerkamp04_2, Paananen06, Zhai07, Cherng07, Capponi08, Inaba09, Yip11, Inaba12, Guan13, Okanami14, Koga17, Rapp07, Rapp08, Ozawa10, Ohara11, Nishida12, Niemann12, Tajima19, Tajima21, Yoshida22}. Therefore,  the terms ``spin,'' ``color,'' and ``component'' will be used interchangeably for convenience of explanation.

\section{Exact eigenstates and underlying algebraic structures\label{sec_construction}}

\subsection{$N$-color $\eta$ pairing\label{sec_N-color}}

To construct exact eigenstates of the MHM \eqref{eq_Hasym}, we generalize the $\eta$-pairing mechanism \cite{Yang89} to $N$-color fermion systems. First, we define generalized $\eta$ operators by
\begin{align}
\eta_{\alpha,\beta}^\dag&\equiv\sum_j e^{i\bm{Q}\cdot\bm{R}_j}c_{j,\alpha}^\dag c_{j,\beta}^\dag\ \ (\alpha,\beta=1,\cdots,N;\alpha\neq\beta)\notag\\
&=\sum_j \eta_{j,\alpha,\beta}^\dag,
\label{eq_eta_op}
\end{align}
where $\bm{R}_j\in\{1,\cdots,L\}^d$ denotes the coordinate of site $j$, the lattice constant is set to unity, and $\bm{Q}=(\pi,\pi,\cdots,\pi)$. Note that $e^{i\bm{Q}\cdot\bm{R}_j}$ is either $+1$ or $-1$. In the second line of Eq.~\eqref{eq_eta_op}, we have introduced a local $\eta$ operator by $\eta_{j,\alpha,\beta}^\dag\equiv e^{i\bm{Q}\cdot\bm{R}_j}c_{j,\alpha}^\dag c_{j,\beta}^\dag$. The operator $\eta_{\alpha,\beta}^\dag$ creates an on-site pair of two fermions with color $\alpha$ and $\beta$ \footnote{A different generalization of $\eta$ operators is considered in Ref.~\cite{Yoshida22}, in which a generalized $\eta$ pair is formed by $N$ particles. Our $\eta$ operator creates a pair of two particles and thus is different from the construction in Ref.~\cite{Yoshida22}.}. In the case of two-component systems, Eq.~\eqref{eq_eta_op} reduces to the $\eta$ operator introduced by Yang \cite{Yang89}.

To diagnose whether the MHM possesses an $\eta$-pairing symmetry, we calculate the commutation relations between the generalized $\eta$ operators and each term of the Hamiltonian. 
The commutation relation between the generalized $\eta$ operators and the kinetic term of the Hamiltonian \eqref{eq_Hasym} is calculated as
\begin{align}
[\eta_{\alpha,\beta}^\dag,T]=&-\sum_{\langle i,j\rangle}\sum_\gamma t_{i,j}[\eta_{i,\alpha,\beta}^\dag+\eta_{j,\alpha,\beta}^\dag,c_{i,\gamma}^\dag c_{j,\gamma}+c_{j,\gamma}^\dag c_{i,\gamma}]\notag\\
=&-\sum_{\langle i,j\rangle}t_{i,j}(e^{i\bm{Q}\cdot\bm{R}_i}c_{j,\beta}^\dag c_{i,\alpha}^\dag-e^{i\bm{Q}\cdot\bm{R}_i}c_{j,\alpha}^\dag c_{i,\beta}^\dag\notag\\
&+e^{i\bm{Q}\cdot\bm{R}_j}c_{i,\beta}^\dag c_{j,\alpha}^\dag-e^{i\bm{Q}\cdot\bm{R}_j}c_{i,\alpha}^\dag c_{j,\beta}^\dag)\notag\\
=&0,
\label{eq_comm_etaT}
\end{align}
since $e^{i\bm{Q}\cdot\bm{R}_j}=-e^{i\bm{Q}\cdot\bm{R}_i}$ holds for any pair of nearest-neighbor sites $i,j$. 
In contrast, the generalized $\eta$ operators do not commute with the interaction term:
\begin{align}
[\eta_{\alpha,\beta}^\dag,V]=&\frac{1}{2}\sum_j\sum_{\gamma\neq\delta}U_{\gamma,\delta}e^{i\bm{Q}\cdot\bm{R}_j}\notag\\
&\times(c_{j,\gamma}^\dag c_{j,\alpha}^\dag n_{j,\delta}\delta_{\beta,\gamma}-c_{j,\gamma}^\dag c_{j,\beta}^\dag n_{j,\delta}\delta_{\alpha,\gamma}\notag\\
&+n_{j,\gamma}c_{j,\delta}^\dag c_{j,\alpha}^\dag\delta_{\beta,\delta}-n_{j,\gamma}c_{j,\delta}^\dag c_{j,\beta}^\dag \delta_{\alpha,\delta})\notag\\
=&-U_{\alpha,\beta}\eta_{\alpha,\beta}^\dag+R_{\alpha,\beta},
\label{eq_comm_etaV}
\end{align}
where
\begin{equation}
R_{\alpha,\beta}\equiv -\sum_j\sum_{\gamma(\neq\alpha,\beta)}(U_{\alpha,\gamma}+U_{\beta,\gamma})e^{i\bm{Q}\cdot\bm{R}_j}c_{j,\alpha}^\dag c_{j,\beta}^\dag n_{j,\gamma}.
\end{equation}
Equations \eqref{eq_comm_etaT} and \eqref{eq_comm_etaV} are combined to give
\begin{equation}
[\eta_{\alpha,\beta}^\dag,H]=-U_{\alpha,\beta}\eta_{\alpha,\beta}^\dag+R_{\alpha,\beta}.
\label{eq_comm_etaH}
\end{equation}
In the two-component case, $R_{1,2}=0$ and Eq.~\eqref{eq_comm_etaH} gives an $\eta$-pairing symmetry of the Hubbard model if the chemical potential is appropriately tuned \cite{YangZhang90}. When $R_{\alpha,\beta}=0$, the relation \eqref{eq_comm_etaH} shows the presence of a dynamical symmetry \cite{Buca19}. 
In contrast to the two-component case, the $N$-component Hubbard model \eqref{eq_Hasym} for $N\geq 3$ does not possess a (dynamical) $\eta$-pairing symmetry due to the residual term $R_{\alpha,\beta}$, except for the special case with $U_{\alpha,\gamma}+U_{\beta,\gamma}=0$ for all $\gamma\neq\alpha,\beta$ \footnote{We note that the generalized $\eta$ operators can give a (dynamical) symmetry of the MHM if sites occupied by more than two particles are prohibited \cite{Ying01}. However, we do not impose such restriction on the model.}. 

Because of the absence of the $\eta$-pairing symmetry, $N$-color $\eta$-pairing states created by the generalized $\eta$ operators are not eigenstates of the Hamiltonian \eqref{eq_Hasym} in general. Here, as a special family of $N$-color $\eta$-pairing states, we introduce \textit{primary $N$-color} $\eta$\textit{-pairing states} by
\begin{equation}
\ket{\psi_{M_2,M_3,\cdots,M_N}}\equiv(\eta_{2,1}^\dag)^{M_2}(\eta_{3,1}^\dag)^{M_3}\cdots(\eta_{N,1}^\dag)^{M_N}\ket{0},
\label{eq_SUNeta}
\end{equation}
where $\ket{0}$ is the vacuum state of fermions and $M_2,M_3,\cdots,M_N$ are non-negative integers. We assume $M_2+M_3+\cdots+M_N\leq N_{\mathrm{s}}$ so that the state \eqref{eq_SUNeta} does not vanish. 
The total number of fermions contained in the state $\ket{\psi_{M_2,M_3,\cdots,M_N}}$ is $N_{\mathrm{f}}=2M_2+2M_3+\cdots+2M_N$. 
These states reduce to Yang's $\eta$-pairing states \cite{Yang89} for $N=2$. 

The $\eta$ pairs that appear in the state \eqref{eq_SUNeta}, created by $\eta_{\alpha,1}^\dag\ (\alpha=2,\cdots,N)$, are such that they contain fermions of the same color. While this special color shared by all the $\eta$ pairs may be any one from $1$ to $N$, here we choose the color 1 without loss of generality.

We now show that the primary $N$-color $\eta$-pairing states \eqref{eq_SUNeta} are exact eigenstates of the Hamiltonian \eqref{eq_Hasym}:
\begin{equation}
H\ket{\psi_{M_2,\cdots,M_N}}=(M_2U_{2,1}+\cdots+M_NU_{N,1})\ket{\psi_{M_2,\cdots,M_N}}.
\label{eq_SUNeta_energy}
\end{equation}
The proof is done separately for the kinetic term and the interaction term. 
First, the commutation relation \eqref{eq_comm_etaT} ensures that the $\eta$-pairing states \eqref{eq_SUNeta} are eigenstates of the kinetic term, i.e., $T\ket{\psi_{M_2,\cdots,M_N}}=0$. 
To prove that those states are also eigenstates of the interaction term $V$, we rewrite them using the Fock states as
\begin{align}
&\ket{\psi_{M_2,\cdots,M_N}}\notag\\
=&\sum_{j_1,\cdots,j_{N_{\mathrm{f}}/2}}\eta_{j_1,2,1}^\dag\cdots\eta_{j_{M_2},2,1}^\dag\eta_{j_{M_2+1},3,1}^\dag\cdots\eta_{j_{M_2+M_3},3,1}^\dag\cdots\notag\\
&\times\eta_{j_{N_{\mathrm{f}}/2-M_N+1},N,1}^\dag\cdots\eta_{j_{N_{\mathrm{f}}/2},N,1}^\dag\ket{0},
\label{eq_SUNeta_siterep}
\end{align}
and use the following property of the primary $N$-color $\eta$-pairing states: they only contain doubly occupied sites or empty sites since $\eta_{j,\alpha,1}^\dag \eta_{j,\beta,1}^\dag=0$ for all $\alpha,\beta$. Therefore, no two site indices in the sum in Eq.~\eqref{eq_SUNeta_siterep} can be the same. Therefore, each term on the right-hand side of Eq.~\eqref{eq_SUNeta_siterep} is an eigenstate of the interaction term and its eigenvalue is given by the sum of the numbers of $\eta$ pairs multiplied by their respective interaction coefficients. 
Thus, we obtain Eq.~\eqref{eq_SUNeta_energy}.

We note that the residual term $R_{\alpha,\beta}$ in the commutation relation \eqref{eq_comm_etaH} vanishes when acting on the primary $N$-color $\eta$-pairing states:
\begin{equation}
R_{\alpha,1}\ket{\psi_{M_2,\cdots,M_N}}=0\ \ (\alpha=2,\cdots,N).
\label{eq_eta_res}
\end{equation}
Equation \eqref{eq_SUNeta_energy} is consistent with this property; 
in fact, from Eq.~\eqref{eq_eta_res} and $H\ket{0}=0$, Eq.~\eqref{eq_SUNeta_energy} is obtained by repeatedly using the commutation relation \eqref{eq_comm_etaH}.

\begin{figure}
\includegraphics[width=8.6cm]{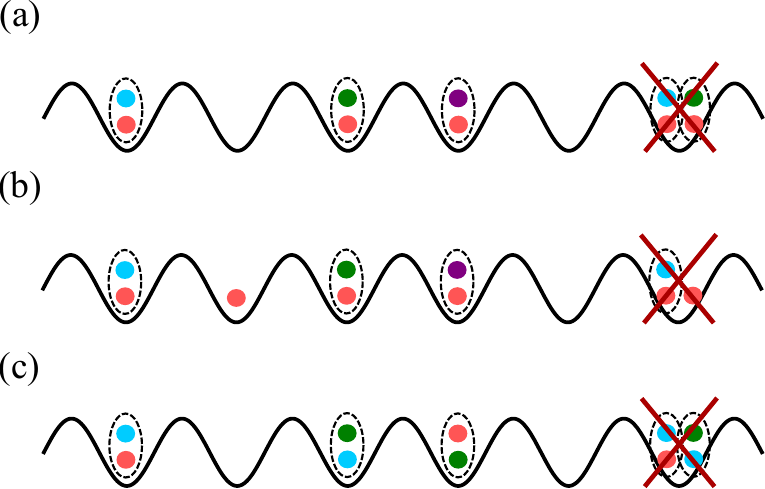}
\caption{Schematic illustrations of (a) the primary $N$-color $\eta$-pairing state [Eq.~\eqref{eq_SUNeta}], (b) the generalized $N$-color $\eta$-pairing state [Eq.~\eqref{eq_SUNeta_ex}], and (c) the three-color $\eta$-pairing state [Eq.~\eqref{eq_SU3eta}]. Lattice potentials are shown by black curves. Dots with different colors represent fermions in different components. As shown in the rightmost sites in panels (a) and (c), any two pairs (shown by dashed ellipses) cannot occupy the same site because of the Pauli exclusion between fermions of the same color. In panel (b), no pair can occupy a site with another pair or an unpaired fermion of red color which is the common color for every pair.}
\label{fig_schematic}
\end{figure}

Physically, the Pauli exclusion principle plays a key role in the $\eta$-pairing mechanism in the MHM. In fact, since any $\eta$ pairs in the primary $N$-color $\eta$-pairing states \eqref{eq_SUNeta} contain a color-1 fermion, the Pauli exclusion principle dictates that multiple $\eta$ pairs cannot occupy the same site [see Fig.~\ref{fig_schematic} (a) for a schematic illustration]. The absence of multiple $\eta$ pairs at the same site is the reason why the primary $\eta$-pairing states become eigenstates of the interaction term.

We can also show that the following state is an eigenstate of the Hamiltonian \eqref{eq_Hasym}:
\begin{align}
(\eta_{2,1}^\dag)^{M_2}(\eta_{3,1}^\dag)^{M_3}\cdots(\eta_{N,1}^\dag)^{M_N}c_{n_1,1}^\dag c_{n_2,1}^\dag \cdots c_{n_r,1}^\dag\ket{0},
\label{eq_SUNeta_ex}
\end{align}
where
\begin{equation}
c_{n,\alpha}^\dag \equiv\sum_j v_j^{(n)}c_{j,\alpha}^\dag
\end{equation}
is a creation operator of a single-particle eigenstate of the kinetic term $T$, for which the coefficients $v_j^{(n)}$ are given by a solution of an eigenvalue equation
\begin{equation}
\sum_jt_{i,j}v_j^{(n)}=\epsilon_{n}v_i^{(n)},
\end{equation}
with the normalization condition $\sum_j|v_j^{(n)}|^2=1$. The state \eqref{eq_SUNeta_ex} contains $r$ unpaired fermions in addition to $\eta$ pairs. Note that the primary $N$-color $\eta$-pairing state \eqref{eq_SUNeta} is a special case of the generalized $\eta$-pairing state \eqref{eq_SUNeta_ex} with no unpaired fermions (i.e., $r=0$). To show that the state \eqref{eq_SUNeta_ex} is indeed an eigenstate, we first note that a state $c_{n_1,1}^\dag c_{n_2,1}^\dag \cdots c_{n_r,1}^\dag\ket{0}$ is an eigenstate of the kinetic term with eigenvalue $\epsilon_{n_1}+\epsilon_{n_2}+\cdots+\epsilon_{n_r}$. Then, it follows from the commutation relation \eqref{eq_comm_etaT} that the state \eqref{eq_SUNeta_ex} is also an eigenstate of the kinetic term. To show that this state is also an eigenstate of the interaction term, we expand the state using the real-space Fock basis as in Eq.~\eqref{eq_SUNeta_siterep}. Since $\eta_{j,\alpha,1}^\dag \eta_{j,\beta,1}^\dag=0$ and $\eta_{j,\alpha,1}^\dag c_{j,1}^\dag=0$, the state \eqref{eq_SUNeta_ex} cannot contain those sites that are occupied by more than two particles and is therefore an eigenstate of the interaction term [see also Fig.~\ref{fig_schematic} (b)]. Thus, the state \eqref{eq_SUNeta_ex} is an eigenstate of the Hamiltonian \eqref{eq_Hasym} with eigenvalue $\epsilon_{n_1}+\cdots+\epsilon_{n_r}+M_2U_{2,1}+\cdots+M_NU_{N,1}$.

Notably, the three-component case ($N=3$) is distinct from the case with $N\geq 4$. In this case, in addition to the state \eqref{eq_SUNeta_ex} with $N=3$, \textit{arbitrary three-color} $\eta$\textit{-pairing states}
\begin{equation}
\ket{\psi_{l,m,n}^{(3)}}\equiv(\eta_{1,2}^\dag)^{l}(\eta_{2,3}^\dag)^{m}(\eta_{3,1}^\dag)^{n}\ket{0}
\label{eq_SU3eta}
\end{equation}
are also eigenstates of the three-component Hubbard model:
\begin{align}
H\ket{\psi_{l,m,n}^{(3)}}=(lU_{1,2}+mU_{2,3}+nU_{3,1})\ket{\psi_{l,m,n}^{(3)}}.
\label{eq_SU3eta_eigen}
\end{align}
Here, $l,m,n$ are non-negative integers that satisfy $l+m+n\leq N_{\mathrm{s}}$. 
The key feature of the three-component system is that the Pauli exclusion principle works for arbitrary two $\eta$ pairs, i.e., $\eta_{j,\alpha,\beta}^\dag\eta_{j,\gamma,\delta}^\dag=0$, where $(\alpha,\beta)$ and $(\gamma,\delta)$ are taken from $(1,2)$, $(2,3)$, and $(3,1)$. Then, the proof of Eq.~\eqref{eq_SU3eta_eigen} can be done in a manner similar to that for the state \eqref{eq_SUNeta}. 
However, we note that one cannot add unpaired fermions to the state \eqref{eq_SU3eta} unlike the case of Eq.~\eqref{eq_SUNeta_ex}, because $\eta_{j,\alpha,\beta}^\dag c_{j,\gamma}^\dag\neq 0$ for $\gamma\neq\alpha,\beta$.

If the model has additional symmetries such as SU($N$), then a broader class of $N$-color $\eta$-pairing states becomes eigenstates of the Hamiltonian as shown in Appendix~\ref{sec_SUN_Hubbard}.

\subsection{SO($2N$) symmetry of the kinetic term\label{sec_SO2N}}

As shown in Eq.~\eqref{eq_comm_etaT}, the generalized $\eta$ operators commute with the kinetic term, implying that the symmetry of the kinetic term is larger than that of the total Hamiltonian. Here, we discuss the symmetry of the kinetic term, which can be used to understand the algebraic structure underlying the $N$-color $\eta$ pairing. 
The generalized $\eta$ operators satisfy the commutation relation
\begin{align}
[\eta_{\alpha,\beta}^\dag,\eta_{\gamma,\delta}]
=&\delta_{\alpha,\gamma}F_{\beta,\delta}+\delta_{\beta,\delta}F_{\alpha,\gamma}-\delta_{\alpha,\delta}F_{\beta,\gamma}-\delta_{\beta,\gamma}F_{\alpha,\delta}\notag\\
&-N_{\mathrm{s}}(\delta_{\alpha,\gamma}\delta_{\beta,\delta}-\delta_{\alpha,\delta}\delta_{\beta,\gamma}),
\label{eq_etaalg}
\end{align}
where
\begin{equation}
F_{\alpha,\beta}\equiv\sum_j c_{j,\alpha}^\dag c_{j,\beta}\ \ (\alpha,\beta=1,\cdots,N).
\label{eq_SUN_generator}
\end{equation}
The operators $F_{\alpha,\beta}$ are the generalization of the ordinary spin operators to the SU($N$) group and therefore will be referred to as SU($N$) spin operators or simply spin operators in the following discussions \cite{Cazalilla14}. 
The SU($N$) spin operators satisfy the commutation relation
\begin{align}
[F_{\alpha,\beta},F_{\gamma,\delta}]
&=\delta_{\beta,\gamma}F_{\alpha,\delta}-\delta_{\alpha,\delta}F_{\gamma,\beta}.
\label{eq_SUNalg}
\end{align}
The commutation relation between the SU($N$) spin operators and the $\eta$ operators is given by
\begin{align}
[\eta_{\alpha,\beta}^\dag,F_{\gamma,\delta}]
=&\delta_{\alpha,\delta}\eta_{\beta,\gamma}^\dag-\delta_{\beta,\delta}\eta_{\alpha,\gamma}^\dag.
\label{eq_etaFalg}
\end{align}
The kinetic term $T$ commutes with the SU($N$) spin operators: 
\begin{gather}
[F_{\alpha,\beta},T]=0.
\label{eq_comm_FT}
\end{gather}
Since $T$ also commutes with the generalized $\eta$ operators, the symmetry of the kinetic term is, in fact, larger than SU($N$).

The symmetry of the kinetic term is made manifest when one employs a Majorana representation of the operators \cite{Yang91, Yoshida21}. We introduce Majorana operators by
\begin{gather}
\gamma_{j,2\alpha-1}=c_{j,\alpha}+c_{j,\alpha}^\dag,\ \gamma_{j,2\alpha}=i(c_{j,\alpha}-c_{j,\alpha}^\dag)
\end{gather}
for sites that satisfy $e^{i\bm{Q}\cdot\bm{R}_j}=1$, and
\begin{gather}
\gamma_{j,2\alpha-1}=i(c_{j,\alpha}^\dag -c_{j,\alpha}),\ \gamma_{j,2\alpha}=c_{j,\alpha}+c_{j,\alpha}^\dag
\end{gather}
for sites that satisfy $e^{i\bm{Q}\cdot\bm{R}_j}=-1$. The Majorana operators are Hermitian, i.e., $\gamma_{j,n}^\dag =\gamma_{j,n}$, and obey the anticommutation relation
\begin{equation}
\{\gamma_{i,m},\gamma_{j,n}\}=2\delta_{i,j}\delta_{m,n}.
\label{eq_Clifford}
\end{equation}
In terms of the Majorana operators, the kinetic term can be expressed as
\begin{align}
T=&-\frac{i}{2}\sum_{\langle i,j\rangle}\sum_{n=1}^{2N}t_{i,j}\gamma_{i,n}\gamma_{j,n},
\end{align}
which is manifestly invariant under the linear transformation of the Majorana operators
\begin{equation}
\gamma_{i,n}\mapsto\gamma_{i,n}'=\sum_{m=1}^{2N}O_{n,m}\gamma_{i,m},\ O=\{ O_{n,m}\}\in \mathrm{SO}(2N).
\end{equation}
Therefore, the kinetic term is SO($2N$) symmetric. The generators of the so($2N$) Lie algebra are given by \cite{Georgi_eng}
\begin{equation}
M_{m,n}=-\frac{i}{4}\sum_j[\gamma_{j,m},\gamma_{j,n}].
\end{equation}

The generators of the so($2N$) algebra can be expressed in terms of the spin operators and the $\eta$ operators as
\begin{subequations}
\begin{align}
\frac{i}{4}\sum_j\gamma_{j,2\alpha-1}\gamma_{j,2\beta-1}=&-\frac{1}{4i}(F_{\alpha,\beta}-F_{\beta,\alpha}+\eta_{\alpha,\beta}^\dag-\eta_{\alpha,\beta}),\label{eq_Maj_etaF1}\\
\frac{i}{4}\sum_j\gamma_{j,2\alpha}\gamma_{j,2\beta}=&-\frac{1}{4i}(F_{\alpha,\beta}-F_{\beta,\alpha}-\eta_{\alpha,\beta}^\dag+\eta_{\alpha,\beta}),\\
\frac{i}{4}\sum_j\gamma_{j,2\alpha-1}\gamma_{j,2\beta}=&\frac{1}{4}(-F_{\alpha,\beta}-F_{\beta,\alpha}+\eta_{\alpha,\beta}^\dag+\eta_{\alpha,\beta}),\\
\frac{i}{4}\sum_j\gamma_{j,2\alpha}\gamma_{j,2\beta-1}=&\frac{1}{4}(F_{\alpha,\beta}+F_{\beta,\alpha}+\eta_{\alpha,\beta}^\dag+\eta_{\alpha,\beta}),\\
\frac{i}{4}\sum_j\gamma_{j,2\alpha-1}\gamma_{j,2\alpha}=&\frac{1}{4}N_{\mathrm{s}}-\frac{1}{2}F_{\alpha,\alpha},\label{eq_Maj_etaF4}
\end{align}
\end{subequations}
where $\beta\neq\alpha$. 
Thus, the symmetries implied by the commutation relations \eqref{eq_comm_etaT} and \eqref{eq_comm_FT} are included in the SO($2N$) symmetry of the kinetic term.
The commutation relations \eqref{eq_etaalg}, \eqref{eq_SUNalg}, and \eqref{eq_etaFalg} are integrated into the so($2N$) algebra of the Majorana operators through Eqs.~\eqref{eq_Maj_etaF1}-\eqref{eq_Maj_etaF4}.

The interaction term $V$ is not SO($2N$) symmetric as inferred from the commutation relation \eqref{eq_comm_etaV}. In terms of the Majorana representation, we can express an SO($2N$)-symmetric interaction as
\begin{align}
V_{\mathrm{SO}(2N)}=&\left(\frac{i}{2}\right)^N U\sum_j\gamma_{j,1}\gamma_{j,2}\cdots\gamma_{j,2N}\notag\\
=&U\sum_j\left(n_{j,1}-\frac{1}{2}\right)\cdots\left(n_{j,N}-\frac{1}{2}\right),
\label{eq_Nbody_int}
\end{align}
which involves $k$-body interactions with $k=2,\cdots,N$. For $N=2$, Eq.~\eqref{eq_Nbody_int} reduces to the ordinary Hubbard interaction up to a constant shift of the chemical potential, leading to the SO(4) symmetry of the two-component Hubbard model \cite{Yang91}. 
However, for $N\geq 3$, Eq.~\eqref{eq_Nbody_int} is not equivalent to the standard two-body Hubbard interaction. This highlights the distinction between the $N=2$ and $N\geq 3$ Hubbard models, in the latter of which the $\eta$-pairing symmetry is naturally broken by the interaction.

Because of the SO($2N$) symmetry of the kinetic term, arbitrary $N$-color $\eta$-pairing states (which are created by the generalized $\eta$ operators acting on the vacuum state) are eigenstates of the non-interacting ($V=0$) Hamiltonian. Among those states, the primary $N$-color $\eta$-pairing states \eqref{eq_SUNeta} and the three-color $\eta$-pairing states \eqref{eq_SU3eta} are left to be eigenstates when the interaction term $V$ is added to the free-fermion Hamiltonian. 
This structure is reminiscent of the construction of quantum many-body scar states discussed in literature \cite{Serbyn21, Papic21, Moudgalya21, Moudgalya18_1, Moudgalya18_2, Moudgalya20, Mark20, Schecter19, Shibata20, Mark20_2, ODea20, Bull20, Ren21, Pakrouski20, Pakrouski21}. 
For example, in Refs.~\cite{Moudgalya20, Mark20}, it was shown that the $\eta$-pairing symmetry of the two-component Hubbard model can be broken by adding some tailored perturbations to the Hamiltonian so that Yang's $\eta$-pairing states remain to be eigenstates, which can then be regarded as quantum many-body scar states. 
Similar constructions apply to many other quantum many-body scar states \cite{Schecter19, Shibata20, Mark20_2, ODea20, Bull20, Ren21, Pakrouski20, Pakrouski21}. 
In our case, the generalized $\eta$-pairing symmetry is naturally broken by the Hubbard interaction between multicomponent fermions, while some of the $N$-color $\eta$-pairing states remain to be eigenstates.  
In fact, the generalized $\eta$-pairing states can be considered as quantum many-body scar states of the MHM. 
This point will be discussed in Sec.~\ref{sec_scar}.

\subsection{Spectrum generating algebra\label{sec_SGA}}

Although the generalized $\eta$ operators do not give a (dynamical) symmetry of the MHM, they obey a special type of closed algebraic relations when their action is restricted to a certain Hilbert subspace. 
To see this, let $\mathcal{W}$ be a Hilbert subspace spanned by a basis set
\begin{equation}
\Bigl\{\prod_{j=1}^{N_{\mathrm{s}}}o_j\ket{0}|o_j\in \mathcal{A}_j;j=1,\cdots,N_{\mathrm{s}}\Bigr\},
\label{eq_basisset}
\end{equation}
where 
\begin{align}
\mathcal{A}_j=&\{I,c_{j,1}^\dag,c_{j,2}^\dag c_{j,1}^\dag,\cdots,c_{j,N}^\dag c_{j,1}^\dag\},
\label{eq_subsp_basis}
\end{align}
and $I$ denotes the identity operator. 
Then, the generalized $\eta$ operators satisfy
\begin{equation}
([\eta_{\alpha,1}^\dag,H]+U_{\alpha,1}\eta_{\alpha,1}^\dag)\mathcal{W}=0\ (\alpha=2,\cdots,N),
\label{eq_etaSGA}
\end{equation}
since $R_{\alpha,1}\mathcal{W}=0$. 
The relation \eqref{eq_etaSGA} has the form of a (restricted) spectrum generating algebra \cite{Mark20_2, Moudgalya20} introduced in the studies of quantum many-body scars \footnote{Similar algebraic structures are discussed in Refs.~\cite{Batista09, Wouters18}.}. 
The spectrum generating algebra \eqref{eq_etaSGA} provides a clear understanding of the underlying algebraic structure behind the generalized $\eta$-pairing states. 
In fact, from Eq.~\eqref{eq_etaSGA}, if a state $\ket{\psi_0}$ in the subspace $\mathcal{W}$ is an eigenstate of the MHM with eigenenergy $E_0$, then a state $\eta_{\alpha,1}^\dag\ket{\psi_0}$ is also an eigenstate with eigenenergy $E_0+U_{\alpha,1}$. By repeating this argument with $\ket{\psi_0}=\ket{0}$, one can see that the primary $N$-color $\eta$-pairing states \eqref{eq_SUNeta} are eigenstates of the MHM. Also, by taking
\begin{equation}
\ket{\psi_0}=c_{n_1,1}^\dag c_{n_2,1}^\dag \cdots c_{n_r,1}^\dag\ket{0},
\end{equation}
one can obtain more general eigenstates shown in Eq.~\eqref{eq_SUNeta_ex}.

In the three-component case, the relation
\begin{equation}
([\eta_{\alpha,\beta}^\dag,H]+U_{\alpha,\beta}\eta_{\alpha,\beta}^\dag)\mathcal{W}_3=0\ [(\alpha,\beta)=(1,2),(2,3),(3,1)]
\label{eq_etaSGA_SU(3)}
\end{equation}
holds for a Hilbert subspace $\mathcal{W}_3$ that is spanned by a basis set
\begin{equation}
\Bigl\{\prod_{j=1}^{N_{\mathrm{s}}}o_j\ket{0}|o_j\in \mathcal{A}_j^{(3)};j=1,\cdots,N_{\mathrm{s}}\Bigr\}
\label{eq_basisset_3}
\end{equation}
with
\begin{equation}
\mathcal{A}_j^{(3)}=\{I,c_{j,1}^\dag c_{j,2}^\dag,c_{j,2}^\dag c_{j,3}^\dag,c_{j,3}^\dag c_{j,1}^\dag\}.
\label{eq_subsp_basis_SU(3)}
\end{equation}
The spectrum generating algebra \eqref{eq_etaSGA_SU(3)} ensures that the three-color $\eta$-pairing states \eqref{eq_SU3eta} are exact eigenstates of the three-component Hubbard model.

\subsection{Duality between primary $N$-color $\eta$-pairing states and SU($N$) ferromagnetic states\label{sec_Shiba}}

In the two-component case, the $\eta$-pairing symmetry is closely related to the duality between the repulsive and attractive Hubbard models by the Shiba transformation \cite{LiebWu68, Shiba72, YangZhang90, 1dHubbard_book}. 
In the multicomponent case, we define generalized Shiba transformations by
\begin{equation}
S_{\alpha}c_{j,\beta}S_{\alpha}^\dag=\delta_{\alpha,\beta}e^{i\bm{Q}\cdot\bm{R}_j}c_{j,\alpha}^\dag+(1-\delta_{\alpha,\beta})c_{j,\beta},
\label{eq_Shiba}
\end{equation}
where  
\begin{equation}
S_\alpha=\prod_j (c_{j,\alpha}^\dag+e^{-i\bm{Q}\cdot\bm{R}_j}c_{j,\alpha})
\end{equation}
$(\alpha=1,\cdots,N)$ is unitary \cite{1dHubbard_book} (see also Ref.~\cite{Ying01}). 
The generalized Shiba transformation $S_\alpha$ is the particle-hole transformation of the $\alpha$th component combined with adding the site-dependent phase factor $e^{i\bm{Q}\cdot\bm{R}_j}$.
The generalized Shiba transformations map the $\eta$ operators to the spin operators as follows:
\begin{equation}
S_\alpha \eta_{\alpha,\beta}^\dag S_\alpha^\dag = -F_{\beta,\alpha},\ \  S_\beta \eta_{\alpha,\beta}^\dag S_\beta^\dag = F_{\alpha,\beta}.
\label{eq_eta_Shiba}
\end{equation}
Using this transformation, we can map the primary $N$-color $\eta$-pairing state to
\begin{align}
\ket{\phi_{M_2,\cdots,M_N}}\equiv& S_1\ket{\psi_{M_2,\cdots,M_N}}\notag\\
=&(F_{2,1})^{M_2}\cdots(F_{N,1})^{M_N}\ket{\mathrm{FP}_1},
\label{eq_SUNferro}
\end{align}
where
\begin{equation}
\ket{\mathrm{FP}_1}=S_1\ket{0}=\prod_{j=1}^{N_{\mathrm{s}}}c_{j,1}^\dag\ket{0}
\end{equation}
is a fully polarized state. 
The state \eqref{eq_SUNferro} corresponds to a ferromagnetic state in SU($N$)-symmetric Fermi systems, and therefore we shall call it an SU($N$) ferromagnetic state \cite{Katsura13, Tamura19, Tamura21}. 

The SU($N$) ferromagnetic state \eqref{eq_SUNferro} is an exact eigenstate of the MHM \eqref{eq_Hasym}. The proof can be done in a manner similar to that for the generalized $\eta$-pairing states. First, since the kinetic term has the SU($N$) symmetry \eqref{eq_comm_FT} and $T\ket{\mathrm{FP}_1}=0$, the state \eqref{eq_SUNferro} is an eigenstate of the kinetic term: $T\ket{\phi_{M_2,\cdots,M_N}}=0$. Second, we can also show $V\ket{\phi_{M_2,\cdots,M_N}}=0$, since the state \eqref{eq_SUNferro} does not contain sites occupied by more than one particle, as one can explicitly confirm this from the definition. 

The existence of the $\eta$-pairing eigenstates in the MHM can be understood from the mapping between the primary $N$-color $\eta$-pairing state and the SU($N$) ferromagnetic state. We can see this by applying the generalized Shiba transformation to the Hamiltonian, which yields
\begin{align}
S_1HS_1^\dag=&-\sum_{\langle i,j\rangle}\sum_{\alpha=1}^N t_{i,j}(c_{i,\alpha}^\dag c_{j,\alpha}+\mathrm{H.c.})\notag\\
&+\sum_j \sum_{\alpha=2}^N U_{\alpha,1}n_{j,\alpha}\notag\\
&+\frac{1}{2}\sum_j\sum_{\alpha\neq\beta}U_{\alpha,\beta}^{\prime}n_{j,\alpha}n_{j,\beta},
\label{eq_H_Shiba}
\end{align}
where
\begin{align}
U_{\alpha,\beta}^{\prime}=
\begin{cases}
-U_{\alpha,\beta} & (\alpha=1\ \mathrm{or}\ \beta=1); \\
U_{\alpha,\beta} & (\text{otherwise}).
\end{cases}
\end{align}
The Hamiltonian \eqref{eq_H_Shiba} after the mapping is also a MHM, except for the second term which gives a constant in a given sector of the Hilbert space with a definite number of particles in each component. Then, we obtain another proof of Eq.~\eqref{eq_SUNeta_energy} from Eq.~\eqref{eq_SUNferro} and
\begin{align}
&S_1HS_1^\dag\ket{\phi_{M_2,\cdots,M_2}}\notag\\
&=(M_2U_{2,1}+\cdots+M_NU_{N,1})\ket{\phi_{M_2,\cdots,M_2}},
\end{align}
where the first and third terms on the right-hand side of Eq.~\eqref{eq_H_Shiba} vanish.

The mapping \eqref{eq_SUNferro} suggests that the primary $N$-color $\eta$-pairing state \eqref{eq_SUNeta} is regarded as a ferromagnetic state of SU($N$) pseudospins 
\begin{equation}
\ket{0_j}, \ket{1_j}\equiv c_{j,2}^\dag c_{j,1}^\dag\ket{0_j},\cdots,\ket{(N-1)_j}\equiv c_{j,N}^\dag c_{j,1}^\dag\ket{0_j},
\label{eq_pseudospin}
\end{equation}
where $\ket{0_j}$ denotes an empty state at site $j$. 
The SU($N$) pseudospin degrees of freedom are mapped to the ordinary SU($N$) spin degrees of freedom
\begin{align}
&c_{j,1}^\dag\ket{0_j}\propto S_{j,1}\ket{0_j}, c_{j,2}^\dag\ket{0_j}\propto S_{j,1}\ket{1_j},\notag\\
&\cdots,c_{j,N}^\dag\ket{0_j}\propto S_{j,1}\ket{(N-1)_j}
\end{align}
by a local Shiba transformation with $S_{j,1}\equiv c_{j,1}^\dag+e^{-i\bm{Q}\cdot\bm{R}_j}c_{j,1}$. 
The $\eta$ operators $\eta_{\alpha,1}, \eta_{\alpha,1}^\dag\ (\alpha=2,\cdots,N)$ and a subset $F_{\alpha,\beta}\ (\alpha,\beta=2,\cdots,N)$ of the spin operators form an su($N$) algebra, which follows from Eqs.~\eqref{eq_SUNalg} and \eqref{eq_eta_Shiba}. 
The SU($N$) pseudospin \eqref{eq_pseudospin} is a natural generalization of the SU(2) pseudospin defined by a doublet of the empty state and the doubly occupied state in the two-component Hubbard model \cite{YangZhang90, Zhang90}.

\section{SU($N$) magnetic fragmented fermionic condensates\label{sec_magSF}}

\subsection{Multicomponent off-diagonal long-range order\label{sec_ODLRO}}

The primary $N$-color $\eta$-pairing states \eqref{eq_SUNeta} can be regarded as a Bose-Einstein condensate of multicomponent $\eta$ pairs created by $\eta_{\alpha,1}^\dag\ (\alpha=2,\cdots,N)$. Accordingly, the generalized $\eta$-pairing states show multicomponent ODLRO. 
As detailed in Appendix \ref{sec_corr_calc}, the pair correlation functions of the primary $N$-color $\eta$-pairing states are calculated as
\begin{align}
&\frac{\bra{\psi_{M_2,M_3,\cdots,M_N}}c_{i,\alpha}^\dag c_{i,1}^\dag c_{j,1}c_{j,\alpha}\ket{\psi_{M_2,M_3,\cdots,M_N}}}{\braket{\psi_{M_2,M_3,\cdots,M_N}|\psi_{M_2,M_3,\cdots,M_N}}}\notag\\
=&\frac{M_\alpha(N_{\mathrm{s}}-N_{\mathrm{f}}/2)}{N_{\mathrm{s}}(N_{\mathrm{s}}-1)}e^{i\bm{Q}\cdot(\bm{R}_i-\bm{R}_j)},
\label{eq_paircorr_SU(N)_norm}
\end{align}
where $i\neq j$ and $\alpha=2,\cdots,N$, and $N_{\mathrm{f}}=2M_2+\cdots+2M_N$ is the total number of particles. 
The correlation function \eqref{eq_paircorr_SU(N)_norm} has a constant magnitude that does not depend on the distance between sites $i,j$, indicating the ODLRO. However, the magnitude of the pair correlation functions may depend on $\alpha$. 
Equation \eqref{eq_paircorr_SU(N)_norm} generalizes the result for the $\eta$-pairing state of the two-component Hubbard model in Ref.~\cite{Yang89}. 
In contrast to the two-component case, here the ODLRO extends over $N-1$ different pairs. 
The right-hand side of Eq.~\eqref{eq_paircorr_SU(N)_norm} does not vanish in the thermodynamic limit $N_{\mathrm{s}}\to\infty$ if $M_\alpha$ and $(N_{\mathrm{s}}-N_{\mathrm{f}}/2)$ are of the order of $N_{\mathrm{s}}$. 

Similarly, the pair correlation functions of the three-color $\eta$-pairing state \eqref{eq_SU3eta} are given by (see Appendix \ref{sec_corr_calc})
\begin{align}
\frac{\bra{\psi_{l,m,n}^{(3)}}\eta_{i,\alpha,\beta}^\dag\eta_{j,\alpha,\beta}\ket{\psi_{l,m,n}^{(3)}}}{\braket{\psi_{l,m,n}^{(3)}|\psi_{l,m,n}^{(3)}}}
=&\frac{r_{\alpha,\beta}(N_{\mathrm{s}}-N_{\mathrm{f}}^{(3)}/2)}{N_{\mathrm{s}}(N_{\mathrm{s}}-1)},
\end{align}
where $i\neq j$, $r_{1,2}=l, r_{2,3}=m,r_{3,1}=n$, and $N_{\mathrm{f}}^{(3)}=2(l+m+n)$. Hence, the ODLRO exists individually for the three-component $\eta$ pairs in the state \eqref{eq_SU3eta} if $l,m,n$, and $(N_{\mathrm{s}}-N_{\mathrm{f}}^{(3)}/2)$ are of the order of $N_{\mathrm{s}}$.

The pair correlation can experimentally be measured through the momentum distribution of on-site pairs, which we will refer to as doublons \cite{Nakagawa21}. We define the creation operator of an $(\alpha,\beta)$ doublon by
\begin{equation}
d_{\bm{k},\alpha,\beta}^\dag\equiv\frac{1}{\sqrt{N_{\mathrm{s}}}}\sum_j c_{j,\alpha}^\dag c_{j,\beta}^\dag e^{i\bm{k}\cdot\bm{R}_j}.
\end{equation}
Note that $d_{\bm{Q},\alpha,\beta}^\dag=\eta_{\alpha,\beta}^\dag/\sqrt{N_{\mathrm{s}}}$. The momentum distribution of $(\alpha,1)$ doublons at $\bm{k}=\bm{Q}$ for the primary $N$-color $\eta$-pairing state \eqref{eq_SUNeta} is given by
\begin{align}
&\langle d_{\bm{Q},\alpha,1}^\dag d_{\bm{Q},\alpha,1}\rangle\notag\\
=&\frac{1}{N_{\mathrm{s}}}\sum_j\langle n_{j,1}n_{j,\alpha}\rangle+\frac{1}{N_{\mathrm{s}}}\sum_{i\neq j}\langle \eta_{i,\alpha,1}^\dag\eta_{j,\alpha,1}\rangle\notag\\
=&\frac{M_\alpha}{N_{\mathrm{s}}}+\frac{1}{N_{\mathrm{s}}}\frac{M_\alpha(N_{\mathrm{s}}-N_{\mathrm{f}}/2)}{N_{\mathrm{s}}(N_{\mathrm{s}}-1)}\times N_{\mathrm{s}}(N_{\mathrm{s}}-1)\notag\\
=&\frac{M_\alpha(N_{\mathrm{s}}-N_{\mathrm{f}}/2+1)}{N_{\mathrm{s}}},
\label{eq_SUN_doublon_occ}
\end{align}
where $\alpha=2,\cdots,N$. 
Thus, if $M_\alpha$ and $(N_{\mathrm{s}}-N_{\mathrm{f}}/2)$ are of the order of $N_{\mathrm{s}}$, then a macroscopic number of $(\alpha,1)$ doublons undergo Bose-Einstein condensation at momentum $\bm{Q}$, as expected. 

While Yang's $\eta$-pairing state (i.e., $N=2$) saturates a bound on the momentum distribution of doublons \cite{Yang62, Nakagawa21}, the generalized $\eta$-pairing states with $N\geq 3$ components do not saturate a similar bound. See Appendix \ref{sec_ODLRO_bound} for details.

\subsection{Coexistence of SU($N$) magnetism and ODLRO\label{sec_magnetism}}

Another unique feature of the $N$-color $\eta$-pairing states can be seen from spin correlation functions $\langle F_{i,\alpha,\beta}F_{j,\beta,\alpha}\rangle\ (\alpha\neq\beta;\alpha,\beta\neq 1)$. Here,
\begin{equation}
F_{i,\alpha,\beta}\equiv c_{i,\alpha}^\dag c_{i,\beta}
\end{equation}
is a local SU($N$) spin operator. 
The spin correlation functions of the primary $N$-color $\eta$-pairing state can be calculated similarly to the pair correlation functions as (see Appendix \ref{sec_corr_calc} for the derivation)
\begin{align}
&\frac{\bra{\psi_{M_2,M_3,\cdots,M_N}}F_{i,\alpha,\beta}F_{j,\beta,\alpha}\ket{\psi_{M_2,M_3,\cdots,M_N}}}{\braket{\psi_{M_2,M_3,\cdots,M_N}|\psi_{M_2,M_3,\cdots,M_N}}}\notag\\
=&\frac{M_\alpha M_\beta}{N_{\mathrm{s}}(N_{\mathrm{s}}-1)},
\label{eq_spincorr_SU(N)_norm}
\end{align}
where $i\neq j$, $\alpha\neq\beta$, and $\alpha,\beta\neq 1$. The spin correlation function \eqref{eq_spincorr_SU(N)_norm} does not depend on the distance between sites $i,j$ and indicates magnetic long-range order. Thus, in the primary $N$-color $\eta$-pairing states, SU($N$) magnetic order and multicomponent ODLRO coexist. 
The magnetic long-range order is enabled by the internal structure of $\eta$ pairs; while an $\eta$ pair in the two-component system forms a spin singlet and thus cannot show magnetism, the multicomponent nature of the generalized $\eta$ pairs allows SU($N$) magnetism in the $\eta$-pairing states. 
Thus, the coexistence of magnetism and ODLRO is a unique feature of the multicomponent system. In fact, the spin correlation function \eqref{eq_spincorr_SU(N)_norm} is absent in the two-component case. 

The generalized Shiba transformation \eqref{eq_eta_Shiba} relates the spin correlations of the primary $N$-color $\eta$-pairing state to the SU($N$) ferromagnetic state as 
\begin{align}
&\frac{\bra{\psi_{M_2,M_3,\cdots,M_N}}F_{i,\alpha,\beta}F_{j,\beta,\alpha}\ket{\psi_{M_2,M_3,\cdots,M_N}}}{\braket{\psi_{M_2,M_3,\cdots,M_N}|\psi_{M_2,M_3,\cdots,M_N}}}\notag\\
=&\frac{\bra{\phi_{M_2,M_3,\cdots,M_N}}F_{i,\alpha,\beta}F_{j,\beta,\alpha}\ket{\phi_{M_2,M_3,\cdots,M_N}}}{\braket{\phi_{M_2,M_3,\cdots,M_N}|\phi_{M_2,M_3,\cdots,M_N}}},
\end{align}
where $\alpha\neq\beta$ and $\alpha,\beta\neq 1$. 
Therefore, the magnetic long-range order \eqref{eq_spincorr_SU(N)_norm} of the primary $N$-color $\eta$-pairing state is analogous to that of the SU($N$) ferromagnetic state. However, they are not equivalent since the other spin correlation functions $\langle F_{i,\alpha,1}F_{j,1,\alpha}\rangle\ (\alpha\neq1)$ are not the same. 
In fact, we have
\begin{align}
\frac{\bra{\psi_{M_2,M_3,\cdots,M_N}}F_{i,\alpha,1}F_{j,1,\alpha}\ket{\psi_{M_2,M_3,\cdots,M_N}}}{\braket{\psi_{M_2,M_3,\cdots,M_N}|\psi_{M_2,M_3,\cdots,M_N}}}=0,
\label{eq_spincorr_vanish}
\end{align}
for the primary $N$-color $\eta$-pairing state, whereas
\begin{align}
&\frac{\bra{\phi_{M_2,M_3,\cdots,M_N}}F_{i,\alpha,1}F_{j,1,\alpha}\ket{\phi_{M_2,M_3,\cdots,M_N}}}{\braket{\phi_{M_2,M_3,\cdots,M_N}|\phi_{M_2,M_3,\cdots,M_N}}}\notag\\
=&\frac{\bra{\psi_{M_2,M_3,\cdots,M_N}}\eta_{i,\alpha,1}^\dag\eta_{j,\alpha,1}\ket{\psi_{M_2,M_3,\cdots,M_N}}}{\braket{\psi_{M_2,M_3,\cdots,M_N}|\psi_{M_2,M_3,\cdots,M_N}}}\notag\\
=&\frac{M_\alpha(N_{\mathrm{s}}-N_{\mathrm{f}}/2)}{N_{\mathrm{s}}(N_{\mathrm{s}}-1)}\ (i\neq j),
\end{align}
for the SU($N$) ferromagnetic state. Equation \eqref{eq_spincorr_vanish} is consistent with the fact that Yang's $\eta$-pairing states do not show magnetism.

For the three-color $\eta$-pairing states \eqref{eq_SU3eta}, we similarly obtain the spin correlation functions as 
\begin{align}
\frac{\bra{\psi_{l,m,n}^{(3)}}F_{i,\alpha,\beta}F_{j,\beta,\alpha}\ket{\psi_{l,m,n}^{(3)}}}{\braket{\psi_{l,m,n}^{(3)}|\psi_{l,m,n}^{(3)}}}=\frac{\tilde{r}_{\alpha,\beta}}{N_{\mathrm{s}}(N_{\mathrm{s}}-1)},
\end{align}
where $i\neq j$, $\alpha\neq\beta$, $\tilde{r}_{1,2}=mn$, $\tilde{r}_{2,3}=nl$, $\tilde{r}_{3,1}=lm$, and $\tilde{r}_{\beta,\alpha}=\tilde{r}_{\alpha,\beta}$. Thus, the three-color $\eta$-pairing states also show the magnetic long-range order.

\subsection{Generalized $\eta$-pairing states as fragmented condensates\label{sec_fragmented}}

Condensation of fermion pairs is characterized by the presence of an extensive (i.e., of the order of $N_{\mathrm{s}}$) eigenvalue of a two-particle reduced density matrix \cite{Penrose56, Yang62}
\begin{equation}
(\rho_2)_{(i,\alpha_1,j,\alpha_2),(k,\alpha_3,l,\alpha_4)}\equiv\mathrm{Tr}[c_{j,\alpha_2}^\dag c_{i,\alpha_1}^\dag c_{k,\alpha_3} c_{l,\alpha_4}\rho],
\label{eq_rho_2}
\end{equation}
where $\rho$ is a density matrix of the system. 
In the primary $N$-color $\eta$-pairing state, $N-1$ different $\eta$ pairs, which are created by $\eta_{\alpha,1}^\dag\ (\alpha=2,\cdots,N)$, are simultaneously condensed. Accordingly, the wave functions of the pairs,
\begin{equation}
f_{(i,\beta_1,j,\beta_2)}^{(\alpha)}\equiv\frac{1}{\sqrt{2N_{\mathrm{s}}}}\delta_{i,j}e^{-i\bm{Q}\cdot\bm{R}_j}(\delta_{\beta_1,\alpha}\delta_{\beta_2,1}-\delta_{\beta_1,1}\delta_{\beta_2,\alpha}),
\end{equation}
where $\alpha=2,\cdots,N$, constitute $N-1$ eigenvectors of the two-particle density matrix $\rho_2$ of the primary $N$-color $\eta$-pairing state. 
In fact, by setting $\rho=\ket{\psi_{M_2,\cdots,M_N}}\bra{\psi_{M_2,\cdots,M_N}}/\braket{\psi_{M_2,\cdots,M_N}|\psi_{M_2,\cdots,M_N}}$, we have
\begin{align}
&(\rho_2f^{(\alpha)})_{(i,\beta_1,j,\beta_2)}\notag\\
=&\sum_{k,l}\sum_{\beta_3,\beta_4}(\rho_2)_{(i,\beta_1,j,\beta_2),(k,\beta_3,l,\beta_4)}f_{(k,\beta_3,l,\beta_4)}^{(\alpha)}\notag\\
=&\frac{\sqrt{2}}{\sqrt{N_{\mathrm{s}}}}\sum_le^{-i\bm{Q}\cdot\bm{R}_l}\mathrm{Tr}[c_{j,\beta_2}^\dag c_{i,\beta_1}^\dag c_{l,\alpha}c_{l,1}\rho]\notag\\
=&\frac{2M_\alpha(N_{\mathrm{s}}-N_{\mathrm{f}}/2+1)}{N_{\mathrm{s}}}f_{(i,\beta_1,j,\beta_2)}^{(\alpha)},
\end{align}
where we have used $\mathrm{Tr}[c_{j,1}^\dag c_{j,\alpha}^\dag c_{j,\alpha}c_{j,1}\rho]=M_\alpha/N_{\mathrm{s}}$ and Eq.~\eqref{eq_paircorr_SU(N)_norm}. The $N-1$ eigenvalues
\begin{equation}
\Lambda_\alpha=\frac{2M_\alpha(N_{\mathrm{s}}-N_{\mathrm{f}}/2+1)}{N_{\mathrm{s}}}\ \ (\alpha=2,\cdots,N)
\end{equation}
are of the order of $N_{\mathrm{s}}$ if $M_\alpha\ (\alpha=2,\cdots,N)$ and $N_{\mathrm{s}}-N_{\mathrm{f}}/2$ are of the order of $N_{\mathrm{s}}$. This means that the primary $N$-color $\eta$-pairing state is a fragmented fermionic condensate \cite{Mueller06}, in which the two-particle density matrix has multiple extensive eigenvalues. 
Similarly, in the three-color $\eta$-pairing state, three different $\eta$ pairs are simultaneously Bose-Einstein condensed, forming a fragmented condensate.

It should be noted that fragmented Bose-Einstein condensates are based on a certain symmetry of the Hamiltonian and hence generally fragile against perturbations that break that symmetry \cite{Mueller06}. In the present case, the relevant symmetry of the Hamiltonian is $\mathrm{U}(1)^N$ (see Sec.~\ref{sec_model}) and the symmetry-breaking perturbation is the one that breaks it by transferring particles in one component to the other of the condensates. 
We note that such symmetry-breaking term is negligible for alkaline-earth-like atoms, in which nuclear spins are decoupled from electronic degrees of freedom \cite{Gorshkov10, Cazalilla14}. 
The nuclear-spin-changing process can be achieved by using a Raman process \cite{Tusi21}. 
A crucial observation here is that such a process must be implemented intentionally and in a fine-tuned manner; otherwise fragmented condensates can be stable. 
In particular, for $\eta$ pairing formed between $^{171}$Yb and $^{173}$Yb (see Sec.~\ref{sec_model}), such a mixing process is precluded due to the superselection rule, and therefore this $\eta$-paired fragmented condensate is absolutely stable. This is in sharp contrast to usual fragmented condensates such as spinor condensates \cite{Mueller06} and constitutes a very unique feature of $\eta$-paired multicomponent fermionic condensates.

\section{Weak ergodicity breaking\label{sec_scar}}

\subsection{Absence of $\eta$-pairing symmetry, its physical consequence, and the ETH}

The absence of the $\eta$-pairing symmetry leads to an important consequence on thermalization in the MHM. The two-component Hubbard model has the $\eta$-pairing symmetry and the $\eta$ operators obey an su(2) algebra $[\eta_{2,1}^\mu,\eta_{2,1}^\nu]=i\epsilon_{\mu\nu\lambda}\eta_{2,1}^\lambda\ (\mu,\nu,\lambda=x,y,z)$ \cite{YangZhang90}, where
\begin{subequations}
\begin{align}
\eta_{2,1}^x\equiv&\frac{1}{2}(\eta_{2,1}^\dag+\eta_{2,1}),\\
\eta_{2,1}^y\equiv&\frac{1}{2i}(\eta_{2,1}^\dag-\eta_{2,1}),\\
\eta_{2,1}^z\equiv&\frac{1}{2}\sum_j(n_{j,1}+n_{j,2}-1).
\end{align}
\end{subequations}
Accordingly, eigenstates of the two-component Hubbard model can be labeled by the quantum numbers of the $\eta$-pairing SU(2) symmetry and the spin SU(2) symmetry \cite{YangZhang90, 1dHubbard_book}. The $\eta$-pairing SU(2) symmetry is related to the conservation of the total particle number $2\eta_{2,1}^z+N_{\mathrm{s}}$ and the conservation of the number of doublons at momentum $\bm{Q}$:
\begin{align}
\frac{1}{N_{\mathrm{s}}}\eta_{2,1}^\dag \eta_{2,1}=d_{\bm{Q},2,1}^\dag d_{\bm{Q},2,1}.
\label{eq_etapm}
\end{align}
Since conserved quantities do not change during the time evolution, thermalization of an isolated quantum system should be discussed separately in each sector specified by the values of the conserved quantities. In a given sector, it can be shown that arbitrary initial states will thermalize after a sufficiently long time if this sector satisfies the strong ETH, which states that every energy eigenstate is thermal, i.e., indistinguishable from the microcanonical ensemble concerning local or few-body observables \cite{Rigol08, DAlessio16, Gogolin16, Mori18, Ueda20}. 
Yang's $\eta$-pairing state, which is the $N=2$ case of Eq.~\eqref{eq_SUNeta}, satisfies the ETH of the two-component Hubbard model; in fact, this state is unique in a sector specified by the quantum numbers of the $\eta$-pairing symmetry  \cite{Vafek17}. In Refs.~\cite{Moudgalya20, Mark20}, it is shown that Yang's $\eta$-pairing state can be a quantum many-body scar state which violates the ETH if the model is deformed by special terms that break the $\eta$-pairing symmetry but leave the $\eta$-pairing state as an eigenstate. 
However, this construction requires tailored perturbations that alter the model from the simple Hubbard model.

The absence of the $\eta$-pairing symmetry in the MHM implies that the primary $N$-color $\eta$-pairing eigenstates \eqref{eq_SUNeta} [and the three-color $\eta$-pairing eigenstates \eqref{eq_SU3eta}] are quantum many-body scar states. In fact, since the generic internal symmetry of the MHM is U(1)$^N$ (see Sec.~\ref{sec_model}), the Hilbert space of the model is divided into subsectors labeled by the number of particles in each component, and the generalized $\eta$-pairing eigenstates belong to one of the subsectors which may contain many thermal eigenstates that are essentially distinct from the $\eta$-pairing eigenstates. 
We will corroborate that the generalized $\eta$-pairing eigenstates do not  obey the ETH, by calculating the entanglement entropy and non-thermalizing dynamics in the following subsections. This indicates a weak breakdown of ergodicity in the MHM. We note that special attention needs to be paid if the model has additional internal symmetry such as SU($N$). See Appendix \ref{sec_SUN_Hubbard} for this point.

In the MHM, the Heisenberg equation of motion for the number of doublons at momentum $\bm{Q}$ is given by
\begin{align}
&\frac{d}{dt} d_{\bm{Q},\alpha,\beta}^\dag(t) d_{\bm{Q},\alpha,\beta}(t)
\notag\\
=&\frac{i}{N_{\mathrm{s}}}\sum_{i,j}\sum_{\gamma(\neq \alpha,\beta)}(U_{\alpha,\gamma}+U_{\beta,\gamma}) e^{i\bm{Q}\cdot(\bm{R}_i-\bm{R}_{j})}\notag\\
&\times n_{i,\gamma}(c_{i,\alpha}^\dag c_{i,\beta}^\dag c_{j,\beta}c_{j,\alpha}-\mathrm{H.c.}),
\label{eq_non_conservation}
\end{align}
and this quantity is not conserved due to the absence of the $\eta$-pairing symmetry. The right-hand side of Eq.~\eqref{eq_non_conservation} indicates that the number of doublons at momentum $\bm{Q}$ can be relaxed due to the correlations between particles in three different components. Clearly, such correlations are absent in the two-component case.

The conservation of the number of doublons at momentum $\bm{Q}$ for $N=2$ and the non-conservation thereof for $N\geq 3$ can experimentally be tested, offering a method for diagnosing the existence (or absence) of the $\eta$-pairing symmetry. 
The change in this quantity also provides a signature of the three-particle correlations appearing on the right-hand side of Eq.~\eqref{eq_non_conservation}.

\subsection{Sub-volume law entanglement entropy\label{sec_EE}}

A useful indicator of the violation of the ETH is entanglement entropy, since the ETH implies that the entanglement entropy of a thermal eigenstate obeys a volume law if it lies in the bulk of the eigenspectrum \cite{Abanin19, Serbyn21, Papic21, Moudgalya21}. 
To calculate the entanglement entropy of the $N$-color $\eta$-pairing states, we divide the lattice into two parts $A$ and $B$. The number of sites of subsystem $A\ (B)$ is denoted by $N_{\mathrm{s},A}$ ($N_{\mathrm{s},B}=N_{\mathrm{s}}-N_{\mathrm{s},A}$). We consider the entanglement entropy of the primary $N$-color $\eta$-pairing state:
\begin{align}
S_A=-\mathrm{Tr}_A[\rho_A\log\rho_A],
\end{align}
where 
\begin{align}
\rho_A=\mathrm{Tr}_B\left[\frac{\ket{\psi_{M_2,\cdots,M_N}}\bra{\psi_{M_2,\cdots,M_N}}}{\braket{\psi_{M_2,\cdots,M_N}|\psi_{M_2,\cdots,M_N}}}\right]
\end{align}
is the reduced density matrix in subsystem $A$. 
As detailed in Appendix~\ref{sec_EE_calc}, the entanglement entropy is calculated to give
\begin{align}
S_A=&\frac{N-1}{2}\log N_{\mathrm{s},A}+\mathrm{const.},
\label{eq_EE_SUNeta}
\end{align}
which generalizes the result previously obtained for the two-component case \cite{Vafek17}.
For the three-color $\eta$-pairing state \eqref{eq_SU3eta}, the entanglement entropy is given by setting $N-1=3$ in Eq.~\eqref{eq_EE_SUNeta}. 
The entanglement entropy \eqref{eq_EE_SUNeta} is not proportional to $N_{\mathrm{s},A}$ and therefore does not obey the volume law. 
Furthermore, it can be shown that the $N$-color $\eta$-pairing eigenstates lie in the bulk of the eigenspectrum if the interaction parameters are appropriately tuned (see Appendix \ref{sec_scar_app}). 
The sub-volume law entanglement entropy thus indicates that the primary $N$-color $\eta$-pairing states are non-thermal eigenstates of the MHM, and that the three-color $\eta$-pairing states are also non-thermal in the three-component Hubbard model.

\begin{figure}
    \includegraphics[width=8.5cm]{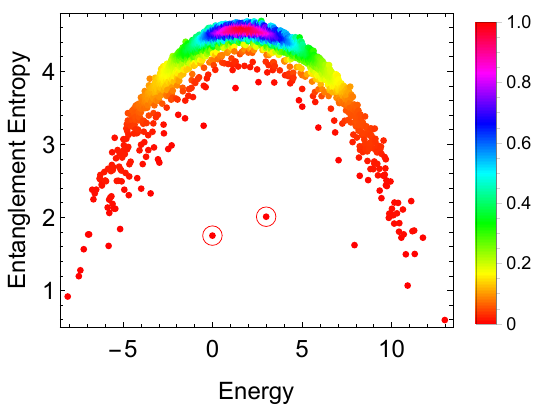}
    \caption{Entanglement entropy of all energy eigenstates of a one-dimensional three-component Hubbard model. The system size is $N_{\mathrm{s}}=6$, the size of the subsystem is $N_{\mathrm{s},A}=3$, and the particle number of each component is given by $F_{1,1}=F_{2,2}=F_{3,3}=2$. The hopping amplitude is set to $t_{j,j+1}=1+\delta t_j$, where $\delta t_j\in [-0.6,0.6]$ is randomly chosen. The interaction parameters are set to $U_{1,2}=0.8,U_{1,3}=1.0$, and $U_{2,3}=1.2$. The normalized density of data points is color coded. The two red circles indicate an SU(3) ferromagnetic state with eigenenergy $E=0$ and a three-color $\eta$-pairing state with eigenenergy $E=3$.}
    \label{fig_EE_su3}
    \end{figure}

\begin{figure}
    \includegraphics[width=8.5cm]{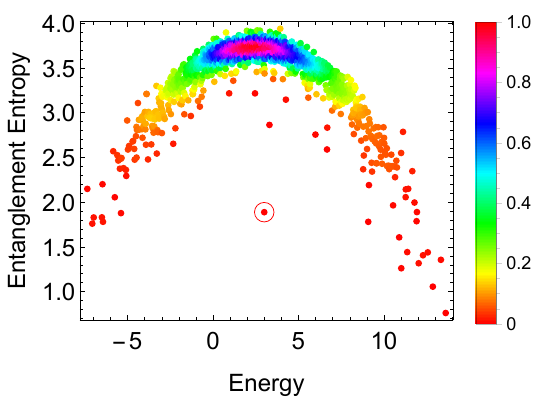}
    \caption{Entanglement entropy of all energy eigenstates of a one-dimensional four-component Hubbard model. The system size is $N_{\mathrm{s}}=5$, the size of the subsystem is $N_{\mathrm{s},A}=2$, and the particle number of each component is given by $F_{1,1}=3$ and $F_{2,2}=F_{3,3}=F_{4,4}=1$. The hopping amplitude is set to $t_{j,j+1}=1+\delta t_j$, where $\delta t_j\in [-0.6,0.6]$ is randomly chosen. The interaction parameters are set to $U_{1,2}=0.8,U_{1,3}=1.0,U_{1,4}=1.2,U_{2,3}=1.4,U_{2,4}=1.6$, and $U_{3,4}=1.8$. The normalized density of data points is color coded. The red circle indicates a primary four-color $\eta$-pairing state with eigenenergy $E=3$.}
    \label{fig_EE_su4}
    \end{figure}

In Fig.~\ref{fig_EE_su3}, we show a numerical result of the entanglement entropy of the three-component ($N=3$) Hubbard model \eqref{eq_Hasym} in one dimension. Here, we consider a six-site chain with open boundary conditions and two particles in each component (i.e., six particles in total). In this sector, the model has a three-color $\eta$-pairing state [Eq.~\eqref{eq_SU3eta} with $l=m=n=2$] and an SU(3) ferromagnetic state [Eq.~\eqref{eq_SUNferro} with $M_2=M_3=2$] as energy eigenstates. These energy eigenstates are identifed in Fig.~\ref{fig_EE_su3} by their eigenenergies and the values of the entanglement entropy, which can analytically be calculated (see Appendix \ref{sec_EE_calc}). The numerical result shows that these energy eigenstates have significantly lower entanglement entropy than thermal eigenstates. We also show a numerical result for a four-component ($N=4$) case in Fig.~\ref{fig_EE_su4}. We study a five-site chain with particle numbers $F_{1,1}=3$ and $F_{2,2}=F_{3,3}=F_{4,4}=1$. In this case, the model does not have an SU(4) ferromagnetic eigenstate (since the total particle number is greater than the number of sites) but has a primary four-color $\eta$-pairing eigenstate [Eq.~\eqref{eq_SUNeta} with $M_2=M_3=M_4=1$]. The $\eta$-pairing eigenstate can clearly be distinguished from thermal eigenstates in Fig.~\ref{fig_EE_su4}. 

\subsection{Non-thermalizing dynamics: Persistent oscillations\label{sec_oscil}}

The weak breakdown of ergodicity, or the violation of the strong ETH, implies that some initial states do not thermalize. 
To observe a non-thermalizing dynamics of the primary $N$-color $\eta$-pairing states, it is convenient to apply an external field to the MHM:
\begin{equation}
H_{\mathrm{dyn}}\equiv H+\frac{\Omega}{2}(F_{2,3}+F_{3,2})+\frac{\delta}{2}(F_{2,2}-F_{3,3}).
\label{eq_Hubbard_field}
\end{equation}
Here, the two components coupled by an external field may be any two from $2,\cdots,N$. 
The external field can experimentally be realized with a Raman coupling that induces transition between two hyperfine states of atoms \cite{Tusi21}.

One can confirm that the action of the Hamiltonian $H_{\mathrm{dyn}}$ is closed within a subspace spanned by primary $N$-color $\eta$-pairing states
\begin{equation}
\ket{m}\equiv\frac{\ket{\psi_{S+m,S-m,M_4,\cdots,M_N}}}{\sqrt{\braket{\psi_{S+m,S-m,M_4,\cdots,M_N}|\psi_{S+m,S-m,M_4,\cdots,M_N}}}},
\label{eq_ketm}
\end{equation}
where $S=(M_2+M_3)/2$ and $m=-S,-S+1,\cdots,S-1,S$. The matrix representation of the external-field operators $S^x\equiv (F_{2,3}+F_{3,2})/2, S^y\equiv (F_{2,3}-F_{3,2})/(2i),$ and $S^z\equiv(F_{2,2}-F_{3,3})/2$ in this subspace gives a spin-$S$ representation of the su(2) algebra since
\begin{subequations}
\begin{gather}
F_{2,3}\ket{m}=\sqrt{S(S+1)-m(m+1)}\ket{m+1},\\
F_{3,2}\ket{m}=\sqrt{S(S+1)-m(m-1)}\ket{m-1},\\
\frac{1}{2}(F_{2,2}-F_{3,3})\ket{m}=m\ket{m}.
\end{gather}
\end{subequations}
Therefore, we can map the states $\ket{m}\ (m=-S,\cdots,S)$ to the eigenstates of the $z$ component of a single collective pseudospin $\bm{S}=(S^x,S^y,S^z)$ with magnitude $S$. The Hamiltonian after this mapping is expressed as
\begin{equation}
H_{\mathrm{dyn}}=\Omega S^x+(U_{2,1}-U_{3,1}+\delta)S^z
\label{eq_H_spinS}
\end{equation}
up to a constant shift. 
The Heisenberg equations of motion for the collective pseudospin operators are given by
\begin{subequations}
\begin{align}
\frac{d}{dt}S^x=&-(U_{2,1}-U_{3,1}+\delta)S^y,\label{eq_spinSHeis_x}\\
\frac{d}{dt}S^y=&-\Omega S^z+(U_{2,1}-U_{3,1}+\delta)S^x,\label{eq_spinSHeis_y}\\
\frac{d}{dt}S^z=&\Omega S^y.\label{eq_spinSHeis_z}
\end{align}
\end{subequations}

We take a primary $N$-color $\eta$-pairing state $\ket{\psi_{M_2,\cdots,M_N}}$ as an initial state and consider the dynamics of the normalized numbers of doublons
\begin{subequations}
\begin{align}
m_2(t)\equiv&\frac{1}{M_2+M_3}\sum_j\langle n_{j,1}(t)n_{j,2}(t)\rangle,\label{eq_n12}\\
m_3(t)\equiv&\frac{1}{M_2+M_3}\sum_j\langle n_{j,1}(t)n_{j,3}(t)\rangle,\label{eq_n13}
\end{align}
\end{subequations}
where $n_{j,\alpha}(t)=e^{iH_{\mathrm{dyn}}t}n_{j,\alpha}e^{-iH_{\mathrm{dyn}}t}$ is the Heisenberg representation of the density operator of the $\alpha$th component. 
We note $\sum_j\langle n_{j,1}(t)n_{j,2}(t)\rangle=\sum_j\langle n_{j,2}(t)\rangle$ and $\sum_j\langle n_{j,1}(t)n_{j,3}(t)\rangle=\sum_j\langle n_{j,3}(t)\rangle$ in the subspace spanned by the $\eta$-pairing states \eqref{eq_ketm}, and therefore we have $M_2=\sum_j\langle n_{j,1}(0)n_{j,2}(0)\rangle$ and $M_3=\sum_j\langle n_{j,1}(0)n_{j,3}(0)\rangle$. From the solution of the Heisenberg equations of motion \eqref{eq_spinSHeis_x}-\eqref{eq_spinSHeis_z}, the time evolution of the numbers of doublons is given by
\begin{subequations}
\begin{align}
m_2(t)=&\frac{S+\langle S^z(t)\rangle}{M_2+M_3}\notag\\
=&\frac{1}{2}+\left(m_2(0)-\frac{1}{2}\right)\notag\\
&\times\frac{(U_{2,1}-U_{3,1}+\delta)^2+\Omega^2\cos(\omega t)}{(U_{2,1}-U_{3,1}+\delta)^2+\Omega^2},\label{eq_spinSHeis_sol}\\
m_3(t)=&1-m_2(t),\label{eq_spinSHeis_sol2}
\end{align}
\end{subequations}
where
\begin{equation}
\omega=\sqrt{(U_{2,1}-U_{3,1}+\delta)^2+\Omega^2}.
\label{eq_oscil_period}
\end{equation}
Note that the solutions \eqref{eq_spinSHeis_sol} and \eqref{eq_spinSHeis_sol2} do not depend on the total number $M_2+M_3$ of $(1,2)$ and $(1,3)$ doublons.

The solutions \eqref{eq_spinSHeis_sol} and \eqref{eq_spinSHeis_sol2} show that the numbers of doublons oscillate without any relaxation. The persistent oscillations can be interpreted as a precession of the collective pseudospin $\bm{S}$. 
The local parts $F_{j,2,3},F_{j,3,2}$, and $(F_{j,2,2}-F_{j,3,3})/2$ of the external field act as SU(2) spin operators on the two components $\ket{1_j},\ket{2_j}$ of the SU($N$) pseudospin \eqref{eq_pseudospin}. 
More generally, two components $\ket{l_j}$ and $\ket{k_j}$ of the SU($N$) pseudospin \eqref{eq_pseudospin} for $l,k\geq 1$ can be coupled by adding an external field $\propto (F_{l+1,k+1}+F_{k+1,l+1})$ to the Hamiltonian. 
When more components of the SU($N$) pseudospin are coupled, the $N$-color $\eta$-pairing states may show more complex collective dynamics. 
We note that the non-thermalizing dynamics presented in this subsection does not have its counterpart in the conventional $N=2$ case, where Yang's $\eta$-pairing state is spin-singlet and does not have any internal degrees of freedom.

\begin{figure}
    \includegraphics[width=8.5cm]{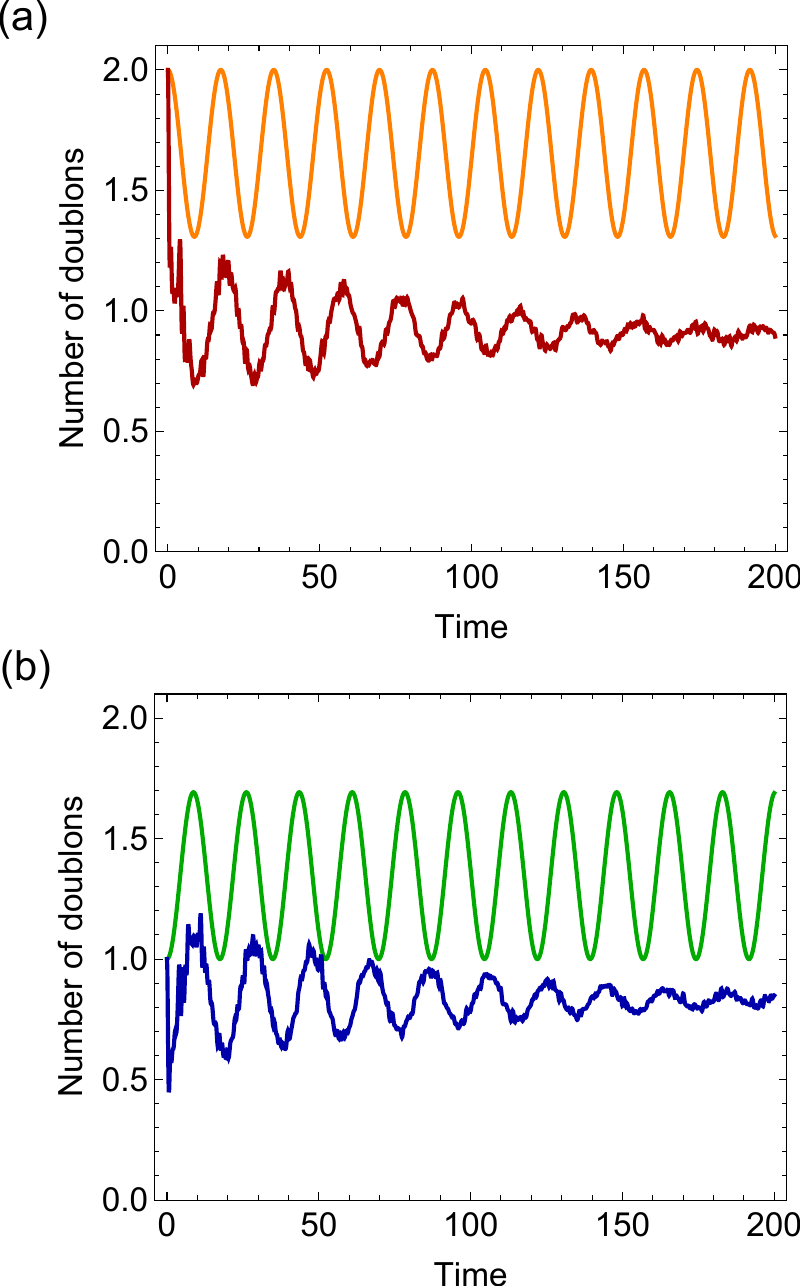}
    \caption{Dynamics of the number of doublons in the driven three-component Hubbard model in one dimension. The system size is $N_{\mathrm{s}}=6$ and the total number of particles is six. The hopping amplitude is set to $t_{j,j+1}=1$ and the interaction parameters are $U_{1,2}=0.8,U_{1,3}=1.0$, and $U_{2,3}=1.2$. (a) Dynamics of $\sum_j\langle n_{j,1}(t)n_{j,2}(t)\rangle$ from the $\eta$-pairing initial state $\ket{\psi_{\mathrm{ini}}^{(\eta)}}$ (orange) and the Fock initial state $\ket{\psi_{\mathrm{ini}}^{(\mathrm{F})}}$ (red). (b) Dynamics of $\sum_j\langle n_{j,1}(t)n_{j,3}(t)\rangle$ from the $\eta$-pairing initial state $\ket{\psi_{\mathrm{ini}}^{(\eta)}}$ (green) and the Fock initial state $\ket{\psi_{\mathrm{ini}}^{(\mathrm{F})}}$ (blue). The unit of time is the inverse of the hopping amplitude.}
    \label{fig_doublon_oscil}
    \end{figure}

To compare the persistent oscillations of the $\eta$-pairing states with thermalization dynamics from a generic initial state, we numerically calculate the dynamics of the number of doublons for the driven three-component ($N=3$) Hubbard model \eqref{eq_Hubbard_field} in a six-site chain with open boundary conditions. We take two different initial states: a primary $\eta$-pairing state $\ket{\psi_{\mathrm{ini}}^{(\eta)}}\equiv\ket{\psi_{2,1}}/\sqrt{\braket{\psi_{2,1}|\psi_{2,1}}}$ and a Fock state $\ket{\psi_{\mathrm{ini}}^{(\mathrm{F})}}\equiv c_{1,1}^\dag c_{1,2}^\dag c_{3,1}^\dag c_{3,2}^\dag c_{5,1}^\dag c_{5,3}^\dag\ket{0}$. As shown in Fig.~\ref{fig_doublon_oscil}, the dynamics of the number of doublons from the Fock initial state shows a damped oscillation, while that from the $\eta$-pairing state does not decay. This result is consistent with the weak ergodicity breaking of the multicomponent Hubbard model, supporting that the generalized $\eta$-pairing states are quantum many-body scar states. We note that the oscillations of the number of doublons from the Fock initial state are related to a nonzero overlap $|\braket{\psi_{\mathrm{ini}}^{(\mathrm{F})}|\psi_{\mathrm{ini}}^{(\eta)}}|=2/\sqrt{240}\simeq 0.13$ with the $\eta$-pairing state, while the period of the oscillation is slightly shifted due to hopping of particles which can break or create doublons (note that the period \eqref{eq_oscil_period} of the oscillation of the $\eta$-pairing states does not depend on the hopping amplitude).

\section{Partial integrability and non-thermalization in Hubbard models\label{sec_partial_integrability}}

\subsection{Integrable sectors in the multicomponent Hubbard model and generalization to other Hubbard models\label{sec_sector}}

The MHM allows the construction of exact eigenstates although it is not integrable. The exact eigenstates are categorized into two families: the $\eta$-pairing states and the SU($N$) ferromagnetic states. 
A general form of the $\eta$-pairing eigenstates is given by Eq.~\eqref{eq_SUNeta_ex}, which may contain unpaired fermions. 
Similarly, a general form of the SU($N$) ferromagnetic eigenstates is written as
\begin{equation}
(F_{2,1})^{M_2}\cdots(F_{N,1})^{M_N}c_{n_1,1}^\dag \cdots c_{n_r,1}^\dag\ket{0}.\label{eq_SUNferro_ex}
\end{equation}
In fact, the state \eqref{eq_SUNferro_ex} is an eigenstate of the MHM in Eq.~\eqref{eq_Hasym}, which can be shown similarly to special case \eqref{eq_SUNferro}. That is, the state \eqref{eq_SUNferro_ex} is an eigenstate of the kinetic term $T$ with eigenvalue $\epsilon_{n_1}+\cdots+\epsilon_{n_r}$ due to its SU($N$) symmetry, and the eigenstate is annihilated by the interaction term $V$ since the state \eqref{eq_SUNferro_ex} does not contain sites occupied by more than one fermion. 
The previously shown SU($N$) ferromagnetic state \eqref{eq_SUNferro} corresponds to the case with $r=N_{\mathrm{s}}$ in Eq.~\eqref{eq_SUNferro_ex}. 

The $\eta$-pairing eigenstates \eqref{eq_SUNeta_ex} and the SU($N$) ferromagnetic eigenstates \eqref{eq_SUNferro_ex} share a common structure; they are generated by acting with the $\eta$ or spin operators on the spin-polarized states of the form
\begin{equation}
c_{n_1,1}^\dag \cdots c_{n_r,1}^\dag\ket{0}.
\label{eq_polarized}
\end{equation}
The eigenstates \eqref{eq_polarized} are identical to those of the non-interacting case, in which the model is integrable. 
Therefore, the $\eta$-pairing eigenstates and the SU($N$) ferromagnetic eigenstates of the MHM can be mapped to the eigenstates of the integrable free-fermion model, although the MHM itself is not integrable in the entire Hilbert space. 
We shall refer to this structure as partial integrability of the MHM (see also Refs.~\cite{Sato95, Sato96} for this terminology). Similar structures have been found in some models that exhibit Hilbert space fragmentation \cite{Moudgalya19arXiv, Yang20, Bastianello21, Yoshinaga21}.

The $\eta$-pairing eigenstates with unpaired fermions in Eq.~\eqref{eq_SUNeta_ex} for $N=2$ have been well known in the two-component Hubbard model \cite{YangZhang90, Vafek17}. Similar free-fermion-like eigenstates have also been discussed in Refs.~\cite{Znidaric13, Iadecola19}. 
However, the MHM has a unique property that is absent in the conventional two-component case; the Hubbard interaction term $V$ has, in general, neither the $\eta$ symmetry nor the SU($N$) symmetry. Therefore, the eigenstates \eqref{eq_SUNeta_ex} and \eqref{eq_SUNferro_ex} do not require the full symmetry of the Hamiltonian but only the SU($N$) symmetry [or the SO($2N$) symmetry (see Sec.~\ref{sec_SO2N})] of the kinetic term is required. This leads to a crucial difference in thermalization in the MHM, as shown in Sec.~\ref{sec_nontherm_integrability}. 

The above discussion can be generalized to other Hubbard models and their variants. For example, since the SU($N$) ferromagnetic state \eqref{eq_SUNferro_ex} does not contain multiply occupied sites, it is an exact eigenstate of a Hamiltonian
\begin{align}
H_{\mathrm{gen}}^{(2)}=&-\sum_{\langle i,j\rangle}\sum_{\alpha=1}^N t_{i,j}(c_{i,\alpha}^\dag c_{j,\alpha}+\mathrm{H.c.})\notag\\
&+\sum_j\sum_{\alpha,\beta,\gamma,\delta} U_{j;\alpha,\beta,\gamma,\delta}c_{j,\alpha}^\dag c_{j,\beta}^\dag c_{j,\gamma} c_{j,\delta}
\label{eq_Hubbard_2int}
\end{align}
for arbitrary two-body on-site interactions with coefficients $U_{j;\alpha,\beta,\gamma,\delta}$. Note that this model with $N=2$ is unique and equivalent to the ordinary Hubbard model, since the on-site interaction automatically satisfies the spin SU(2) symmetry. 
However, in the multicomponent ($N\geq 3$) case, the model \eqref{eq_Hubbard_2int} includes several important models that are not covered by the conventional Hubbard model \eqref{eq_Hasym}. 
For example, a two-orbital Hubbard model with degenerate bands \cite{Roth66, KugelKhomskii73, KugelKhomskii82, Koga04}
\begin{align}
H_{\mathrm{orb}}=&-\sum_{\langle i,j\rangle}\sum_{m=1,2}\sum_{\sigma=\up,\down}t_{i,j}(c_{i,m,\sigma}^\dag c_{j,m,\sigma}+\mathrm{H.c.})\notag\\
&+U\sum_{j,m} c_{j,m,\up}^\dag c_{j,m,\up}c_{j,m,\down}^\dag c_{j,m,\down}\notag\\
&+U'\sum_{j,\sigma,\sigma'}c_{j,1,\sigma}^\dag c_{j,1,\sigma} c_{j,2,\sigma'}^\dag c_{j,2,\sigma'}\notag\\
&+J_H\sum_{j,\sigma,\sigma'}c_{j,1,\sigma}^\dag c_{j,1,\sigma'} c_{j,2,\sigma'}^\dag c_{j,2,\sigma}\notag\\
&+J_H'\sum_j(c_{j,1,\up}^\dag c_{j,1,\down}^\dag c_{j,2,\up}c_{j,2,\down}+\mathrm{H.c.}),
\label{eq_H_orb}
\end{align}
belongs to the $N=4$ case of Eq.~\eqref{eq_Hubbard_2int}. 
Here, the orbital and spin degrees of freedom, which are denoted by $m$ and $\sigma$, respectively, constitute four components. 
Another example is a two-orbital SU($M$) Hubbard model of ultracold alkaline-earth-like atoms in a magic-wavelength optical lattice \cite{Gorshkov10} 
\begin{align}
H_{\mathrm{AEA}}=&-\sum_{\langle i,j\rangle}\sum_{m=e,g}\sum_{\sigma=1}^M t_{i,j}(c_{i,m,\sigma}^\dag c_{j,m,\sigma}+\mathrm{H.c.})\notag\\
&+\sum_{j,m}\frac{U_m}{2} n_{j,m}(n_{j,m}-1)
+U_{eg}\sum_j n_{j,e}n_{j,g}\notag\\
&+V_{\mathrm{ex}}\sum_{j,\sigma,\sigma'}c_{j,g,\sigma}^\dag c_{j,e,\sigma'}^\dag c_{j,g,\sigma'} c_{j,e,\sigma},
\label{eq_H_AEA}
\end{align}
where $n_{j,m}\equiv \sum_\sigma c_{j,m,\sigma}^\dag c_{j,m,\sigma}$. Here, the orbital ($m=e,g$) and nuclear spin ($\sigma=1,\cdots,M$) degrees of freedom provide $N=2M$ component systems. 
Besides the multiorbital models, a spin-3/2 Hubbard model 
\begin{align}
H_{3/2}=&-\sum_{\langle i,j\rangle}\sum_{\sigma=\pm 3/2,\pm1/2}t_{i,j}(c_{i,\sigma}^\dag c_{j,\sigma}+\mathrm{H.c.})\notag\\
&+U_0\sum_j P_{0,0}^\dag(j)P_{0,0}(j)\notag\\
&+U_2\sum_j\sum_{m=\pm2,\pm1,0}P_{2,m}^\dag(j)P_{2,m}(j),
\label{eq_H_3_2}
\end{align}
where $P_{0,0}(j)$ and $P_{2,m}(j)$ are the singlet and quintet pairing operators, has also been proposed for ultracold fermionic atoms \cite{Wu03, Wu06}. This model is an $N=4$ case of Eq.~\eqref{eq_Hubbard_2int}. 
The models \eqref{eq_H_orb}-\eqref{eq_H_3_2} have the SU($N$) ferromagnetic eigenstates \eqref{eq_SUNferro_ex} despite the absence of the full SU($N$) symmetry. 
As shown later in Sec.~\ref{sec_nontherm_integrability}, this property leads to the weak breakdown of ergodicity in the models \eqref{eq_H_orb}-\eqref{eq_H_3_2}. 

The $N$-color $\eta$-pairing eigenstates can also be generalized to eigenstates of a certain class of Hubbard-like models. In Appendix \ref{sec_PAM}, we provide an extension of the MHM to the case with site-dependent potentials and interactions strengths, including a periodic Anderson model. 

In the discussion so far, the eigenstates in the integrable sector of the MHM can be mapped to those of non-interacting models. While our primary focus is the MHM \eqref{eq_Hasym}, we here note that we can also embed eigenstates of some integrable models with interactions by perturbing the MHM. To see this, we consider a one-dimensional model
\begin{align}
H_{\mathrm{gen}}^{(\mathrm{NN})}=&-t_{\mathrm{h}}\sum_j\sum_{\alpha=1}^N (c_{j,\alpha}^\dag c_{j+1,\alpha}+\mathrm{H.c.})\notag\\
&+U^{\prime}\sum_j\sum_{\alpha,\beta}n_{j,\alpha}n_{j+1,\beta}\notag\\
&+\sum_j\sum_{\alpha,\beta,\gamma,\delta} U_{j;\alpha,\beta,\gamma,\delta}c_{j,\alpha}^\dag c_{j,\beta}^\dag c_{j,\gamma} c_{j,\delta},
\label{eq_exHubbard_2int}
\end{align}
where $U_{j;\alpha,\beta,\gamma,\delta}$ is an arbitrary two-body interaction coefficient. Here, an SU($N$)-symmetric nearest-neighbor interaction term with strength $U^{\prime}$ is added to the model \eqref{eq_Hubbard_2int} in one dimension, while the SU($N$) symmetry of the entire Hamiltonian \eqref{eq_exHubbard_2int} can be broken by the on-site interaction. The eigenstates in a spin-polarized sector of Eq.~\eqref{eq_exHubbard_2int} with a single component $\alpha=1$ are obtained by those of a Hamiltonian
\begin{equation}
H_{\mathrm{NN}}=-t_{\mathrm{h}}\sum_j(c_{j,1}^\dag c_{j+1,1}+\mathrm{H.c.})+U^{\prime}\sum_jn_{j,1}n_{j+1,1},
\label{eq_spinless_NNint}
\end{equation}
since the on-site interaction does not act on particles in the same component. The Hamiltonian \eqref{eq_spinless_NNint} is integrable and equivalent to the XXZ chain (under an appropriate boundary condition) via the Jordan-Wigner transformation \cite{SamajBajnok_book}. Given an eigenstate $\ket{\phi_1}$ of the integrable Hamiltonian \eqref{eq_spinless_NNint}, a family of states
\begin{equation}
(F_{2,1})^{M_2}\cdots(F_{N,1})^{M_N}\ket{\phi_1}
\label{eq_exHubbard_embed}
\end{equation}
become eigenstates of the Hamiltonian in Eq.~\eqref{eq_exHubbard_2int}, since they do not contain doubly occupied sites and are annihilated by any on-site interaction. In this case, the eigenstates \eqref{eq_exHubbard_embed} of the model \eqref{eq_exHubbard_2int} can be mapped to those of the integrable model \eqref{eq_spinless_NNint} with interactions.

Another embedding of an integrable model is obtained by considering the following Hamiltonian:
\begin{align}
H_{\mathrm{gen}}^{(3)}
&=-t_{\mathrm{h}}\sum_{j}\sum_{\alpha=1}^N(c_{j,\alpha}^\dag c_{j+1,\alpha}+\mathrm{H.c.})\notag\\
&+U\sum_j\sum_{\alpha<\beta}n_{j,\alpha}n_{j,\beta}\notag\\
&+\sum_j\sum_{\alpha_1,\cdots,\alpha_6}U^{(3)}_{j;\alpha_1,\cdots,\alpha_6}c_{j,\alpha_1}^\dag c_{j,\alpha_2}^\dag c_{j,\alpha_3}^\dag c_{j,\alpha_4} c_{j,\alpha_5} c_{j,\alpha_6}.
\label{eq_Hubbard_3int}
\end{align}
Here, a three-body interaction term with coefficient $U^{(3)}_{j;\alpha_1,\cdots,\alpha_6}$ that breaks the SU($N$) symmetry is added to the one-dimensional SU($N$) Hubbard model. 
In a Hilbert subspace $\mathcal{H}_{1,2}$ that contains only two components $\alpha=1,2$, eigenstates of the model \eqref{eq_Hubbard_3int} reduce to those of the one-dimensional two-component Hubbard model, which is integrable \cite{LiebWu68, 1dHubbard_book}. Then, a family of states
\begin{equation}
(F_{3,1})^{M_3}\cdots(F_{N,1})^{M_N}(F_{3,2})^{M_3^\prime}\cdots(F_{N,2})^{M_N^\prime}\ket{\phi_{1,2}},
\label{eq_Hubbard_embed}
\end{equation}
where $\ket{\phi_{1,2}}$ is an eigenstate in the subsector $\mathcal{H}_{1,2}$, become eigenstates of the Hamiltonian \eqref{eq_Hubbard_3int}, since these states, by construction, do not contain sites occupied by more than two particles. In this case, the partial integrability of the Hamiltonian \eqref{eq_Hubbard_3int} is due to the subsector that can be mapped to the one-dimensional two-component Hubbard model, which is a genuinely interacting integrable model.

\subsection{Non-thermalizing dynamics due to partial integrability\label{sec_nontherm_integrability}}

Now we discuss the consequence of partial integrability on thermalization in the MHM and its generalized models. 
The existence of integrable subsectors implies that some initial states in the subsectors do not thermalize. 
Such initial states that fail to thermalize take the following forms:
\begin{align}
\ket{\psi_\eta}=&(\eta_{2,1}^\dag)^{n_2}\cdots(\eta_{N,1}^\dag)^{n_N}\ket{\phi_1},\label{eq_ini_eta}\\
\ket{\psi_{\mathrm{FM}}}=&(F_{2,1})^{n_2}\cdots(F_{N,1})^{n_N}\ket{\phi_1},\label{eq_ini_ferro}
\end{align}
where $\ket{\phi_1}$ is an $n$-particle state that only contains component 1. 
We define the Hilbert subspaces spanned by states in the form \eqref{eq_ini_eta} or \eqref{eq_ini_ferro} as
\begin{align}
\mathcal{H}_\eta^{\{ n_\alpha\}}\equiv&\{(\eta_{2,1}^\dag)^{n_2}\cdots(\eta_{N,1}^\dag)^{n_N}\ket{\phi_1}|\ket{\phi_1}\in\mathcal{H}_1^{(n)}\},\label{eq_sp_eta}\\
\mathcal{H}_{\mathrm{FM}}^{\{ n_\alpha\}}\equiv&\{(F_{2,1})^{n_2}\cdots(F_{N,1})^{n_N}\ket{\phi_1}|\ket{\phi_1}\in\mathcal{H}_1^{(n)}\},\label{eq_sp_ferro}
\end{align}
where 
\begin{equation}
\mathcal{H}_1^{(n)}\equiv\mathrm{span}[\{ c_{j_1,1}^\dag\cdots c_{j_n,1}^\dag\ket{0}\}_{j_1,\cdots,j_n}]
\end{equation}
is a Hilbert subspace of spin-polarized states with $n$ particles. 
The superscript $\{ n_\alpha\}$ in $\mathcal{H}_\eta^{\{ n_\alpha\}}$ and $\mathcal{H}_{\mathrm{FM}}^{\{ n_\alpha\}}$ indicates the number $n_\alpha$ of particles in the $\alpha$th component ($\alpha=1,\cdots,N$). For $\alpha=1$, we define $n_1=n+n_2+\cdots +n_N$ in Eq.~\eqref{eq_sp_eta} and $n_1=n-n_2-\cdots -n_N$ in Eq.~\eqref{eq_sp_ferro}. 
In each subspace, the Hamiltonian \eqref{eq_Hasym} acts on states as
\begin{align}
H\ket{\psi_\eta}=&(T+C)\ket{\psi_\eta}\notag\\
=&(\eta_{2,1}^\dag)^{n_2}\cdots(\eta_{N,1}^\dag)^{n_N}(T+C)\ket{\phi_1},
\label{eq_H_ini_eta}\\
H\ket{\psi_{\mathrm{FM}}}=&T\ket{\psi_{\mathrm{FM}}}\notag\\
=&(F_{2,1})^{n_2}\cdots(F_{N,1})^{n_N}T\ket{\phi_1},
\label{eq_H_ini_ferro}
\end{align}
where $C=n_2U_{2,1}+\cdots+n_NU_{N,1}$ is a constant. 
Note that the action of the Hamiltonian is closed in each subspace since $T\ket{\phi_1}\in\mathcal{H}_1^{(n)}$. Therefore, an initial state starting from the subspace $\mathcal{H}_\eta^{\{ n_\alpha\}}$ or $\mathcal{H}_{\mathrm{FM}}^{\{ n_\alpha\}}$ remain in the same subspace during the time evolution and these subspaces are dynamically disconnected from the rest of the Hilbert space. In other words, the Krylov subspaces spanned by $\{\ket{\psi_0},H\ket{\psi_0},H^2\ket{\psi_0}\cdots\}\ (\ket{\psi_0}=\ket{\psi_\eta}\ \mathrm{or}\ \ket{\psi_{\mathrm{FM}}})$ do not cover the entire Hilbert space, similarly as in systems with (weak) Hilbert space fragmentation \cite{Moudgalya19arXiv}.

For simplicity, here we assume translation invariance $t_{i,j}=t_{\mathrm{h}}$ and calculate the time evolution of the spin-resolved momentum distribution
\begin{equation}
O_{\bm{k},\alpha}=c_{\bm{k},\alpha}^\dag c_{\bm{k},\alpha},
\label{eq_mom_dist}
\end{equation}
where $c_{\bm{k},\alpha}=\frac{1}{\sqrt{N_{\mathrm{s}}}}\sum_j c_{j,\alpha}e^{-i\bm{k}\cdot\bm{R}_j}$. 
The observable $O_{\bm{k},\alpha}$ commutes with the kinetic term,
\begin{equation}
[T,O_{\bm{k},\alpha}]=0,
\label{eq_T_O}
\end{equation}
while it does not commute with the Hamiltonian if $U_{\alpha,\beta}\neq 0$ for some $\beta$:
\begin{align}
[H,O_{\bm{k},\alpha}]=&\frac{1}{2N_{\mathrm{s}}}\sum_{j,j'}\sum_{\beta(\neq\alpha)}U_{\alpha,\beta}e^{i\bm{k}\cdot(\bm{R}_j-\bm{R}_{j'})}\notag\\
&\times c_{j,\alpha}^\dag c_{j',\alpha}(n_{j,\beta}-n_{j',\beta})-\mathrm{H.c.}.\label{eq_H_O}
\end{align}
For the initial states $\ket{\psi_{\mathrm{S}}}\ (\mathrm{S}=\eta,\mathrm{FM})$, we have
\begin{align}
\langle O_{\bm{k},\alpha}(t)\rangle_{\mathrm{S}}\equiv&\frac{\bra{\psi_{\mathrm{S}}}e^{iHt}O_{\bm{k},\alpha}e^{-iHt}\ket{\psi_{\mathrm{S}}}}{\braket{\psi_{\mathrm{S}}|\psi_{\mathrm{S}}}}\notag\\
=&\frac{\bra{\psi_{\mathrm{S}}}e^{iTt}O_{\bm{k},\alpha}e^{-iTt}\ket{\psi_{\mathrm{S}}}}{\braket{\psi_{\mathrm{S}}|\psi_{\mathrm{S}}}}\notag\\
=&\frac{\bra{\psi_{\mathrm{S}}}O_{\bm{k},\alpha}\ket{\psi_{\mathrm{S}}}}{\braket{\psi_{\mathrm{S}}|\psi_{\mathrm{S}}}}=\langle O_{\bm{k},\alpha}(0)\rangle_{\mathrm{S}},
\label{eq_O_evol}
\end{align}
from Eqs.~\eqref{eq_H_ini_eta}, \eqref{eq_H_ini_ferro}, and \eqref{eq_T_O}. The initial distributions of the first component are given by (see Appendix \ref{sec_mom_dist_integ})
\begin{align}
\langle O_{\bm{k},1}(0)\rangle_{\eta}=&\langle O_{\bm{k},1}\rangle_1+\sum_{\alpha=2}^N\frac{M_\alpha}{N_{\mathrm{s}}-n}\langle 1-O_{\bm{k},1}\rangle_1,\label{eq_mom_1_eta}\\
\langle O_{\bm{k},1}(0)\rangle_{\mathrm{FM}}=&\langle O_{\bm{k},1}\rangle_1-\sum_{\alpha=2}^N\frac{M_\alpha}{n}\langle O_{\bm{k},1}\rangle_1,\label{eq_mom_1_ferro}
\end{align}
and for $\alpha\neq1$ by
\begin{align}
\langle O_{\bm{k},\alpha}(0)\rangle_{\eta}=&\frac{M_\alpha}{N_{\mathrm{s}}-n}\langle 1-O_{\bm{Q}-\bm{k},1}\rangle_1,\label{eq_mom_alpha_eta}\\
\langle O_{\bm{k},\alpha}(0)\rangle_{\mathrm{FM}}=&\frac{M_\alpha}{n}\langle O_{\bm{k},1}\rangle_1,\label{eq_mom_alpha_ferro}
\end{align}
where
\begin{equation}
\langle \cdots\rangle_1\equiv\frac{\bra{\phi_1}\cdots\ket{\phi_1}}{\braket{\phi_1|\phi_1}}.
\end{equation}

The observable \eqref{eq_O_evol} is clearly time-independent and retains the memory of the initial states, indicating the absence of thermalization. We note that the observables are not conserved quantities of the Hamiltonian because of the interaction term [see Eqs.~\eqref{eq_H_O}]. 
Physically, this means that the interaction can lead to the relaxation of these observables, but it cannot in the integrable subsectors where the interaction term becomes a constant. 

The existence of the non-thermalizing subsectors implies that special initial states taken from the subsectors $\mathcal{H}_\eta^{\{ n_\alpha\}}$ and $\mathcal{H}_{\mathrm{FM}}^{\{ n_\alpha\}}$ may defy thermalization, despite the fact that generic initial states are expected to show thermalization due to the non-integrable nature of the MHM. 
Since the MHM has, in general, neither the $\eta$ symmetry nor the SU($N$) symmetry, energy eigenstates of this model cannot be distinguished by the quantum numbers of these symmetries. 
In fact, when the internal symmetry of the MHM is U(1)$^N$ (see Sec.~\ref{sec_model}), generic energy eigenstates after resolving all the conserved quantities can only be labeled by the energy and the number of each component, and the subsectors $\mathcal{H}_\eta^{\{ n_\alpha\}}$ and $\mathcal{H}_{\mathrm{FM}}^{\{ n_\alpha\}}$ are embedded in the Hilbert subspace of eigenstates having the same particle number for each component. 
Since the SU($N$) ferromagnetic states \eqref{eq_SUNferro_ex} are eigenstates of the generalized Hubbard models \eqref{eq_Hubbard_2int}-\eqref{eq_H_3_2}, these models show the non-ergodic time evolution as well \footnote{Like the primary $N$-color $\eta$-pairing states of the MHM, the SU($N$) ferromagnetic states can be regarded as quantum many-body scar states of the MHM and the generalized Hubbard models that do not have the SU($N$) symmetry.}. 

We note that the non-ergodic time evolution described in this subsection is distinct from the persistent oscillations in Sec.~\ref{sec_oscil}, which require the coupling between the $\eta$-pairing states with different particle numbers in some components. Thus, the MHM exhibits two non-ergodic features: persistent oscillations between quantum many-body scar states and non-thermalization due to the presence of integrable subsectors.

\subsection{Dissipation-induced non-thermalization\label{sec_diss}}

The weak ergodicity breaking in the MHM suggests the possibility of observing non-thermalizing dynamics in experiments. 
However, to observe non-thermalization due to partial integrability, an initial state should belong to the integrable subspaces (or at least should have sufficiently large weights on those subspaces). This may imply that the non-thermalizing dynamics requires fine-tuning of an initial state and that experimentally accessible initial states would thermalize in most cases. 
To circumvent this problem, here we propose to utilize dissipation for the observation of non-thermalization caused by partial integrability.

Suppose that the system described by the MHM is coupled to a Markovian environment. The dynamics of the density matrix $\rho(t)$ of the system is  governed by the quantum master equation \cite{Lindblad76, GKS76, BreuerPetruccione}
\begin{equation}
\frac{d\rho}{dt}=-i[H,\rho]+\mathcal{D}_{s}[\rho].
\label{eq_master}
\end{equation}
Here, we consider two types of dissipators $\mathcal{D}_s$ specified by the index $s=1,2$.  
One dissipator is given by
\begin{align}
\mathcal{D}_{1}[\rho]=&\sum_{j}\sum_{\alpha=2}^N\Gamma_{j,\alpha}\notag\\
&\times\left(F_{j,1,\alpha}\rho F_{j,1,\alpha}^\dag -\frac{1}{2}\{ F_{j,1,\alpha}^\dag F_{j,1,\alpha},\rho\}\right)\notag\\
&+\sum_j\sum_{\alpha<\beta\ (\alpha\neq1)}\tilde{\Gamma}_{j,\alpha,\beta}\notag\\
&\times\left(L_{j,\alpha,\beta}\rho L_{j,\alpha,\beta}^\dag -\frac{1}{2}\{ L_{j,\alpha,\beta}^\dag L_{j,\alpha,\beta},\rho\}\right),
\end{align}
with $\Gamma_{j,\alpha}>0$ and $\tilde{\Gamma}_{j,\alpha,\beta}>0$, 
and the other dissipator is given by
\begin{align}
\mathcal{D}_{2}[\rho]=&\sum_{j}\sum_{\alpha<\beta}\tilde{\Gamma}_{j,\alpha,\beta}\notag\\
&\times\left(L_{j,\alpha,\beta}\rho L_{j,\alpha,\beta}^\dag -\frac{1}{2}\{ L_{j,\alpha,\beta}^\dag L_{j,\alpha,\beta},\rho\}\right),
\end{align}
with $\tilde{\Gamma}_{j,\alpha,\beta}>0$. 
The first Lindblad operator 
\begin{equation}
F_{j,1,\alpha}= c_{j,1}^\dag c_{j,\alpha}
\label{eq_Lindblad_spinflip}
\end{equation}
induces a spin-flip process, which can be realized by, e.g., the decay of atomic states due to spontaneous emission of a  photon \cite{Nakagawa21}. 
The second Lindblad operator
\begin{equation}
L_{j,\alpha,\beta}= c_{j,\alpha}c_{j,\beta}
\label{eq_Lindblad_loss}
\end{equation}
describes two-body loss of particles. Such two-body loss occurs in ultracold atoms that experience two-body inelastic collisions \cite{FossFeig12, Nakagawa19, Nakagawa20, Rosso22}. 

For the quantum master equation \eqref{eq_master} with the dissipator $\mathcal{D}_s$, we define the Hilbert subspace $\mathcal{H}_{\mathrm{D},s}^{\{ n_\alpha\}}$ by
\begin{align}
\mathcal{H}_{\mathrm{D},1}^{\{ n_\alpha\}}\equiv\mathrm{span}[\{& \ket{E^{\{ n_\alpha\}}}|\ H\ket{E^{\{ n_\alpha\}}}=E^{\{ n_\alpha\}}\ket{E^{\{ n_\alpha\}}},\notag\\
&\sum_jn_{j,\alpha}\ket{E^{\{ n_\alpha\}}}=n_\alpha\ket{E^{\{ n_\alpha\}}}\ (\forall \alpha),\notag\\
&F_{j,1,\alpha}\ket{E^{\{ n_\alpha\}}}=0\ (\forall j,\alpha=2,\cdots,N), \notag\\
&L_{j,\alpha,\beta}\ket{E^{\{ n_\alpha\}}}=0\ (\forall j,\alpha<\beta,\alpha\neq1)\}],\\
\mathcal{H}_{\mathrm{D},2}^{\{ n_\alpha\}}\equiv\mathrm{span}[\{& \ket{E^{\{ n_\alpha\}}}|\ H\ket{E^{\{ n_\alpha\}}}=E^{\{ n_\alpha\}}\ket{E^{\{ n_\alpha\}}},\notag\\
&\sum_jn_{j,\alpha}\ket{E^{\{ n_\alpha\}}}=n_\alpha\ket{E^{\{ n_\alpha\}}}\ (\forall \alpha),\notag\\
&L_{j,\alpha,\beta}\ket{E^{\{ n_\alpha\}}}=0\ (\forall j,\alpha<\beta)\}].
\end{align}
We shall call the subspace $\mathcal{H}_{\mathrm{D},s}^{\{ n_\alpha\}}$ the dark space, as it is spanned by dark states \cite{Diehl08, Kraus08} which are immune to dissipation. 
The dark space is also a decoherence-free subspace \cite{Lidar98, Lidar03} and a state $\ket{\psi}\in\mathcal{H}_{D,s}^{\{ n_\alpha\}}$ undergoes unitary time evolution since $\mathcal{D}_s[\ket{\psi}\bra{\psi}]=0$. 

It can easily be shown that the integrable subsectors $\mathcal{H}_\eta^{\{ n_\alpha\}}$ [Eq.~\eqref{eq_sp_eta}] and $\mathcal{H}_{\mathrm{FM}}^{\{ n_\alpha\}}$ [Eq.~\eqref{eq_sp_ferro}] are included in the dark space:
\begin{align}
\mathcal{H}_{\eta}^{\{ n_\alpha\}}\subset \mathcal{H}_{\mathrm{D},1}^{\{ n_\alpha\}},\label{eq_etaDFS}\\
\mathcal{H}_{\mathrm{FM}}^{\{ n_\alpha\}}\subset \mathcal{H}_{\mathrm{D},2}^{\{ n_\alpha\}}.\label{eq_ferroDFS}
\end{align}
In other words, the generalized $\eta$-pairing states \eqref{eq_SUNeta_ex} and the SU($N$) ferromagnetic states \eqref{eq_SUNferro_ex} are dark states of the quantum master equation for the dissipators $\mathcal{D}_1$ and $\mathcal{D}_2$, respectively. 
In particular, this means that the primary $N$-color $\eta$-pairing states \eqref{eq_SUNeta} can be prepared as dark states [this can be shown by setting $\mathcal{H}_1^{(0)}\equiv\mathrm{span}[\ket{0}]$ in Eq.~\eqref{eq_sp_eta}].
To show \eqref{eq_etaDFS} and \eqref{eq_ferroDFS}, it suffices to show that the eigenstates \eqref{eq_SUNeta_ex} and \eqref{eq_SUNferro_ex} are annihilated by the Lindblad operators \eqref{eq_Lindblad_spinflip} and \eqref{eq_Lindblad_loss}, respectively. 
This can be confirmed from the Fock-state representation [as in Eq.~\eqref{eq_SUNeta_siterep}] of the eigenstates \eqref{eq_SUNeta_ex}, $F_{j,1,\alpha}\eta_{j,\alpha,1}^\dag\ket{0}=L_{j,\alpha,\beta}\eta_{j,\alpha,1}^\dag\ket{0}=F_{j,1,\alpha}c_{j,1}^\dag\ket{0}=L_{j,\alpha,\beta}c_{j,1}^\dag\ket{0}=0\ (\alpha\neq 1;\alpha<\beta)$, and the fact that the SU($N$) ferromagnetic states do not have sites with multiple occupancy. 
This result is a natural generalization of that for the two-component dissipative Hubbard models studied in Refs.~\cite{FossFeig12, Nakagawa20, Nakagawa21}. 
See also a recent study of an SU(3) dissipative Hubbard model \cite{Rosso22}.

The dark space is invariant under the time evolution and thus a state in the dark space evolves in time within the same space. 
Since the time evolution in the decoherence-free subspace is unitary, an initial state in the integrable subsectors $\mathcal{H}_\eta^{\{ n_\alpha\}}$ or $\mathcal{H}_{\mathrm{FM}}^{\{ n_\alpha\}}$ undergoes the same non-thermalizing dynamics as in the case without dissipation. 
Moreover, if an initial state is not included in the integrable subsectors, the dynamics of the density matrix $\rho(t)$ is confined in the dark space due to dissipation after a sufficiently long time. This suggests that the non-thermalizing dynamics may be observed for a wide range of initial states without fine-tuning. 

Specifically, the density matrix $\rho_s(t)$ of the system after a sufficiently long time evolution under the quantum master equation with the dissipator $\mathcal{D}_s$ can be written as
\begin{align}
\rho_s(t)=&\sum_{n_1+\cdots+n_N\leq n_{\mathrm{ini}}}p_s(\{n_\alpha\})\rho_s^{\{n_\alpha\}}(t),\label{eq_SS}
\end{align}
where $n_{\mathrm{ini}}$ is the particle number in the initial state,
\begin{align}
\rho_s^{\{n_\alpha\}}(t)=&\sum_{l,m}c_{s,l,m}({\{n_\alpha\}})\exp[i(E_{s,m}^{\{n_\alpha\}}-E_{s,l}^{\{n_\alpha\}})t]\notag\\
&\times\ket{E_{s,l}^{\{n_\alpha\}}}\bra{E_{s,m}^{\{n_\alpha\}}}
\end{align}
is the density matrix satisfying $\mathrm{Tr}[\rho_s^{\{n_\alpha\}}(t)]=1$, 
and $\ket{E_{s,l}^{\{n_\alpha\}}}$ denotes an energy eigenstate in the dark space $\mathcal{H}_{\mathrm{D},s}^{\{ n_\alpha\}}$ with $n_\alpha$ particles in the $\alpha$th component $(\alpha=1,\cdots,N)$ and eigenvalue $E_{s,l}^{\{n_\alpha\}}$.  
In Eq.~\eqref{eq_SS}, we have ignored the exponentially small contribution from the outside of the dark space. 
Here, a weak U(1)$^N$ symmetry of the quantum master equation (i.e., the invariance of Eq.~\eqref{eq_master} under $c_{j,\alpha}\to e^{i\theta_\alpha}c_{j,\alpha}$) ensures that the density matrix \eqref{eq_SS} does not have off-diagonal components between sectors with different $\{n_\alpha\}$ \cite{Buca12, Albert14}. 
The coefficient $p_s(\{ n_\alpha\})$ corresponds to the probability of $n_\alpha$ particles being found in the $\alpha$th component $(\alpha=1,\cdots,N)$ in the state $\rho_s(t)$. 
The coefficients $p_s(\{ n_\alpha\})$ and $c_{s,l,m}(\{ n_\alpha\})$ depend on the initial condition.

The non-thermalizing dynamics in the dark space can now be explicitly shown. We first note that
\begin{align}
F_{j,1,\alpha}^\dag F_{j,1,\alpha}=&n_{j,\alpha}(1-n_{j,1}),\\
L_{j,\alpha,\beta}^\dag L_{j,\alpha,\beta}=&n_{j,\alpha}n_{j,\beta},
\end{align}
and that the interaction term acts on states in the dark spaces as
\begin{align}
V\ket{\psi_1}=&\sum_{\alpha=2}^N U_{1,\alpha}n_{\alpha}\ket{\psi_1}\ \mathrm{for}\ \ket{\psi_1}\in\mathcal{H}_{\mathrm{D},1}^{\{ n_\alpha\}},\label{eq_Vpsi1}\\
V\ket{\psi_2}=&0\ \mathrm{for}\ \ket{\psi_2}\in\mathcal{H}_{\mathrm{D},2}^{\{ n_\alpha\}}.\label{eq_Vpsi2}
\end{align}
Then, for the translationally invariant case $t_{i,j}=t_{\mathrm{h}}$, the time evolution of the spin-resolved momentum distribution [Eq.~\eqref{eq_mom_dist}] is calculated from Eqs.~\eqref{eq_T_O}, \eqref{eq_Vpsi1}, and \eqref{eq_Vpsi2} as
\begin{align}
\langle O_{\bm{k},\alpha}(t)\rangle_s^{\{n_\alpha\}}\equiv&\mathrm{Tr}[O_{\bm{k},\alpha}\rho_s^{\{ n_\alpha\}}(t)]\notag\\
=&\mathrm{Tr}[O_{\bm{k}}e^{-iHt}\rho_s^{\{ n_\alpha\}}(0)e^{iHt}]\notag\\
=&\mathrm{Tr}[O_{\bm{k},\alpha}e^{-iTt}\rho_s^{\{ n_\alpha\}}(0)e^{iTt}]\notag\\
=&\mathrm{Tr}[O_{\bm{k},\alpha}\rho_s^{\{ n_\alpha\}}(0)].
\label{eq_O_evol_DFS}
\end{align}
Thus, the spin-resolved momentum distribution does not relax in the dark space. Since
\begin{equation}
n_\alpha=\sum_{\bm{k}}\langle O_{\bm{k},\alpha}(t)\rangle_s^{\{n_\alpha\}},
\end{equation}
the subsector of the dark space can be identified from the measured single-shot values of $n_\alpha$ within experimental accuracy, and the non-thermalizing dynamics in the subsector may be observed experimentally. Furthermore, the quantum gas microscopy \cite{Gross21} may facilitate the identification of the subsector due to high-precision measurement of $n_\alpha$.

Several remarks are in order here. First, the dissipators should appropriately be chosen so that the dark spaces are dominated by the integrable subsectors. For example, if some of the coefficients $\Gamma_{j,\alpha}$ and $\tilde{\Gamma}_{j,\alpha,\beta}$ of the dissipators are set to zero, the dark spaces may contain more states that do not belong to the integrable subsectors. 
In particular, the Lindblad operators $L_{j,\alpha,\beta}$ are included in the dissipator $\mathcal{D}_1$ to eliminate such unwanted states and may be unnecessary for reaching the $\eta$-pairing states from some initial conditions. 
Second, the Lindblad operators \eqref{eq_Lindblad_spinflip} and \eqref{eq_Lindblad_loss} may still admit the existence of dark states that lie outside the integrable subspaces. For example, some special eigenstates with no double occupancy are known to exist in the two-component Hubbard model with translation invariance \cite{Yang89, Ye18}. 
However, since the particle density of these eigenstates vanishes in the thermodynamic limit, they do not contribute to the blocks in the dark space with sufficiently large particle numbers. These states do not affect the non-thermalizing dynamics as shown in Eq.~\eqref{eq_O_evol_DFS}.
Third, the dissipation-induced non-thermalization proposed here is distinct from the dissipation-induced non-stationary dynamics of dissipative Hubbard models studied in Refs.~\cite{Buca19, Booker20, Chinzei20, Buca21}. While dissipation in the previous works induces coherent oscillations in the dark space, here we utilize dissipation to distill the integrable dynamics of effectively non-interacting particles in the MHM.

Experimentally, an SU($N$) Hubbard model with two-body loss has been realized with ultracold $^{173}$Yb atoms \cite{Sponselee18,Honda23}. The experimentally observed state with a constant but nonzero particle number is expected to be described by the density matrix \eqref{eq_SS} with $s=2$, which involves the SU($N$) ferromagnetic dark states. 
Furthermore, ultracold alkaline-earth-like atoms described by the two-orbital SU($M$) Hubbard model \eqref{eq_H_AEA} are another promising candidate for the observation of dissipation-induced non-thermalization. In this system, inelastic collision processes involving an excited state of atoms lead to two-body losses described by Lindblad operators $c_{j,e,\sigma}c_{j,e,\sigma'}$ and $c_{j,e,\sigma}c_{j,g,\sigma'}\pm c_{j,e,\sigma'}c_{j,g,\sigma}$ \cite{Sponselee18, Riegger18, Nakagawa18}. In addition, two-body loss processes between ground-state atoms with Lindblad operators $c_{j,g,\sigma}c_{j,g,\sigma'}$ can be induced by photoassociation \cite{Tomita17,Honda23}. The two-orbital SU($M$) Hubbard model \eqref{eq_H_AEA} with such two-body losses has SU($2M$) ferromagnetic states as dark states.

\section{Conclusion and outlook\label{sec_conclusion}}

We have constructed the $N$-color $\eta$-pairing eigenstates of the MHM. While the previous studies on superfluid pairing in multicomponent systems are mostly based on mean-field theory \cite{Modawi97, Ho99, Honerkamp04_2, Paananen06, Zhai07, Cherng07, Inaba09, Yip11, Inaba12, Guan13, Okanami14, Koga17}, the $\eta$-pairing mechanism allows the construction of exact eigenstates, revealing that the MHM can exhibit ODLRO coexisting with SU($N$) magnetism. The SU($N$) magnetic $\eta$ pairing is unique to multicomponent systems and absent in the ordinary spin-1/2 systems. The primary $N$-color $\eta$-pairing states are realized as dark states of the dissipator with the Lindblad operator \eqref{eq_Lindblad_spinflip}. The dissipation-induced $\eta$ pairing may provide a possible nonequilibrium pairing mechanism of multicomponent fermions that does not require low temperatures \cite{Nakagawa21}. Extension of various methods for nonequilibrium realization of $\eta$ pairing \cite{Rosch08, Kantian10, Kaneko19, Kaneko20, Ejima20, Werner19, Li20, Kitamura16, Peronaci20, Cook20, Tindall21, Tindall21_2, Diehl08, Kraus08, Bernier13, Buca19, Tindall19, Murakami21} to multicomponent systems is an important direction for future study. 
It is also worthwhile to search for conditions under which the $N$-color $\eta$-pairing states are ground states of the MHM or other Hamiltonians, as in the two-component case \cite{Essler92, Essler93, Arrachea94, Scadschneider95, deBoer95, deBoer95_2}. 
Since Yang's $\eta$-pairing states show anomalous superfluid response \cite{Tsuji21}, the $N$-color $\eta$-pairing states may realize novel SU($N$) magnetic superfluidity.

The $N$-color $\eta$-pairing eigenstates do not rely on symmetry of the Hamiltonian but arise from the spectrum generating algebras defined in Hilbert subspaces. 
This result uncovers that $\eta$ pairing in the previously studied two-component case is rather special; the Hilbert subspace in which the spectrum generating algebra is defined covers the entire Hilbert space only if the number of components is two.

We have also shown that the $N$-color $\eta$-pairing eigenstates are quantum many-body scars which exhibit weak ergodicity breaking in the MHM. The crucial difference between the $N=2$ and $N\geq 3$ cases is that the Hubbard interaction for multicomponent fermions can break the $\eta$-pairing symmetry and the SU($N$) symmetry (see also Appendix \ref{sec_SUN_Hubbard} for the importance of breaking the latter symmetry). 
As the $\eta$-pairing scar states require tailored perturbations in the two-component case \cite{Mark20_2, Moudgalya20}, 
the present result suggests that the interaction in the two-component Hubbard model is too restrictive in this respect; if the number of components is increased at least by one, the Hubbard interaction term acquires sufficient flexibility so that the $\eta$-pairing states can be true quantum many-body scars. 

The exact eigenstates constitute integrable subsectors of the MHM, leading to non-thermalization in those sectors. While generic initial states in a non-integrable system are expected to relax to the (micro-)canonical ensemble, a state inside an integrable sector will relax to a generalized Gibbs ensemble \cite{Rigol07, Vidmar16}. Therefore, the partial integrability raises an interesting possibility of relaxation to an ensemble that is neither canonical nor generalized Gibbs ensembles if an initial state has comparable weights on integrable sectors and non-integrable sectors. 
To realize such unconventional relaxation, the dissipation-induced non-thermalization discussed in Sec.~\ref{sec_diss} can be exploited, since dissipation can selectively decrease the weights on the non-integrable sectors. Because the dissipative processes in Sec.~\ref{sec_diss} can be controllablly introduced in cold atoms \cite{Tomita17}, one may smoothly change the steady-state ensemble of the MHM from the canonical ensemble to the generalized Gibbs ensemble by changing the time during which the system is subject to dissipation.

Non-ergodic dynamics in the Hubbard model has been observed in cold-atom quantum simulators with sufficiently strong disordered or tilted potentials
\cite{Schreiber15, Bordia17, Scherg21, Kohlert21}. 
The quantum many-body scars and the partial integrability in the MHM provide the mechanisms of weak ergodicity breaking in clean and translationally invariant Hubbard systems. Given the high controllability of quantum simulators, including the techniques of dissipation engineering, the MHM offers a unique platform for the study of many-body physics and quantum statistical mechanics.

\textit{Note added}.---\ 
After the initial submission of this manuscript, a related work \cite{Sun23} appeared in which the $\eta$-pairing states in multicomponent systems are discussed from a different perspective.

\begin{acknowledgments}
We thank Ryusuke Hamazaki, Takashi Mori, and Hironobu Yoshida for helpful discussions. 
M.N. was supported by KAKENHI Grant No.~JP20K14383 and No.~JP24K16989 from the Japan Society for the Promotion of Science (JSPS). 
H.K. was supported by JSPS KAKENHI Grant No.~JP18K03445, No.~JP21H05191, No.~JP23K25790, and the Inamori Foundation. 
M.U. was supported by JSPS KAKENHI Grant No.~JP22H01152. 
\end{acknowledgments}

\appendix

\section{$\eta$-pairing states of the SU($N$) Hubbard model\label{sec_SUN_Hubbard}}

If the interaction coefficients in the MHM in Eq.~\eqref{eq_Hasym} are independent of components, i.e., $U_{\alpha,\beta}=U$ for all $\alpha,\beta=1,\cdots, N\ (\alpha\neq\beta)$, then the model has the SU($N$) [in fact, U($N$)] symmetry since
\begin{equation}
[F_{\alpha,\beta},H]=0,
\label{eq_SUNsym}
\end{equation}
where $\{ F_{\alpha,\beta}\}_{\alpha,\beta=1,\cdots,N}$ are the SU($N$) spin operators defined in Eq.~\eqref{eq_SUN_generator}. 
For the SU($N$) Hubbard model, the primary $N$-color $\eta$-pairing eigenstate can be constructed with the help of the SU($N$) symmetry. 
To see this, we rewrite the primary $N$-color $\eta$-pairing state \eqref{eq_SUNeta} as
\begin{align}
\ket{\psi_{M_2,\cdots,M_N}}=&\frac{M_2!}{(M_2+\cdots+M_N)!}(F_{3,2})^{M_3}\cdots(F_{N,2})^{M_N}\notag\\
&\times(\eta_{2,1}^\dag)^{M_2+\cdots+M_N}\ket{0}.
\label{eq_SUNeta_reduced}
\end{align}
Here the state
\begin{equation}
(\eta_{2,1}^\dag)^{M_2+\cdots+M_N}\ket{0}
\label{eq_SU2eta}
\end{equation}
is equivalent to Yang's $\eta$-pairing state \cite{Yang89} and an eigenstate of the SU($N$) Hubbard model, since the Hamiltonian reduces to the two-component Hubbard model for this state. 
Combining this fact with the SU($N$) symmetry \eqref{eq_SUNsym}, we can readily show that the state \eqref{eq_SUNeta_reduced} is an eigenstate of the SU($N$) Hubbard model. 
We note that the SU($N$) symmetry exists in the MHM only for the special case of $U_{\alpha,\beta}=U$. 
The key property of the primary $N$-color $\eta$-pairing state \eqref{eq_SUNeta_reduced} is that it remains as an eigenstate if the interaction term does not have the SU($N$) symmetry.

Because of the SU($N$) symmetry, all the states generated by the action of the spin operators $\{ F_{\alpha,\beta}\}$ on Yang's $\eta$-pairing state \eqref{eq_SU2eta} become eigenstates of the SU($N$) Hubbard model. 
The SU($N$) Hubbard model therefore has more $\eta$-pairing eigenstates that are absent in the general MHM. For example, a state
\begin{align}
F_{4,1}\eta_{2,1}^\dag \eta_{3,1}^\dag\ket{0}=&\eta_{3,4}^\dag\eta_{2,1}^\dag\ket{0}+\eta_{2,4}^\dag\eta_{3,1}^\dag\ket{0}
\label{eq_additional_eta}
\end{align}
is an eigenstate of the SU($N$) Hubbard model. We note that each term on the right-hand side of Eq.~\eqref{eq_additional_eta} is not an eigenstate. Thus, even with the SU($N$) symmetry, not all $N$-color $\eta$-pairing states can be eigenstates of the MHM. 
The $\eta$-pairing eigenstates are characterized by the absence of sites that are occupied by more than two particles. This can be exemplified for the state \eqref{eq_additional_eta} as
\begin{align}
&(\eta_{3,4}^\dag\eta_{2,1}^\dag+\eta_{2,4}^\dag\eta_{3,1}^\dag)\ket{0}\notag\\
=&\sum_{i,j}e^{i\bm{Q}\cdot(\bm{R}_i+\bm{R}_j)}(c_{i,3}^\dag c_{i,4}^\dag c_{j,2}^\dag c_{j,1}^\dag+c_{i,2}^\dag c_{i,4}^\dag c_{j,3}^\dag c_{j,1}^\dag)\ket{0}.
\label{eq_additional_eta_2}
\end{align}
Since
\begin{align}
c_{j,3}^\dag c_{j,4}^\dag c_{j,2}^\dag c_{j,1}^\dag+c_{j,2}^\dag c_{j,4}^\dag c_{j,3}^\dag c_{j,1}^\dag=0
\label{eq_vanishing_four}
\end{align}
due to the anticommutativity of fermionic operators, the terms with $i = j$ on the right-hand side of Eq.~\eqref{eq_additional_eta_2} vanish identically. 
The vanishing probability of quadruply occupied sites ensured by Eq.~\eqref{eq_vanishing_four} is physically understood as complete destructive interference between two $\eta$-pairing states on the right-hand side of Eq.~\eqref{eq_additional_eta}. Such an interference effect is not guaranteed if the interaction does not have the SU($N$) symmetry, since these two $\eta$-pairing states may not have the same interaction energy. 

Testing the ETH for the $\eta$-pairing eigenstates of the SU($N$) Hubbard model needs some cares. In this case, because of the U($N$) symmetry, energy eigenstates should first be labeled by the values of conserved quantities and the ETH should be tested within each symmetry sector. 
Let $\mathcal{H}_{1,2}$ be a Hilbert subspace that has only two components $\alpha=1,2$. In this subspace, the SU($N$) Hubbard model reduces to the two-component Hubbard model and thus
\begin{equation}
[\eta_{2,1}^\dag \eta_{2,1},H]\ket{\psi}=0
\label{eq_etapm_H12}
\end{equation}
holds for arbitrary $\ket{\psi}\in\mathcal{H}_{1,2}$. 
Then, from Eqs.~\eqref{eq_SUNsym}, \eqref{eq_etapm_H12}, and
\begin{equation}
\Biggl[\sum_{\alpha\neq\beta}\eta_{\alpha,\beta}^\dag\eta_{\alpha,\beta},F_{\gamma,\delta}\Biggr]=0,
\end{equation}
it follows that for any such $\ket{\psi}$ and any product $A_F=\prod_{\alpha,\beta}(F_{\alpha,\beta})^{M_{\alpha,\beta}}$ of the spin operators, we have
\begin{equation}
\Biggl[\sum_{\alpha\neq\beta}\eta_{\alpha,\beta}^\dag\eta_{\alpha,\beta},H\Biggr]A_F\ket{\psi}=0.
\label{eq_etapm_AFH}
\end{equation}
Equation \eqref{eq_etapm_AFH} shows that the symmetrized physical quantity $\sum_{\alpha\neq\beta}\eta_{\alpha,\beta}^\dag\eta_{\alpha,\beta}$, which is proportional to the total number of doublons at momentum $\bm{Q}$ [see Eq.~\eqref{eq_etapm}], is conserved in a symmetry sector to which an $\eta$-pairing eigenstate $A_F(\eta_{2,1}^\dag)^n\ket{0}$ belongs. The existence of the emergent conserved quantity in a given symmetry sector can further decompose the sector into subsectors labeled by the value of the conserved quantity. The $\eta$-pairing eigenstates of the SU($N$) Hubbard model may satisfy the ETH in such subsectors, as Yang's $\eta$-pairing state does in the two-component Hubbard model \cite{Vafek17}. In this sense, the SU($N$) symmetry of the interaction term needs to be broken for the weak ergodicity breaking due to the $\eta$-pairing eigenstates. 
If the interaction term of the MHM has a symmetry larger than U(1)$^N$, then a similar discussion can be made for $\eta$-pairing eigenstates that can be obtained from Yang's $\eta$-pairing state by applying appropriate spin operators.

\section{Calculation of correlations functions\label{sec_corr_calc}}

Here, we calculate the pair correlation functions for the $N$-color $\eta$-pairing states. To this end, it is convenient to introduce an unnormalized coherent state
\begin{equation}
\ket{\xi_2,\xi_3,\cdots,\xi_N}\equiv e^{\xi_2\eta_{2,1}^\dag}e^{\xi_3\eta_{3,1}^\dag}\cdots e^{\xi_N\eta_{N,1}^\dag}\ket{0},
\end{equation}
where $\xi_2,\cdots,\xi_N\in\mathbb{C}$. 
With local $\eta$ operators [see Eq.~\eqref{eq_eta_op}], we can rewrite the coherent state as
\begin{align}
\ket{\xi_2,\xi_3,\cdots,\xi_N}=&\exp[\xi_2\eta_{2,1}^\dag+\cdots\xi_N\eta_{N,1}^\dag]\ket{0}\notag\\
=&\prod_j \exp[\xi_2\eta_{j,2,1}^\dag+\cdots\xi_N\eta_{j,N,1}^\dag]\ket{0}\notag\\
=&\prod_j(1+\xi_2\eta_{j,2,1}^\dag+\cdots\xi_N\eta_{j,N,1}^\dag)\ket{0},
\label{eq_generating_state}
\end{align}
where we have used $[\eta_{i,\alpha,\beta}^\dag,\eta_{j,\gamma,\delta}^\dag]=0$ and
\begin{align}
(\xi_2\eta_{j,2,1}^\dag+\cdots\xi_N\eta_{j,N,1}^\dag)^2=&\sum_{\alpha,\beta=2}^N\xi_\alpha\xi_\beta\eta_{j,\alpha,1}^\dag\eta_{j,\beta,1}^\dag
=0.
\end{align}
The squared norm of the coherent state is calculated to be
\begin{widetext}
\begin{align}
\braket{\xi_2,\cdots,\xi_N|\xi_2,\cdots,\xi_N}=&\prod_j\bra{0_j}(1+\xi_2^*\eta_{j,2,1}+\cdots+\xi_N^*\eta_{j,N,1})(1+\xi_2\eta_{j,2,1}^\dag+\cdots\xi_N\eta_{j,N,1}^\dag)\ket{0_j}\notag\\
=&\prod_j\Bigl(\braket{0_j|0_j}+|\xi_2|^2\bra{0_j}\eta_{j,2,1}\eta_{j,2,1}^\dag\ket{0_j}+\cdots+|\xi_N|^2\bra{0_j}\eta_{j,N,1}\eta_{j,N,1}^\dag\ket{0_j}\Bigr)\notag\\
=&(1+|\xi_2|^2+\cdots+|\xi_N|^2)^{N_{\mathrm{s}}}\notag\\
=&\sum_{n_1+\cdots+n_N=N_{\mathrm{s}}}\binom{N_{\mathrm{s}}}{n_1,\cdots,n_N}|\xi_2|^{2n_2}\cdots|\xi_N|^{2n_N},
\label{eq_norm_expansion1}
\end{align}
where $\ket{0_j}$ denotes the empty state at site $j$, and $\binom{n}{n_1,\cdots,n_N}\equiv\frac{n!}{n_1!\cdots n_N!}$.
At the same time, by using
\begin{align}
&\ket{\xi_2,\cdots,\xi_N}
=\sum_{n_2,\cdots,n_N=0}^\infty\frac{\xi_2^{n_2}\cdots\xi_N^{n_N}}{n_2!\cdots n_N!}(\eta_{2,1}^\dag)^{n_2}\cdots (\eta_{N,1}^\dag)^{n_N}\ket{0},
\label{eq_generating_state_2}
\end{align}
the squared norm is written as
\begin{align}
\braket{\xi_2,\cdots,\xi_N|\xi_2,\cdots,\xi_N}
=&\sum_{n_2,\cdots,n_N=0}^\infty\frac{|\xi_2|^{2n_2}\cdots|\xi_N|^{2n_N}}{(n_2!)^2\cdots (n_N!)^2}
\bra{0}(\eta_{N,1})^{n_N}\cdots(\eta_{2,1})^{n_2}(\eta_{2,1}^\dag)^{n_2}\cdots (\eta_{N,1}^\dag)^{n_N}\ket{0}.
\label{eq_norm_expansion2}
\end{align}
By comparing Eqs.~\eqref{eq_norm_expansion1} and \eqref{eq_norm_expansion2}, we obtain the squared norm of the primary $N$-color $\eta$-pairing state [Eq.~\eqref{eq_SUNeta}] as
\begin{align}
\mathcal{N}(M_2,M_3,\cdots,M_N)\equiv&\braket{\psi_{M_2,M_3,\cdots,M_N}|\psi_{M_2,M_3,\cdots,M_N}}\notag\\
=&(M_2!)^2\cdots(M_N!)^2\binom{N_{\mathrm{s}}}{N_{\mathrm{s}}-M_2-\cdots-M_N,M_2,\cdots,M_N}\notag\\
=&\frac{N_{\mathrm{s}}!M_2!M_3!\cdots M_N!}{(N_{\mathrm{s}}-M_2\cdots-M_N)!}.
\label{eq_norm_SU(N)}
\end{align}

The $\eta$-pair correlation function for the coherent state is calculated as
\begin{align}
\bra{\xi_2,\cdots,\xi_N}\eta_{i,\alpha,1}^\dag\eta_{j,\alpha,1}\ket{\xi_2,\cdots,\xi_N}
=&(1+|\xi_2|^2+\cdots+|\xi_N|^2)^{N_{\mathrm{s}}-2}\notag\\
&\times\bra{0_i}(1+\xi_2^*\eta_{i,2,1}+\cdots+\xi_N^*\eta_{i,N,1})\eta_{i,\alpha,1}^\dag(1+\xi_2\eta_{i,2,1}^\dag+\cdots\xi_N\eta_{i,N,1}^\dag)\ket{0_i}\notag\\
&\times\bra{0_j}(1+\xi_2^*\eta_{j,2,1}+\cdots+\xi_N^*\eta_{j,N,1})\eta_{j,\alpha,1}(1+\xi_2\eta_{j,2,1}^\dag+\cdots\xi_N\eta_{j,N,1}^\dag)\ket{0_j}\notag\\
=&(1+|\xi_2|^2+\cdots+|\xi_N|^2)^{N_{\mathrm{s}}-2}\times\bra{0_i}\xi_\alpha^*\eta_{i,\alpha,1}\eta_{i,\alpha,1}^\dag\ket{0_i}\bra{0_j}\xi_\alpha\eta_{j,\alpha,1}\eta_{j,\alpha,1}^\dag\ket{0_j}\notag\\
=&|\xi_\alpha|^2(1+|\xi_2|^2+\cdots+|\xi_N|^2)^{N_{\mathrm{s}}-2}\notag\\
=&|\xi_\alpha|^2\sum_{n_1+\cdots+n_N=N_{\mathrm{s}}-2}\binom{N_{\mathrm{s}}-2}{n_1,\cdots,n_N}|\xi_2|^{2n_2}\cdots|\xi_N|^{2n_N},
\end{align}
where we assume $i\neq j$. 
By using Eq.~\eqref{eq_generating_state_2}, the same quantity is written as
\begin{align}
\bra{\xi_2,\cdots,\xi_N}\eta_{i,\alpha,1}^\dag\eta_{j,\alpha,1}\ket{\xi_2,\cdots,\xi_N}
=&\sum_{n_2,\cdots,n_N=0}^\infty\frac{|\xi_2|^{2n_2}\cdots|\xi_N|^{2n_N}}{(n_2!)^2\cdots (n_N!)^2}\notag\\
&\times\bra{0}(\eta_{N,1})^{n_N}\cdots(\eta_{2,1})^{n_2}\eta_{i,\alpha,1}^\dag\eta_{j,\alpha,1}(\eta_{2,1}^\dag)^{n_2}\cdots (\eta_{N,1}^\dag)^{n_N}\ket{0}.
\end{align}
Thus, we obtain
\begin{align}
&\bra{0}(\eta_{N,1})^{n_N}\cdots(\eta_{2,1})^{n_2}\eta_{i,\alpha,1}^\dag\eta_{j,\alpha,1}(\eta_{2,1}^\dag)^{n_2}\cdots (\eta_{N,1}^\dag)^{n_N}\ket{0}\notag\\
=&(n_2!)^2\cdots (n_N!)^2\binom{N_{\mathrm{s}}-2}{n_1,\cdots,n_{\alpha-1},n_\alpha-1,n_{\alpha+1},\cdots,n_N}\ \ (\mathrm{where}\ n_1\equiv N_{\mathrm{s}}-n_2-\cdots-n_N-1)\notag\\
=&n_\alpha\frac{(N_{\mathrm{s}}-2)!n_2!\cdots n_N!}{(N_{\mathrm{s}}-n_2-\cdots-n_N-1)!},
\end{align}
from which the pair correlation function of the primary $N$-color $\eta$-pairing state is given by
\begin{align}
\frac{\bra{\psi_{M_2,M_3,\cdots,M_N}}c_{i,\alpha}^\dag c_{i,1}^\dag c_{j,1}c_{j,\alpha}\ket{\psi_{M_2,M_3,\cdots,M_N}}}{\braket{\psi_{M_2,M_3,\cdots,M_N}|\psi_{M_2,M_3,\cdots,M_N}}}=\frac{M_\alpha(N_{\mathrm{s}}-M_2-\cdots-M_N)}{N_{\mathrm{s}}(N_{\mathrm{s}}-1)}e^{i\bm{Q}\cdot(\bm{R}_i-\bm{R}_j)}.
\label{eq_paircorr_SU(N)_norm_app}
\end{align}
Equations \eqref{eq_norm_SU(N)} and \eqref{eq_paircorr_SU(N)_norm_app} are generalizations of the results for the $\eta$-pairing state of the two-component Hubbard model obtained in Ref.~\cite{Yang89}. 
\end{widetext}

The norm and the pair correlation functions of the three-color $\eta$-pairing state can similarly be calculated by using the following coherent state
\begin{align}
\ket{a,b,c}\equiv& e^{a\eta_{1,2}^\dag}e^{b\eta_{2,3}^\dag}e^{c\eta_{3,1}^\dag}\ket{0}\notag\\
=&\prod_j(1+a\eta_{j,1,2}^\dag+b\eta_{j,2,3}^\dag+c\eta_{j,3,1}^\dag)\ket{0},
\label{eq_generating_state_SU(3)}
\end{align}
where we use $\eta_{j,\alpha,\beta}^\dag\eta_{j,\gamma,\delta}^\dag=0$ for $(\alpha,\beta),(\gamma,\delta)=(1,2),(2.3),(3,1)$ in deriving the second equality. 
It follows from
\begin{align}
\braket{a,b,c|a,b,c}
=&(1+|a|^2+|b|^2+|c|^2)^{N_{\mathrm{s}}}\notag\\
=&\sum_{k+l+m+n=N_{\mathrm{s}}}\binom{N_{\mathrm{s}}}{k,l,m,n}|a|^{2l}|b|^{2m}|c|^{2n},
\end{align}
and
\begin{align}&
\braket{a,b,c|a,b,c}\notag\\
=&\sum_{l,m,n=0}^\infty\frac{|a|^{2l}|b|^{2m}|c|^{2n}}{(l!)^2(m!)^2(n!)^2}\notag\\
&\times\bra{0}(\eta_{3,1})^{n}(\eta_{2,3})^{m}(\eta_{1,2})^{l}(\eta_{1,2}^\dag)^{l}(\eta_{2,3}^\dag)^{m}(\eta_{3,1}^\dag)^{n}\ket{0}
\end{align}
that the squared norm of the three-color $\eta$-pairing state [Eq.~\eqref{eq_SU3eta}] is given by
\begin{align}
\braket{\psi_{l,m,m}^{(3)}|\psi_{l,m,n}^{(3)}}
=&\frac{N_{\mathrm{s}}!l!m!n!}{(N_{\mathrm{s}}-l-m-n)!}.
\label{eq_SU3eta_norm_app}
\end{align}
The pair correlation function of the three-color $\eta$-pairing state is given by
\begin{align}
\frac{\bra{\psi_{l,m,n}^{(3)}}\eta_{i,\alpha,\beta}^\dag\eta_{j,\alpha,\beta}\ket{\psi_{l,m,n}^{(3)}}}{\braket{\psi_{l,m,n}^{(3)}|\psi_{l,m,n}^{(3)}}}=&\frac{r_{\alpha,\beta}(N_{\mathrm{s}}-l-m-n)}{N_{\mathrm{s}}(N_{\mathrm{s}}-1)},
\end{align}
where $r_{1,2}=l,r_{2,3}=m$, and $r_{3,1}=n$.

Next, we calculate the spin correlation functions for the $N$-color $\eta$-pairing states. For $\alpha\neq\beta$ with $\alpha,\beta\neq1$, we obtain the spin correlation functions for the coherent state 
\begin{widetext}
\begin{align}
\bra{\xi_2,\cdots,\xi_N}F_{i,\alpha,\beta}F_{j,\beta,\alpha}\ket{\xi_2,\cdots,\xi_N}
=&(1+|\xi_2|^2+\cdots+|\xi_N|^2)^{N_{\mathrm{s}}-2}\notag\\
&\times\bra{0_i}(1+\xi_2^*\eta_{i,2,1}+\cdots+\xi_N^*\eta_{i,N,1})F_{i,\alpha,\beta}(1+\xi_2\eta_{i,2,1}^\dag+\cdots\xi_N\eta_{i,N,1}^\dag)\ket{0_i}\notag\\
&\times\bra{0_j}(1+\xi_2^*\eta_{j,2,1}+\cdots+\xi_N^*\eta_{j,N,1})F_{j,\beta,\alpha}(1+\xi_2\eta_{j,2,1}^\dag+\cdots\xi_N\eta_{j,N,1}^\dag)\ket{0_j}\notag\\
=&(1+|\xi_2|^2+\cdots+|\xi_N|^2)^{N_{\mathrm{s}}-2}\notag\\
&\times\bra{0_i}\xi_\alpha^*\xi_\beta\eta_{i,\alpha,1}F_{i,\alpha,\beta}\eta_{i,\beta,1}^\dag\ket{0_i}\bra{0_j}\xi_\alpha\xi_\beta^*\eta_{j,\beta,1}F_{j,\beta,\alpha}\eta_{j,\alpha,1}^\dag\ket{0_j}\notag\\
=&|\xi_\alpha|^2|\xi_\beta|^2(1+|\xi_2|^2+\cdots+|\xi_N|^2)^{N_{\mathrm{s}}-2}\notag\\
=&|\xi_\alpha|^2|\xi_\beta|^2\sum_{n_1+\cdots+n_N=N_{\mathrm{s}}-2}\binom{N_{\mathrm{s}}-2}{n_1,\cdots,n_N}|\xi_2|^{2n_2}\cdots|\xi_N|^{2n_N},
\end{align}
where $i\neq j$. 
Comparison of this with
\begin{align}
\bra{\xi_2,\cdots,\xi_N}F_{i,\alpha,\beta}F_{j,\beta,\alpha}\ket{\xi_2,\cdots,\xi_N}
=&\sum_{n_2,\cdots,n_N=0}^\infty\frac{|\xi_2|^{2n_2}\cdots|\xi_N|^{2n_N}}{(n_2!)^2\cdots (n_N!)^2}\notag\\
&\times\bra{0}(\eta_{N,1})^{n_N}\cdots(\eta_{2,1})^{n_2}F_{i,\alpha,\beta}F_{j,\beta,\alpha}(\eta_{2,1}^\dag)^{n_2}\cdots (\eta_{N,1}^\dag)^{n_N}\ket{0}
\end{align}
yields the result for the spin correlation function of the primary $N$-color $\eta$-pairing state:
\begin{align}
\frac{\bra{\psi_{M_2,M_3,\cdots,M_N}}F_{i,\alpha,\beta}F_{j,\beta,\alpha}\ket{\psi_{M_2,M_3,\cdots,M_N}}}{\braket{\psi_{M_2,M_3,\cdots,M_N}|\psi_{M_2,M_3,\cdots,M_N}}}
=\frac{M_\alpha M_\beta}{N_{\mathrm{s}}(N_{\mathrm{s}}-1)},
\label{eq_spincorr_SU(N)_norm_app}
\end{align}
where $\alpha\neq\beta$ with $\alpha,\beta\neq1$. 
The spin correlation function $\langle F_{i,\alpha,1}F_{j,1,\alpha}\rangle\ (\alpha\neq1)$ can similarly be calculated. However, noticing that $F_{j,1,\alpha}\eta_{j,\beta,1}^\dag=0$, we obtain from Eq.~\eqref{eq_SUNeta_siterep} that
\begin{align}
\frac{\bra{\psi_{M_2,M_3,\cdots,M_N}}F_{i,\alpha,1}F_{j,1,\alpha}\ket{\psi_{M_2,M_3,\cdots,M_N}}}{\braket{\psi_{M_2,M_3,\cdots,M_N}|\psi_{M_2,M_3,\cdots,M_N}}}=0,
\end{align}
for $\alpha\neq 1$.

Finally, we calculate the spin correlation functions of the three-color $\eta$-pairing state. We exemplify the calculation using $\langle F_{i,2,1}F_{j,1,2}\rangle$ without loss of generality. For the coherent state \eqref{eq_generating_state_SU(3)}, we have
\begin{align}
\bra{a,b,c}F_{i,2,1}F_{j,1,2}\ket{a,b,c}=&(1+|a|^2+|b|^2+|c|^2)^{N_{\mathrm{s}}-2}\notag\\
&\times\bra{0_i}(1+a^*\eta_{i,1,2}+b^*\eta_{i,2,3}+c^*\eta_{i,3,1})F_{i,2,1}(1+a\eta_{i,1,2}^\dag+b\eta_{i,2,3}^\dag+c\eta_{i,3,1}^\dag)\ket{0_i}\notag\\
&\times\bra{0_j}(1+a^*\eta_{j,1,2}+b^*\eta_{j,2,3}+c^*\eta_{j,3,1})F_{j,1,2}(1+a\eta_{j,1,2}^\dag+b\eta_{j,2,3}^\dag+c\eta_{j,3,1}^\dag)\ket{0_j}\notag\\
=&(1+|a|^2+|b|^2+|c|^2)^{N_{\mathrm{s}}-2}\notag\\
&\times\bra{0_i}b^*c\eta_{i,2,3}F_{i,2,1}\eta_{i,3,1}^\dag\ket{0_i}\bra{0_j}c^*b\eta_{j,3,1}F_{j,1,2}\eta_{j,2,3}^\dag\ket{0_j}\notag\\
=&|b|^2|c|^2(1+|a|^2+|b|^2+|c|^2)^{N_{\mathrm{s}}-2}\notag\\
=&|b|^2|c|^2\sum_{k+l+m+n=N_{\mathrm{s}}-2}\binom{N_{\mathrm{s}}-2}{k,l,m,n}|a|^{2l}|b|^{2m}|c|^{2n},
\end{align}
and
\begin{align}
\bra{a,b,c}F_{i,2,1}F_{j,1,2}\ket{a,b,c}=&\sum_{l,m,n=0}^\infty\frac{|a|^{2l}|b|^{2m}|c|^{2n}}{(l!)^2(m!)^2(n!)^2}\bra{0}(\eta_{2,1})^n(\eta_{2,3})^m(\eta_{1,2})^lF_{i,2,1}F_{j,1,2}(\eta_{1,2}^\dag)^l(\eta_{2,3}^\dag)^m(\eta_{3,1}^\dag)^n\ket{0}.
\end{align}
Thus, combining the result for the squared norm [Eq.~\eqref{eq_SU3eta_norm_app}], we obtain the spin correlation function
\begin{align}
\frac{\bra{\psi_{l,m,n}^{(3)}}F_{i,2,1}F_{j,1,2}\ket{\psi_{l,m,n}^{(3)}}}{\braket{\psi_{l,m,n}^{(3)}|\psi_{l,m,n}^{(3)}}}=\frac{mn}{N_{\mathrm{s}}(N_{\mathrm{s}}-1)},
\label{eq_SU3eta_spincorr_app}
\end{align}
which does not depend on $l$ because on-site pairs of fermions in color 1 and 2 do not contribute to the spin correlation under consideration.
 
\end{widetext}

\section{Generalized $\eta$-pairing states and a bound on momentum distribution of doublons\label{sec_ODLRO_bound}}

The $\eta$-pairing state of the two-component Hubbard model saturates a bound on the momentum distribution of doublons \cite{Nakagawa21}.
We can derive an analogous bound that universally holds for any $N$-component lattice fermion systems. First, we note the following inequality
\begin{align}
\mathrm{Tr}[F^\dag F\rho]\leq \Lambda_2\leq\frac{N_{\mathrm{f}}(NN_{\mathrm{s}}-N_{\mathrm{f}}+2)}{NN_{\mathrm{s}}},
\label{eq_bound}
\end{align}
where 
\begin{gather}
F\equiv\sum_{i,j}\sum_{\alpha,\beta}f_{(i,\alpha,j,\beta)}c_{i,\alpha}c_{j,\beta}\ \ (f_{(i,j,\alpha,\beta)}\in\mathbb{C}),\label{eq_F1}\\
\sum_{i,j}\sum_{\alpha,\beta}|f_{(i,\alpha,j,\beta)}|^2=1,\label{eq_F2}
\end{gather}
$\rho$ is an arbitrary density matrix with $N_{\mathrm{f}}$ particles, and $\Lambda_2$ is the maximum eigenvalue of the two-particle reduced density matrix defined in Eq.~\eqref{eq_rho_2}. 
The first inequality in Eq.~\eqref{eq_bound} follows from the relation
\begin{align}
&\mathrm{Tr}[F^\dag F\rho]\notag\\
=&\sum_{i,j,k,l}\sum_{\alpha,\beta,\gamma,\delta}(f_{(i,\alpha,j,\beta)})^*(\rho_2)_{(i,\alpha,j,\beta),(k,\gamma,l,\delta)}f_{(k,\gamma,l,\delta)}\notag\\
=&\langle f,\rho_2f\rangle,
\end{align}
where $\langle v_1,v_2\rangle$ denotes the canonical inner product of two $(N_{\mathrm{s}}^2N^2)$-dimensional vectors $v_1$ and $v_2$. 
Note that the vector $f=\{ f_{(i,\alpha,j,\beta)}\}$ satisfies the normalization condition \eqref{eq_F2}. 
The second inequality in Eq.~\eqref{eq_bound} was derived by Yang \cite{Yang62}. The inequality \eqref{eq_bound} places an upper bound on the expectation value of a physical quantity of the form $F^\dag F$ with Eq.~\eqref{eq_F1}. 
However, for the momentum distribution of doublons, we can derive a more strict bound. 
To see this, we note that the momentum distribution $\langle d_{\bm{k},1,2}^\dag d_{\bm{k},1,2}\rangle$ of (1,2) doublons is obtained by setting
\begin{equation}
f_{(i,\alpha,j,\beta)}=\frac{1}{\sqrt{2N_{\mathrm{s}}}}\delta_{i,j}e^{-i\bm{k}\cdot\bm{R}_j}(\delta_{\alpha,2}\delta_{\beta,1}-\delta_{\alpha,1}\delta_{\beta,2}).
\end{equation}
Then, since $f_{(i,\alpha,j,\beta)}=0$ for $\alpha,\beta\neq 1,2$, we have
\begin{align}
\mathrm{Tr}[F^\dag F\rho]=\mathrm{Tr}_{1,2}[F^\dag F\rho^{(1,2)}],
\end{align}
where $\mathrm{Tr}_{\alpha}$ denotes the trace over component $\alpha$, and
\begin{equation}
\rho^{(1,2)}\equiv\mathrm{Tr}_{3,4,\cdots,N}[\rho]
\end{equation}
is a reduced density matrix of the components 1 and 2. Since $\rho^{(1,2)}$ can be regarded as a density matrix of a two-component system, we have
\begin{align}
\langle d_{\bm{k},1,2}^\dag d_{\bm{k},1,2}\rangle=\frac{1}{2}\mathrm{Tr}[F^\dag F\rho]\leq\frac{N_{\mathrm{f}}^{(1,2)}(2N_{\mathrm{s}}-N_{\mathrm{f}}^{(1,2)}+2)}{4N_{\mathrm{s}}}
\label{eq_bound_doublon}
\end{align}
for $\rho$ with $N_{\mathrm{f}}^{(1,2)}$ particles in the components 1 and 2. 
For the primary $N$-color $\eta$-pairing state \eqref{eq_SUNeta}, using Eq.~\eqref{eq_SUN_doublon_occ} and $N_{\mathrm{f}}^{(1,2)}=2M_2+M_3+\cdots+M_N$, we have
\begin{widetext}
\begin{align}
\langle d_{\bm{Q},1,2}^\dag d_{\bm{Q},1,2}\rangle=&\frac{M_2(N_{\mathrm{s}}-M_2-\cdots-M_N+1)}{N_{\mathrm{s}}}\notag\\
=&\frac{2M_2(2N_{\mathrm{s}}-2M_2-\cdots-2M_N+2)}{4N_{\mathrm{s}}}\notag\\
\leq&\frac{(2M_2+M_3+\cdots+M_N)(2N_{\mathrm{s}}-2M_2-M_3-\cdots-M_N+2)}{4N_{\mathrm{s}}}.
\end{align}
\end{widetext}
Thus, in contrast to the two-component case, the primary $N$-color $\eta$-pairing states do not saturate the bound \eqref{eq_bound_doublon} on the momentum distribution of doublons except for the case of $M_3=\cdots=M_N=0$. This is because the Pauli exclusion principle applies to different pairs due to the shared component 1.

\section{Calculation of entanglement entropy\label{sec_EE_calc}}

Here we detail how to calculate the entanglement entropy [Eq.~\eqref{eq_EE_SUNeta}] of the primary $N$-color $\eta$-pairing state $\ket{\psi_{M_2,\cdots,M_N}}$. To this end, we divide the entire system into subsystem $A$ with $N_{\mathrm{s},A}$ sites and subsystem $B$ with the remaining $N_{\mathrm{s},B}=N_{\mathrm{s}}-N_{\mathrm{s},A}$ sites. Accordingly, the $\eta$ operators are decomposed as
\begin{align}
\eta_{\alpha,\beta}^\dag=&\eta_{A,\alpha,\beta}^\dag + \eta_{B,\alpha,\beta}^\dag,
\end{align}
where
\begin{align}
\eta_{A,\alpha,\beta}^\dag\equiv&\sum_{j\in A}\eta_{j,\alpha,\beta}^\dag,\\
\eta_{B,\alpha,\beta}^\dag\equiv&\sum_{j\in B}\eta_{j,\alpha,\beta}^\dag.
\end{align}
With this decomposition, the primary $N$-color $\eta$-pairing state is written as
\begin{widetext}
\begin{align}
\ket{\psi_{M_2,\cdots,M_N}}=&(\eta_{A,2,1}^\dag+\eta_{B,2,1}^\dag)^{M_2}\cdots(\eta_{A,N,1}^\dag+\eta_{B,N,1}^\dag)^{M_N}\ket{0}\notag\\
=&\sum_{m_2=0}^{M_2}\cdots\sum_{m_N=0}^{M_N}\binom{M_2}{m_2}\cdots\binom{M_N}{m_N}(\eta_{A,2,1}^\dag)^{m_2}(\eta_{B,2,1}^\dag)^{M_2-m_2}\cdots(\eta_{A,N,1}^\dag)^{m_N}(\eta_{B,N,1}^\dag)^{M_N-m_N}\ket{0}\notag\\
=&\sum_{m_2=0}^{M_2}\cdots\sum_{m_N=0}^{M_N}\binom{M_2}{m_2}\cdots\binom{M_N}{m_N}\sqrt{C_A(m_2,\cdots,m_N)}\sqrt{C_B(M_2-m_2,\cdots,M_N-m_N)}\notag\\
&\times\ket{m_2,\cdots,m_N}_A\otimes\ket{M_2-m_2,\cdots,M_N-m_N}_B,
\end{align}
where $\binom{M}{m}\equiv \frac{M!}{m!(M-m)!}$ is the binomial coefficient,
\begin{equation}
\ket{m_2,\cdots,m_N}_{X}\equiv\frac{1}{\sqrt{C_X(m_2,\cdots,m_N)}}(\eta_{X,2,1}^\dag)^{m_2}\cdots(\eta_{X,N,1}^\dag)^{m_N}\ket{0_X}
\end{equation}
is a normalized state with $\ket{0_X}$ being the vacuum state in subsystem $X=A,B$, and
\begin{align}
C_X(m_2,\cdots,m_N)\equiv&\bra{0_X}(\eta_{X,N,1})^{m_N}\cdots(\eta_{X,2,1})^{m_2}(\eta_{X,2,1}^\dag)^{m_2}\cdots(\eta_{X,N,1}^\dag)^{m_N}\ket{0_X}\notag\\
=&\frac{N_{\mathrm{s},X}!m_2!\cdots m_N!}{(N_{\mathrm{s},X}-m_2-\cdots m_N)!}.
\label{eq_subsystem_norm}
\end{align}
Equation \eqref{eq_subsystem_norm} is obtained similarly as in Eq.~\eqref{eq_norm_SU(N)}. 
Hence, we obtain the Schmidt decomposition of the primary $N$-color $\eta$-pairing state as
\begin{align}
\frac{1}{\sqrt{\mathcal{N}(M_2,\cdots,M_N)}}\ket{\psi_{M_2,\cdots,M_N}}=&
\sum_{m_2+\cdots+m_N\leq N_{\mathrm{s},A}}\lambda(m_2,\cdots,m_N)\ket{m_2,\cdots,m_N}_A\otimes\ket{M_2-m_2,\cdots,M_N-m_N}_B,
\label{eq_Schmidt}
\end{align}
where $\mathcal{N}(M_2,\cdots,M_N)$ is the squared norm of the state [see Eq.~\eqref{eq_norm_SU(N)}] and the Schmidt coefficients are given by
\begin{align}
\lambda(m_2,\cdots,m_N)=&\left[\frac{C_A(m_2,\cdots,m_N)C_B(M_2-m_2,\cdots,M_N-m_N)}{\mathcal{N}(M_2,\cdots,M_N)}\right]^{1/2}\binom{M_2}{m_2}\cdots\binom{M_N}{m_N}\notag\\
=&\left[\binom{M_1}{m_1}\binom{M_2}{m_2}\cdots\binom{M_N}{m_N}\Bigr/\binom{N_{\mathrm{s}}}{N_{\mathrm{s},A}}\right]^{1/2}
\label{eq_Schmidt_coeff}
\end{align}
with $M_1\equiv N_{\mathrm{s}}-M_2-\cdots-M_N$ and $m_1\equiv N_{\mathrm{s},A}-m_2-\cdots-m_N$. Note that $\ket{m_2,\cdots,m_N}_A=0$ if $m_2+\cdots+m_N>N_{\mathrm{s},A}$. 
The Schmidt coefficients satisfy the normalization condition
\begin{equation}
\sum_{m_2+\cdots+m_N\leq N_{\mathrm{s},A}}[\lambda(m_2,\cdots,m_N)]^2=1.
\label{eq_Schmidt_normalization}
\end{equation}
This is confirmed by the formula
\begin{equation}
\sum_{m_1+m_2+\cdots+m_N=N_{\mathrm{s},A}}\binom{M_1}{m_1}\binom{M_2}{m_2}\cdots\binom{M_N}{m_N}=\binom{N_{\mathrm{s}}}{N_{\mathrm{s},A}},
\end{equation}
which can be obtained from the comparison of the coefficients of $x^{N_{\mathrm{s},A}}$ in
\begin{align}
(1+x)^{N_{\mathrm{s}}}=\sum_{n=0}^{N_{\mathrm{s}}}\binom{N_{\mathrm{s}}}{n}x^n
\end{align}
with those of
\begin{align}
(1+x)^{M_1}(1+x)^{M_2}\cdots (1+x)^{M_N}=&\sum_{m_1=0}^{M_1}\sum_{m_2=0}^{M_2}\cdots\sum_{m_N=0}^{M_N}\binom{M_1}{m_1}\binom{M_2}{m_2}\cdots\binom{M_N}{m_N}x^{m_1+m_2+\cdots+m_N}.
\end{align}
Equation \eqref{eq_Schmidt_coeff} is equivalent to the multivariate hypergeometric distribution. We note that the squared Schmidt coefficients can be written as
\begin{align}
[\lambda(m_2,\cdots,m_N)]^2=\binom{N_{\mathrm{s},A}}{m_1,m_2,\cdots,m_N}\binom{N_{\mathrm{s},B}}{l_1,M_2-m_2,\cdots,M_N-m_N}\Bigr/\binom{N_{\mathrm{s}}}{M_1,M_2,\cdots,M_N},
\label{eq_Schmidt_coeff2}
\end{align}
where $l_1\equiv N_{\mathrm{s},B}-(M_2-m_2)-\cdots-(M_N-m_N)$. 
From the Schmidt decomposition \eqref{eq_Schmidt}, the entanglement entropy of the primary $N$-color $\eta$-pairing state is given by
\begin{align}
S_A=&-\mathrm{Tr}_A[\rho_A\log\rho_A]\notag\\
=&-\sum_{m_2+\cdots+m_N\leq N_{\mathrm{s},A}}[\lambda(m_2,\cdots,m_N)]^2\log[\lambda(m_2,\cdots,m_N)]^2,
\label{eq_EE_Schmidt}
\end{align}
where
\begin{align}
\rho_A=\mathrm{Tr}_B\left[\frac{\ket{\psi_{M_2,\cdots,M_N}}\bra{\psi_{M_2,\cdots,M_N}}}{\mathcal{N}(m_2,\cdots,m_N)}\right]
\end{align}
is the reduced density matrix of subsystem $A$.

To examine the scaling law of the entanglement entropy, we first fix $N_{\mathrm{s},A}$ and the densities $\nu_{\alpha}\equiv M_\alpha/N_{\mathrm{s}}\ (\alpha=2,\cdots,N)$ of $\eta$ pairs, and then take the thermodynamic limit $N_{\mathrm{s}}\to\infty$. After that, we examine how the entanglement entropy scales with the size $N_{\mathrm{s},A}$ of the subsystem (see also Ref.~\cite{Vafek17}). By using the Stirling formula, the multinomial coefficient is approximated as
\begin{align}
\binom{M}{n_1,\cdots,n_N}
\sim&
\sqrt{\frac{M}{(2\pi)^{N-1}n_1\cdots n_N}}\left(\frac{M}{n_1}\right)^{n_1}\cdots\left(\frac{M}{n_N}\right)^{n_N},
\end{align}
where $n_1+\cdots+n_N=M$. 
Thus, from Eq.~\eqref{eq_Schmidt_coeff2}, we have
\begin{align}
[\lambda(m_2,\cdots,m_N)]^2
\sim&\sqrt{\frac{N_{\mathrm{s},A}N_{\mathrm{s},B}M_2\cdots M_N}{(2\pi)^{N-1}m_1\cdots m_Nl_1\cdots l_NN_{\mathrm{s}}}}\notag\\
&\times \left(\frac{N_{\mathrm{s},A}M_1}{m_1N_{\mathrm{s}}}\right)^{m_1}\cdots\left(\frac{N_{\mathrm{s},A}M_N}{m_NN_{\mathrm{s}}}\right)^{m_N}\left(\frac{N_{\mathrm{s},B}M_1}{l_1N_{\mathrm{s}}}\right)^{l_1}\cdots\left(\frac{N_{\mathrm{s},B}M_N}{m_NN_{\mathrm{s}}}\right)^{l_N},
\end{align}
where $l_\alpha\equiv M_\alpha-m_\alpha\ (\alpha=2,\cdots,N)$. Then, by taking the limit of $N_{\mathrm{s}}\to\infty$ with $\nu_\alpha$ held fixed, we obtain
\begin{align}
[\lambda(m_2,\cdots,m_N)]^2\sim\sqrt{\frac{N_{\mathrm{s},A}}{(2\pi)^{N-1}m_1\cdots m_N}}\left(\frac{\nu_1 N_{\mathrm{s},A}}{m_1}\right)^{m_1}\cdots\left(\frac{\nu_N N_{\mathrm{s},A}}{m_N}\right)^{m_N},
\end{align}
where $\nu_1\equiv M_1/N_{\mathrm{s}}$. 
Here, we introduce
\begin{equation}
x_\alpha\equiv\frac{m_\alpha-\nu_\alpha N_{\mathrm{s},A}}{\sqrt{N_{\mathrm{s},A}}}\ \ (\alpha=2,\cdots,N).
\end{equation}
If $N_{\mathrm{s},A}\gg1$, then we can expand the logarithm of the Schmidt coefficient up to the second order of $x_\alpha$ as
\begin{align}
\log[\lambda(m_2,\cdots,m_N)]^2\sim-\frac{1}{2}\log\Bigl[(2\pi N_{\mathrm{s},A})^{N-1}\nu_1\nu_2\cdots \nu_N\Bigr]-\sum_{\alpha=2}^N\frac{x_\alpha^2}{2\nu_\alpha}-\sum_{\alpha,\beta=2}^N\frac{x_\alpha x_\beta}{2\nu_1},
\end{align}
and we thus obtain
\begin{equation}
[\lambda(m_2,\cdots,m_N)]^2\sim\frac{1}{\sqrt{(2\pi N_{\mathrm{s},A})^{N-1}\nu_1\nu_2\cdots\nu_N}}\exp\left[-\sum_{\alpha=2}^N\frac{x_\alpha^2}{2\nu_\alpha}-\sum_{\alpha,\beta=2}^N\frac{x_\alpha x_\beta}{2\nu_1}\right].
\end{equation}
After replacing the sum in Eq.~\eqref{eq_EE_Schmidt} with an integral as
\begin{align}
\sum_{m_2+\cdots+m_N\leq N_{\mathrm{s},A}}\to
(N_{\mathrm{s},A})^{(N-1)/2}\int_{-\infty}^\infty dx_2\cdots dx_N,
\end{align}
we arrive at an asymptotic form of the entanglement entropy of the primary $N$-color $\eta$-pairing state:
\begin{align}
S_A
\sim&\frac{1}{2}\log[(2\pi N_{\mathrm{s},A})^{N-1}\nu_1\cdots\nu_N]\int_{-\infty}^\infty dx_2\cdots dx_N\frac{1}{\sqrt{(2\pi)^{N-1}\nu_1\nu_2\cdots\nu_N}}\exp\left[-\sum_{\alpha=2}^N\frac{x_\alpha^2}{2\nu_\alpha}-\sum_{\alpha,\beta=2}^N\frac{x_\alpha x_\beta}{2\nu_1}\right]\notag\\
&+\int_{-\infty}^\infty dx_2\cdots dx_N\frac{\sum_{\alpha=2}^N\frac{x_\alpha^2}{2\nu_\alpha}+\sum_{\alpha,\beta=2}^N\frac{x_\alpha x_\beta}{2\nu_1}}{\sqrt{(2\pi)^{N-1}\nu_1\nu_2\cdots\nu_N}}\exp\left[-\sum_{\alpha=2}^N\frac{x_\alpha^2}{2\nu_\alpha}-\sum_{\alpha,\beta=2}^N\frac{x_\alpha x_\beta}{2\nu_1}\right]\notag\\
=&\frac{N-1}{2}\log N_{\mathrm{s},A}+\mathrm{const.},
\label{eq_EE_app}
\end{align}
which shows a logarithmic sub-volume-law dependence on the size $N_{\mathrm{s},A}$ of the subsystem. In deriving the last equality in Eq.~\eqref{eq_EE_app}, we have used the normalization condition
\begin{align}
1=&\sum_{m_2+\cdots+m_N\leq N_{\mathrm{s},A}}[\lambda(m_2,\cdots,m_N)]^2\notag\\
\sim&\int_{-\infty}^\infty dx_2\cdots dx_N\frac{1}{\sqrt{(2\pi)^{N-1}\nu_1\nu_2\cdots\nu_N}}\exp\left[-\sum_{\alpha=2}^N\frac{x_\alpha^2}{2\nu_\alpha}-\sum_{\alpha,\beta=2}^N\frac{x_\alpha x_\beta}{2\nu_1}\right].
\end{align}
Compared to the result for Yang's $\eta$-pairing state [$N=2$ in Eq.~\eqref{eq_EE_app}] in Ref.~\cite{Vafek17}, we find that the entanglement entropy \eqref{eq_EE_app} is increased by a factor of $N-1$, which corresponds to the number of different pairs in the primary $N$-color $\eta$-pairing state.

The entanglement entropy of the three-color $\eta$-pairing state [Eq.~\eqref{eq_SU3eta}] can similarly be calculated. Since this state contains three different pairs, we obtain the entanglement entropy by setting $N-1=3$ in Eq.~\eqref{eq_EE_app}.

\end{widetext}

\section{Energy condition for quantum many-body scars\label{sec_scar_app}}

Here we derive a sufficient condition under which the primary $N$-color $\eta$-pairing eigenstates lie in the bulk of the energy eigenspectrum of the MHM. This condition, together with the sub-volume law entanglement entropy of these states, ensures that they are quantum many-body scar states. For simplicity, we assume that the internal symmetry of the MHM is U(1)$^N$ (see Sec.~\ref{sec_model}). In this case, the Hilbert space of the MHM is divided into subsectors labeled by particle numbers in each component. As a reference state that belongs to the same subsector as that of the primary $N$-color $\eta$-pairing state $\ket{\psi_{M_2,\cdots,M_N}}$, we take an SU($N$) ferromagnetic eigenstate of the MHM [see Eq.~\eqref{eq_SUNferro_ex}]:
\begin{equation}
(F_{2,1})^{M_2}\cdots(F_{N,1})^{M_N}c_{n_1,1}^\dag \cdots c_{n_{N_{\mathrm{f}}},1}^\dag\ket{0},
\label{eq_SUNferro_ex_app}
\end{equation}
where $N_{\mathrm{f}}=2(M_2+\cdots+M_N)$ is the total number of particles. The energy eigenvalue of the primary $N$-color $\eta$-pairing state is given by $M_2U_{2,1}+\cdots+M_NU_{N,1}$, while that of the SU($N$) ferromagnetic state \eqref{eq_SUNferro_ex_app} is given by $\epsilon_{n_1}+\cdots+\epsilon_{n_{N_{\mathrm{f}}}}$. Let
\begin{equation}
\epsilon_{\mathrm{min}}\equiv\mathrm{min}_{\{ n_k\}_{k=1,\cdots,N_{\mathrm{f}}}}\frac{1}{N_{\mathrm{s}}}\sum_{k=1}^{N_{\mathrm{f}}}\epsilon_{n_k}
\end{equation}
and
\begin{equation}
\epsilon_{\mathrm{max}}\equiv\mathrm{max}_{\{ n_k\}_{k=1,\cdots,N_{\mathrm{f}}}}\frac{1}{N_{\mathrm{s}}}\sum_{k=1}^{N_{\mathrm{f}}}\epsilon_{n_k}
\end{equation}
be the minimum and maximum values of the energy density of the SU($N$) ferromagnetic states, respectively. 
Then, if the energy density of the primary $N$-color $\eta$-pairing state satisfies the inequality
\begin{align}
\epsilon_{\mathrm{min}}<\nu_2U_{2,1}+\cdots+\nu_NU_{N,1}<\epsilon_{\mathrm{max}},
\label{eq_scar_ineq}
\end{align}
in the thermodynamic limit (i.e., $N_{\mathrm{s}}\to\infty$ with $N_{\mathrm{f}}/N_{\mathrm{s}}$ being fixed), this state lies in the bulk of the eigenspectrum. Here, $\nu_\alpha\equiv M_\alpha/N_{\mathrm{s}}$ denotes the density of the pairs. 
The inequality \eqref{eq_scar_ineq} can be satisfied by appropriately tuning the interaction parameters $U_{2,1},\cdots,U_{N,1}$, unless the single-particle spectrum is completely flat (i.e., $\epsilon_{\mathrm{min}}=\epsilon_{\mathrm{max}}$). For the translationally invariant case, explicit forms of $\epsilon_{\mathrm{min}}$ and $\epsilon_{\mathrm{max}}$ can be found in Ref.~\cite{Moudgalya20}. For the three-component Hubbard model, we can similarly place the three-color $\eta$-pairing eigenstate \eqref{eq_SU3eta} with eigenvalue $lU_{1,2}+mU_{2,3}+nU_{3,1}$ in the bulk of the eigenspectrum by appropriately tuning the interaction parameters.

\section{Extension to other Hubbard-like models\label{sec_PAM}}

The $N$-color $\eta$-pairing mechanism can be generalized to some other Hubbard-like models. For example, the MHM \eqref{eq_Hasym} can be generalized as
\begin{align}
\tilde{H}=&\tilde{T}+\tilde{V},\label{eq_tildeH}\\
\tilde{T}=&T,\\
\tilde{V}=& \sum_j\sum_{\alpha}\epsilon_{j,\alpha} n_{j,\alpha}
+\frac{1}{2}\sum_j\sum_{\alpha\neq\beta}U_{j,\alpha,\beta}n_{j,\alpha}n_{j,\beta},
\end{align}
to allow site-dependent potentials $\epsilon_{j,\alpha}$ and site-dependent interaction strengths $U_{j,\alpha,\beta}$ (see also Ref.~\cite{Moudgalya20}). We set $U_{j,\beta,\alpha}=U_{j,\alpha,\beta}$. 
By replacing $V$ with $\tilde{V}$ in the proof given in Sec.~\ref{sec_N-color}, 
one can easily confirm that the $\eta$-pairing state 
\begin{equation}
(\eta_{2,1}^\dag)^{M_2}\cdots(\eta_{N,1}^\dag)^{M_N}\tilde{c}_{n_1,1}^\dag\cdots \tilde{c}_{n_r,1}^\dag\ket{0}
\label{eq_SUNeta_tilde}
\end{equation}
is an eigenstate of the generalized Hubbard model \eqref{eq_tildeH} if
\begin{equation}
U_{j,\alpha,\beta}+\epsilon_{j,\alpha}+\epsilon_{j,\beta}=C_{\alpha,\beta}
\label{eq_tilde_cond}
\end{equation}
holds for $(\alpha,\beta)=(2,1),\cdots,(N,1)$, where $C_{\alpha,\beta}$ is a site-independent constant. 
Here, $\tilde{c}_{n,\alpha}$ denotes an annihilation operator of a single-particle eigenstate of 
\begin{align}
\tilde{H}^{(\alpha)}=&-\sum_{\langle i,j\rangle}t_{i,j}(c_{i,\alpha}^\dag c_{j,\alpha}+\mathrm{H.c.})+\sum_j\epsilon_{j,\alpha} n_{j,\alpha}.
\end{align}
Similarly, in the three-component case, the three-color $\eta$-pairing state \eqref{eq_SU3eta} is an eigenstate of $\tilde{H}$ if $U_{j,\alpha,\beta}+\epsilon_{j,\alpha}+\epsilon_{j,\beta}$ does not depend on $j$ for $(\alpha,\beta)=(1,2), (2,3), (3,1)$.

The SU($N$) ferromagnetic state
\begin{equation}
(F_{2,1})^{M_2}\cdots(F_{N,1})^{M_N}\tilde{c}_{n_1,1}^\dag\cdots \tilde{c}_{n_r,1}^\dag\ket{0}
\label{eq_SUNferro_tilde}
\end{equation}
is an eigenstate of the generalized Hubbard model \eqref{eq_tildeH} only when $\epsilon_{j,\alpha}$ does not depend on $\alpha$. Therefore, the SU($N$) ferromagnetic state \eqref{eq_SUNferro_tilde} and the generalized $\eta$-pairing state \eqref{eq_SUNeta_tilde} do not form simultaneous eigenstates of $\tilde{H}$ except for the SU($N$)-symmetric case. The $\eta$-pairing states in the form of \eqref{eq_SUNeta_tilde} constitute quantum many-body scars and an integrable sector of the generalized Hubbard model \eqref{eq_tildeH} if the SU($N$) symmetry is broken and Eq.~\eqref{eq_tilde_cond} is satisfied. The SU($N$) ferromagnetic states of the form \eqref{eq_SUNferro_tilde} also constitute quantum many-body scars and an integrable sector if the SU($N$) symmetry is broken solely by the interaction strength $U_{j,\alpha,\beta}$.

We next show that a periodic Anderson model (PAM) \cite{Newns87, Millis87}
\begin{align}
H_{\mathrm{PAM}}=&-\sum_{\langle i,j\rangle}\sum_{\alpha=1}^N t_{i,j}^{(c)}(c_{i,\alpha}^\dag c_{j,\alpha}+\mathrm{H.c.})\notag\\
&+\sum_j\sum_\alpha v_j(c_{j,\alpha}^\dag f_{j,\alpha}+\mathrm{H.c.}),\notag\\
&+\sum_j\sum_{\alpha}\epsilon_{f,j,\alpha}f_{j,\alpha}^\dag f_{j,\alpha}+\sum_j\sum_{\alpha<\beta}U_{j,\alpha,\beta}n_{j,\alpha}^{(f)}n_{j,\beta}^{(f)},
\label{eq_PAM}
\end{align}
has generalized $\eta$-pairing eigenstates. Here, $c_{j,\alpha}$ $(f_{j,\alpha})$ is an annihilation operator of itinerant (localized) fermions at site $j$ and spin $\alpha$, and $n_{j,\alpha}^{(f)}=f_{j,\alpha}^\dag f_{j,\alpha}$. We assume the same lattice structure as that of the MHM. The hopping amplitude of itinerant fermions, the energy level of localized fermions, the hybridization strength, and the interaction strength for localized fermions are denoted by $t_{i,j}^{(c)},\epsilon_{f,j,\alpha},v_j$, and $U_{j,\alpha,\beta}$, respectively. We set $U_{j,\beta,\alpha}=U_{j,\alpha,\beta}$. In the case of $N=2$, the PAM \eqref{eq_PAM} has an $\eta$-pairing symmetry and the corresponding $\eta$-pairing eigenstates \cite{Nishino93, Tsunetsugu97}. 
We note that the PAM can be regarded as a special case of Eq.~\eqref{eq_tildeH}. 
We define generalized $\eta$ operators for the PAM by
\begin{equation}
\tilde{\eta}_{\alpha,\beta}^\dag=\sum_je^{i\bm{Q}\cdot\bm{R}_j}(c_{j,\alpha}^\dag c_{j,\beta}^\dag-f_{j,\alpha}^\dag f_{j,\beta}^\dag)
\end{equation}
for $\alpha,\beta=1,\cdots,N$. 
Then, the $\eta$-pairing state
\begin{equation}
(\tilde{\eta}_{2,1}^\dag)^{M_2}\cdots(\tilde{\eta}_{N,1}^\dag)^{M_N}a_{n_1,1}^\dag\cdots a_{n_r,1}^\dag\ket{0}
\label{eq_SUNeta_PAM}
\end{equation}
is an eigenstate of the PAM \eqref{eq_PAM} if
\begin{equation}
U_{j,\alpha,\beta}+\epsilon_{f,j,\alpha}+\epsilon_{f,j,\beta}=0
\label{eq_PAM_cond}
\end{equation}
is satisfied for all $j$ and $(\alpha,\beta)=(2,1),\cdots,(N,1)$. Here, $a_{n,\alpha}$ denotes an annihilation operator of a single-particle eigenstate of the Hamiltonian
\begin{align}
H_{\mathrm{PAM}}^{(\alpha)}=&-\sum_{\langle i,j\rangle}t_{i,j}^{(c)}(c_{i,\alpha}^\dag c_{j,\alpha}+\mathrm{H.c.})\notag\\
&+\sum_j v_j(c_{j,\alpha}^\dag f_{j,\alpha}+\mathrm{H.c.})+\sum_j\epsilon_{f,j,\alpha}f_{j,\alpha}^\dag f_{j,\alpha}.
\end{align}
In the case of $N=3$, a state
\begin{equation}
(\tilde{\eta}_{1,2}^\dag)^l(\tilde{\eta}_{2,3}^\dag)^m(\tilde{\eta}_{3,1}^\dag)^n\ket{0}
\end{equation}
is also an eigenstate of the PAM \eqref{eq_PAM} if Eq.~\eqref{eq_PAM_cond} is satisfied for all $j,\alpha,\beta\ (\alpha\neq\beta)$.

\section{Momentum distribution in the integrable sectors\label{sec_mom_dist_integ}}

Here we calculate the spin-resolved momentum distribution [Eqs.~\eqref{eq_mom_1_eta}-\eqref{eq_mom_alpha_ferro}] for the states in the integrable sectors. We begin with the commutation relation between the observable $O_{\bm{k},\alpha}=c_{\bm{k},\alpha}^\dag c_{\bm{k},\alpha}$ and the $\eta$ operator:
\begin{align}
[\eta_{\alpha,\beta},O_{\bm{k},\gamma}]=&c_{\bm{Q}-\bm{k},\beta}c_{\bm{k},\alpha}\delta_{\alpha,\gamma}+c_{\bm{k},\beta}c_{\bm{Q}-\bm{k},\alpha}\delta_{\beta,\gamma}.
\label{eq_comm_eta_O}
\end{align}
Since the right-hand side of Eq.~\eqref{eq_comm_eta_O} commutes with $\eta_{\alpha,\beta}$, we have
\begin{align}
&[(\eta_{\alpha,\beta})^n,O_{\bm{k},\gamma}]\notag\\
=&n(c_{\bm{Q}-\bm{k},\beta}c_{\bm{k},\alpha}\delta_{\alpha,\gamma}+c_{\bm{k},\beta}c_{\bm{Q}-\bm{k},\alpha}\delta_{\beta,\gamma})(\eta_{\alpha,\beta})^{n-1}.
\label{eq_comm_etan_O}
\end{align}

Let $\ket{\phi_1}$ be a spin-polarized state with $n$ particles in component 1.  
We define the operators 
\begin{align}
A=&(\eta_{N,1})^{M_N}\cdots(\eta_{2,1})^{M_2},\\
A_\alpha=&(\eta_{N,1})^{M_N}\cdots(\eta_{\alpha,1})^{M_\alpha-1}\cdots(\eta_{2,1})^{M_2},
\end{align}
and assume $M_2+\cdots+M_N+n\leq N_{\mathrm{s}}$ to ensure that $A^\dag\ket{\phi_1}$ does not vanish. 
Using Eq.~\eqref{eq_comm_etan_O}, we obtain
\begin{align}
AO_{\bm{k},1}=&O_{\bm{k},1}A+\sum_{\alpha=2}^N M_\alpha c_{\bm{k},1}c_{\bm{Q}-\bm{k},\alpha}A_\alpha,
\end{align}
and
\begin{align}
AO_{\bm{k},\alpha}=O_{\bm{k},\alpha}A+M_\alpha c_{\bm{Q}-\bm{k},1}c_{\bm{k},\alpha}A_\alpha
\end{align}
for $\alpha\neq 1$.

To calculate $A A^\dag\ket{\phi_1}$ and $A_\alpha A^\dag\ket{\phi_1}$, we utilize the fact that the $\eta$ operators satisfy an su(2) subalgebra:
\begin{subequations}
\begin{align}
[\eta_{\alpha,\beta}^\dag,\eta_{\alpha,\beta}]=&2\eta_{\alpha,\beta}^z,
\label{eq_etaSU(2)algpm}\\
[\eta_{\alpha,\beta}^z,\eta_{\alpha,\beta}^\dag]=&\eta_{\alpha,\beta}^\dag,
\label{eq_etaSU(2)algpz}\\
[\eta_{\alpha,\beta}^z,\eta_{\alpha,\beta}]=&-\eta_{\alpha,\beta},
\label{eq_etaSU(2)algmz}
\end{align}
\end{subequations}
where
\begin{equation}
\eta_{\alpha,\beta}^z\equiv\frac{1}{2}\sum_j(n_{j,\alpha}+n_{j,\beta}-1).
\end{equation}
From Eqs.~\eqref{eq_etaSU(2)algpm} and \eqref{eq_etaSU(2)algpz}, we have
\begin{align}
&\eta_{\alpha,\beta}(\eta_{\alpha,\beta}^\dag)^m\notag\\
=&(\eta_{\alpha,\beta}^\dag\eta_{\alpha,\beta}-2\eta_{\alpha,\beta}^z)(\eta_{\alpha,\beta}^\dag)^{m-1}\notag\\
=&\{ \eta_{\alpha,\beta}^\dag(\eta_{\alpha,\beta}^\dag\eta_{\alpha,\beta}-2\eta_{\alpha,\beta}^z)-2\eta_{\alpha,\beta}^\dag(\eta_{\alpha,\beta}^z+1)\}(\eta_{\alpha,\beta}^\dag)^{m-2}\notag\\
=&\{ (\eta_{\alpha,\beta}^\dag)^2\eta_{\alpha,\beta}-\eta_{\alpha,\beta}^\dag(4\eta_{\alpha,\beta}^z+2)\}(\eta_{\alpha,\beta}^\dag)^{m-2}\notag\\
=&\cdots\notag\\
=&(\eta_{\alpha,\beta}^\dag)^m\eta_{\alpha,\beta}-(\eta_{\alpha,\beta}^\dag)^{m-1}\{ 2m\eta_{\alpha,\beta}^z+m(m-1)\}.
\label{eq_etam_etapn}
\end{align}
By repeatedly using Eq.~\eqref{eq_etam_etapn}, we obtain
\begin{align}
(\eta_{2,1})^{M_2}A^\dag\ket{\phi_1}=&\frac{M_2!(N_{\mathrm{s}}-M_3-\cdots-M_N-n)!}{(N_{\mathrm{s}}-M_2-M_3-\cdots-M_N-n)!}\notag\\
&\times(\eta_{3,1}^\dag)^{M_3}\cdots(\eta_{N,1}^\dag)^{M_N}\ket{\phi_1},
\end{align}
which leads to
\begin{align}
A A^\dag\ket{\phi_1}=&\frac{M_2!\cdots M_N!(N_{\mathrm{s}}-n)!}{(N_{\mathrm{s}}-M_2-\cdots-M_N-n)!}\ket{\phi_1},
\label{eq_A_phi1_appendix}
\end{align}
and
\begin{align}
A_\alpha A^\dag\ket{\phi_1}=&\frac{M_2!\cdots M_N!(N_{\mathrm{s}}-n-1)!}{(N_{\mathrm{s}}-M_2-\cdots-M_N-n)!}\eta_{\alpha,1}^\dag\ket{\phi_1}.
\label{eq_A_phi1_appendix2}
\end{align}
Thus, we end up with
\begin{align}
\frac{\bra{\phi_1}AO_{\bm{k},1}A^\dag\ket{\phi_1}}{\bra{\phi_1}AA^\dag\ket{\phi_1}}=&\langle O_{\bm{k},1}\rangle_1+\sum_{\alpha=2}^N\frac{M_\alpha}{N_{\mathrm{s}}-n}\langle 1-O_{\bm{k},1}\rangle_1,
\label{eq_mom_1_eta_appendix}
\end{align}
and
\begin{align}
\frac{\bra{\phi_1}AO_{\bm{k},\alpha}A^\dag\ket{\phi_1}}{\bra{\phi_1}AA^\dag\ket{\phi_1}}=&\frac{M_\alpha}{N_{\mathrm{s}}-n}\langle 1-O_{\bm{Q}-\bm{k},1}\rangle_1
\label{eq_mom_alpha_eta_appendix}
\end{align}
for $\alpha\neq 1$, where
\begin{equation}
\langle \cdots\rangle_1\equiv\frac{\bra{\phi_1}\cdots\ket{\phi_1}}{\braket{\phi_1|\phi_1}}.
\end{equation}
This completes the derivation of Eqs.~\eqref{eq_mom_1_eta} and \eqref{eq_mom_alpha_eta}. 
Physically, the first term on the right-hand side of Eq.~\eqref{eq_mom_1_eta_appendix} is the contribution from unpaired fermions in $\ket{\phi_1}$, and the other terms on the right-hand sides of Eqs.~\eqref{eq_mom_1_eta_appendix} and \eqref{eq_mom_alpha_eta_appendix} are the contributions from fermions in $\eta$ pairs.

Next, we consider the operator
\begin{equation}
B=(F_{1,N})^{M_N}\cdots(F_{1,2})^{M_2},
\end{equation}
for which we assume $M_2+\cdots+M_N\leq n$ to ensure that $B^\dag\ket{\phi_1}$ does not vanish. 
To calculate the momentum distribution for the state $B^\dag\ket{\phi_1}$, we employ the generalized Shiba transformations:
\begin{align}
S_1c_{\bm{k},\alpha}S_1^\dag=&\delta_{\alpha,1}c_{\bm{Q}-\bm{k},1}^\dag+(1-\delta_{\alpha,1})c_{\bm{k},\alpha},\\
S_1O_{\bm{k},\alpha}S_1^\dag=&\delta_{\alpha,1}(1-O_{\bm{Q}-\bm{k},1})+(1-\delta_{\alpha,1})O_{\bm{k},\alpha},\\
S_1BS_1^\dag=&A.
\end{align}
We obtain Eqs.~\eqref{eq_mom_1_ferro} and \eqref{eq_mom_alpha_ferro} as
\begin{align}
\frac{\bra{\phi_1}BO_{\bm{k},1}B^\dag\ket{\phi_1}}{\bra{\phi_1}BB^\dag\ket{\phi_1}}=&\frac{\bra{\phi_1}S_1^\dag A(1-O_{\bm{Q}-\bm{k},1})A^\dag S_1\ket{\phi_1}}{\bra{\phi_1}S_1^\dag AA^\dag S_1\ket{\phi_1}}\notag\\
=&\langle O_{\bm{k},1}\rangle_1-\sum_{\alpha=2}^N\frac{M_\alpha}{n}\langle O_{\bm{k},1}\rangle_1,
\end{align}
and
\begin{align}
\frac{\bra{\phi_1}BO_{\bm{k},\alpha}B^\dag\ket{\phi_1}}{\bra{\phi_1}BB^\dag\ket{\phi_1}}=&\frac{\bra{\phi_1}S_1^\dag AO_{\bm{k},\alpha}A^\dag S_1\ket{\phi_1}}{\bra{\phi_1}S_1^\dag AA^\dag S_1\ket{\phi_1}}\notag\\
=&\frac{M_\alpha}{n}\langle O_{\bm{k},1}\rangle_1,
\end{align}
for $\alpha\neq 1$. Note that $S_1\ket{\phi_1}$ is a spin-polarized state with $N_{\mathrm{s}}-n$ particles. The momentum distribution satisfies
\begin{equation}
\sum_{\alpha=1}^N\frac{\bra{\phi_1}BO_{\bm{k},\alpha}B^\dag\ket{\phi_1}}{\bra{\phi_1}BB^\dag\ket{\phi_1}}=\langle O_{\bm{k},1}\rangle_1.
\end{equation}

\bibliography{SUNeta_ref.bib}

\end{document}